\documentclass[final,5p,times,twocolumn,authoryear]{elsarticle}

\usepackage{amssymb}
\usepackage{amsmath}
\usepackage[caption=true]{subfig}
\usepackage{multirow}
\usepackage{algorithm}
\usepackage{enumerate}
\usepackage{algpseudocode}
\usepackage{float}
\usepackage[colorlinks=true,linkcolor=black, citecolor=blue, urlcolor=blue]{hyperref}
\usepackage{breqn}
\usepackage{widetext}
\usepackage{bbm}
\usepackage{cleveref}
\usepackage{mathtools}
\newcommand{\iu}{\mathrm{i}\mkern1mu}

\numberwithin{equation}{section}

\journal{arXiv}

\begin{document}

\begin{frontmatter}

\title{Fourier instantaneous estimators and the Epps effect}

\author[uct-sta]{Patrick Chang}
\ead{CHNPAT005@myuct.ac.za}
\address[uct-sta]{Department of Statistical Science, University of Cape Town, Rondebosch 7700, South Africa}

\begin{abstract}
We compare the Malliavin-Mancino and Cuchiero-Teichmann Fourier instantaneous estimators to investigate the impact of the Epps effect arising from asynchrony in the instantaneous estimates. We demonstrate the instantaneous Epps effect under a simulation setting and provide a simple method to ameliorate the effect. We find that using the previous tick interpolation in the Cuchiero-Teichmann estimator results in unstable estimates when dealing with asynchrony, while the ability to bypass the time domain with the Malliavin-Mancino estimator allows it to produce stable estimates and is therefore better suited for ultra-high frequency finance. An empirical analysis using Trade and Quote data from the Johannesburg Stock Exchange illustrates the instantaneous Epps effect and how the intraday correlation dynamics can vary between days for the same equity pair.
\end{abstract}


\end{frontmatter}

% \tableofcontents
% \linenumbers

%% main text
\section{Introduction}\label{sec:intro}

The study of the integrated volatility/co-volatility is a well explored field with a plethora of estimators suited for various purposes. To name a few, we have the classical Realised Volatility (RV) for the continuous-time It\^{o} semi-martingale. Jump robust extensions such as the Bi- and Multi-Power variations studied by \cite{BS2004,BGJPS2006,BSW2006}, which has later been generalised by replacing the power function with different specifications (See \cite{JACOD2008,TT2012,PV2009}). Extensions to deal with asynchrony such as the cumulative estimator proposed by \cite{HY2005}, and Fourier estimators such as that proposed by \cite{MM2002,MM2009}. On the other hand, the study of the instantaneous volatility/co-volatility is a relatively new field. Most of the instantaneous estimators rely on the numerical derivative of the integrated covariance (See \cite{APPS2012,BR2018,MZ2008}), which correspond to the {\it local realised volatility estimator}. Due to the differentiation, this can lead to strong numerical instabilities.

Instantaneous estimators based on Fourier transforms present several advantages over differentiation based local RV estimators. First, it relies on the integration of the time series rather than the differentiation which makes it numerically stable. Second, the reconstruction of the instantaneous covariance relies on harmonics. Therefore, the degree of smoothness is determined by the cutting frequency M for the reconstruction. Finally, the methods provide {\it global} estimation of the spot volatility. Meaning the volatilities are estimated with similar accuracy at any time $t$ within the interior of the domain \citep{MRS2017}. 

The spot volatility has a range of applications. The integrated stochastic volatility of volatility can be estimated by using power variation estimators on the reconstructed instantaneous volatility path \citep{CT2015}, or approaches that rely only on integrated quantities \citep{SCM2015}. Another parameter which can also be obtained from the spot estimates is the integrated leverage investigated in \cite{CS2015}. Now both the instantaneous volatility of volatility and spot leverage can be obtained as pointed out in \cite{MRS2017}. Other applications include predicting the intraday Value at Risk (VaR) \cite{MR2015}, calibrating agent based models such as the model in \cite{TIG2012}, in the medical field to approximate heart rate variability \cite{MR2015}, or in the environmental field to studying ecological systems and changes of biodiversity variability in its variation in space and time \citep{CAPP2018,GCPC2013}.

Here we are interested in comparing the Malliavin-Mancino \citep{MM2009} and the Cuchiero-Teichmann \citep{CT2015} Fourier instantaneous estimators and understanding the impact of their cutting frequencies to gain insight into the impact of the Epps effect \citep{EPPS1979} on the instantaneous estimators. The Epps effect is the decay of correlations as the sampling intervals decrease and there is an abundance of literature investigating the causes (See \cite{RENO2001,TK2007,TK2009,MSG2010,MMZ2011}) and corrections (See \cite{MSG2011,HY2005,ZHANG2010,BLPP2019,PCEPTG2020b}), but all in the case of the integrated correlation. To the best of my knowledge, apart from the work by \cite{MI2010}, the impact on the instantaneous correlation caused by the Epps effect has not been explored. This work aims to remedy this by serving as a preliminary investigation into the instantaneous Epps effect as new results in applied harmonic analysis \citep{GS2016,Guar2018}.

To this end, we compare the two estimators for various Stochastic models with and without jumps under synchronous and asynchronous observations to gain insight into the differences between the estimators. We then investigate the impact of the reconstruction frequency $M$ in the estimators, and more importantly, we investigate the impact of the time-scale $\Delta t$ under asynchronous observations to study the instantaneous Epps effect under a simulation setting. We provide an {\it ad hoc} approach to deal with asynchrony based on the understanding of the Epps effect arising from asynchrony and demonstrate that the effect can be ameliorated for the instantaneous estimator. Finally, we demonstrate the instantaneous Epps effect using Trade and Quote data from two banking equities in the Johannesburg Stock Exchange and compare the two estimators after correcting for the Epps effect.

The paper is organised as follows: \Cref{sec:instant} introduces the two Fourier estimators. \Cref{sec:comparison} compares the volatility and co-volatility estimates of the two estimators for the various Stochastic models. \Cref{sec:cutfreq} investigates the impact of the cutting frequencies, the time-scale under asynchrony and demonstrates how to deal with the Epps effect arising from asynchrony. \Cref{sec:empirical} provides the empirical demonstration. Finally, \Cref{sec:conclusion} concludes.

\section{Fourier instantaneous estimators}\label{sec:instant}

\subsection{Malliavin-Mancino}\label{subsec:instantMM}
\cite{MM2009} provided a non-parametric Fourier estimator that is constructed in the frequency domain. It expresses the Fourier coefficients of the volatility process $\Sigma^{ij} (t)$ using the Fourier coefficients of the price process $X^i_t = \ln(P^i_t)$ where $P^i_t$ is a generic asset price at time $t$. By re-scaling the trading times from $[0, T]$ to $[0, 2\pi]$ and using {\it Bohr convolution product} (See Theorem 2.1 of \cite{MM2009}) we have for all $k \in \mathbb{Z}$:
\begin{equation} \label{eq:instantMM:1}
    \begin{aligned}
      \mathcal{F}(\Sigma^{ij})(k) = \lim_{N \rightarrow \infty} \frac{2 \pi}{2N+1} \sum_{|s| \leq N} \mathcal{F}(dX^i)(s) \mathcal{F}(dX^j)(k-s).
    \end{aligned}
\end{equation}
Here $\mathcal{F}(\ast)(\star)$ is the $\star^{\text{th}}$ Fourier coefficient of the $\ast$ process \citep{RS2018,Rag2012}. Using the previous tick interpolation to avoid a downward bias in the estimator \citep{BR2002} and a simple function approximation for the Fourier coefficients, we get:\footnote{Here $\iu \in \mathbb{C}$ in the exponential defining the Fourier transform is such that $\iu = \sqrt{-1}$ and should not be confused with integer indices $i$ for the asset.}
\begin{equation} \label{eq:instantMM:2}
\begin{aligned}
& \mathcal{F}(dX^i)(k) \approx \frac{1}{2\pi} \sum_{h=0}^{n_i-1} \exp(-\iu k t^i_h) \delta_{i}(I_h),    \\
& \mathcal{F}(dX^j)(k) \approx \frac{1}{2\pi} \sum_{\ell=0}^{n_j-1} \exp(-\iu k t^j_{\ell}) \delta_{j}(I_{\ell}),
\end{aligned}
\end{equation}
where $(t^i_h)_{h=0,...,n_i}$ and $(t^j_{\ell})_{\ell=0,...,n_j}$ are the observation times. The price fluctuations are $\delta_{i}(I_h) = X^i_{t_{h+1}^i} - X^i_{t_{h}^i}$ and $\delta_{j}(I_{\ell}) = X^j_{t_{\ell+1}^j} - X^j_{t_{\ell}^j}$ for asset $i$ and $j$ respectively.

The Fourier coefficients of the volatility process is given as:
\begin{equation} \label{eq:instantMM:3}
    \alpha_k\left( \Sigma^{ij}_{n_i,n_j,N} \right) = \frac{2 \pi}{2N+1} \sum_{|s| \leq N} \mathcal{F}(dX^i)(s) \mathcal{F}(dX^j)(k-s),
\end{equation}
for $k \in \{-M,...,M\}$. Therefore, the instantaneous volatility/co-volatility can be reconstructed using \cref{eq:instantMM:3} as:
\begin{equation} \label{eq:instantMM:4}
    \hat{\Sigma}^{ij}_{n_i,n_i,N,M}(t) = \sum_{|k| \leq M} \left( 1 - \frac{|k|}{M} \right) e^{\iu t k} \alpha_k\left( \Sigma^{ij}_{n_i,n_j,N} \right).
\end{equation}
Notice that there are two parameters $N$ and $M$ that require tuning in \cref{eq:instantMM:4}. Here $N$ dictates how many Fourier modes are used in estimating the Fourier coefficients of the volatility. This controls the level of averaging which has a direct implication on the time-scale of investigation (See \cite{PCEPTG2020a}). The second parameter $M$ is the reconstruction frequency. It determines how many Fourier coefficients are used in approximating the spot volatility. \Cref{sec:cutfreq} will investigate the impact of $N$ and $M$ in greater detail. Moreover, notice that $n_i$ and $n_j$ need not be the same. This is the main feature behind the Malliavin-Mancino Fourier estimator. The convolution is performed in the frequency domain, allowing one to bypass the issue of asynchrony in the time domain. This means we do not need to {\it synchronise} the data beforehand by means of imputation. 

The implementation of the instantaneous estimator requires the evaluation of $\{-M-N,...,N+M\}$ Fourier coefficients for asset $i$ and $j$ which can be computationally expensive. Therefore, the implementation relies on non-uniform fast Fourier transforms to computationally speed up the evaluation of \cref{eq:instantMM:2} (See \cite{PCEPTG2020a}).

\subsection{Cuchiero-Teichmann}\label{subsec:instantCT}
\cite{CT2015} provide an extension based on the Malliavin-Mancino Fourier estimator to account for the presence of jumps. This is achieved by modifying jump robust estimators of integrated RVs considered by \cite{BGJPS2006,BSW2006,JACOD2008,PV2009,TT2012} in order to obtain estimates for the Fourier coefficients of the realised path of the instantaneous volatility.

The Cuchiero-Teichmann spot volatility is estimated through three steps. First, we need an estimator for the Fourier coefficients of $\rho_g \left( \Sigma \right)$ from the discrete price observations $X_t$ where $\rho_g (\cdot)$ is a continuous invertible function. Meaning we need an estimator for:
\begin{equation} \label{eq:instantMM:5}
    \mathcal{F}\left( \rho_g \left( \Sigma \right) \right) \left( k \right) = \frac{1}{T} \int_0^T \rho_g \left( \Sigma (t) \right) e^{- \iu \frac{2 \pi}{T} k t}.
\end{equation}
\cite{CT2015} show that the estimator of \cref{eq:instantMM:5} taking the form:
\begin{equation} \label{eq:instantMM:6}
    V(X,g,k)_T^n = \frac{1}{n} \sum_{h=1}^{\lfloor nT \rfloor} e^{- \iu \frac{2 \pi}{T} k t^n_{h-1}} g\left( \sqrt{n} \Delta^n_h X \right),
\end{equation}
converges to the required Fourier coefficients in \cref{eq:instantMM:5} (See Theorem 3.4 of \cite{CT2015}). Here $1/n = \Delta t$ is the discretisation interval, $\Delta^n_h X = X_{t^n_{h}} - X_{t^n_{h-1}}$,\footnote{Note that $\delta_i(I_h)$ and $\Delta^n_h X$ are both price fluctuations. Separate notation is used highlight that the observation times in $\delta_i(I_h)$ need not be equidistant, while $\Delta^n_h X$ must be strictly equidistant.} and the time grid for the observations of $X_t$ in $[0,T]$ are equal and equidistant, {\it i.e.} $t^n_h = h/n$, $h=0,...,\lfloor nT \rfloor$.

Second, we can apply the Fourier-Fej\'{e}r inversion to reconstruct the path of:
\begin{equation} \label{eq:instantMM:7}
    \widehat{\rho_g \left( \Sigma (t) \right)}_{n,M} = \frac{1}{T} \sum_{|k| \leq M} \left( 1 - \frac{|k|}{M} \right) e^{\iu \frac{2 \pi}{T} kt} V(X,g,k)_T^n.
\end{equation}
Finally, the spot volatility can be obtained by inverting \cref{eq:instantMM:7} yielding:
\begin{equation} \label{eq:instantMM:8}
    \hat{\Sigma}_{n,M}(t) = \rho_g^{-1} \left( \widehat{\rho_g \left( \Sigma (t) \right)}_{n,M}  \right).
\end{equation}

Up till now in \Cref{subsec:instantCT}, the use of asset indices $i$ and $j$ have been avoided. This is because the Cuchiero-Teichmann spot volatility estimator does not share the property of the Malliavin-Mancino Fourier estimator where the Fourier coefficients of the volatility are constructed in the frequency domain via a convolution with the Fourier coefficients of the price process (See \cref{eq:instantMM:3}). The Fourier coefficients of the volatility in the Cuchiero-Teichmann estimator is obtained by adapting integrated RVs; meaning direct price observations are used, rather than their Fourier coefficients (See \cref{eq:instantMM:6}). This means that the reconstruction of $\hat{\Sigma}^{12}_{n,M}(t)$ requires the use of the polarisation identity. Therefore, the Cuchiero-Teichmann estimator does not have a method to over come the issue of asynchrony.

Here, we use \cite{TT2012} specification of the function $g(x) = \cos(x)$, therefore $\rho_g \left( \Sigma (t) \right) = e^{-\Sigma (t) / 2}$. Now by using the polarisation identity, we have:
\begin{equation} \label{eq:instantMM:9}
\begin{aligned}
    &\hat{\Sigma}^{ii}_{n,M}(t) = -2 \log\left( \widehat{\rho_{g_{ii}} \left( \Sigma (t) \right)}_{n,M} \right), \quad i \in \{1,2\},    \\
    &\hat{\Sigma}^{12}_{n,M}(t) = \frac{1}{2} \left( -2 \log\left( \widehat{\rho_{g_{12}} \left( \Sigma (t) \right)}_{n,M} \right) - \hat{\Sigma}^{11}_{n,M}(t) - \hat{\Sigma}^{22}_{n,M}(t) \right).
\end{aligned}
\end{equation}

The Cuchiero-Teichmann estimator has a tuning parameter $M$ which is the reconstruction frequency of the spot volatility. There is no parameter $N$ controlling the level of averaging. Therefore, the estimator relies on the discretisation interval $\Delta t = 1/n$ to investigate various time-scales.

\section{Comparison}\label{sec:comparison}

We compare the Malliavin-Mancino (MM) and Cuchiero-Teichmann (CT) spot volatility estimator under the presence of jumps and no jumps for the synchronous and asynchronous case. We consider two types of volatility models: constant volatility and stochastic volatility. The stochastic models considered are the Geometric Brownian Motion (GBM) for constant volatility with no jumps; the Merton Model for constant volatility with jumps; the Heston Model for stochastic volatility with no jumps; and the Bates-type model for stochastic volatility with jumps. The parameters are given for the period $[0,T]$ where $T=1$ can be thought of as a day. 

\subsection*{Geometric Brownian Motion}

The bivariate Geometric Brownian Motion (GBM) satisfies the following system of SDEs
\begin{equation} \label{eq:comp:1}
  \frac{dP^i_t}{P^i_t} = \mu_i dt + \sigma_i dW_i(t), \quad i = 1, 2,
\end{equation}
where $W_i$ are Brownian motions with $\mathrm{Corr}(dW_1, dW_2) = \rho^{12}$. The parameters used in the simulation are given in \Cref{tab:param1}.

\begin{table}[H]
\centering
\begin{tabular}{|c|l|l|}
\hline
Model                       & Parameter                  & Values   \\  \hline
\multirow{3}{*}{GBM/Merton} & $(\mu_1,  \mu_2)$          & (0.01, 0.01)     \\ \cline{2-3} 
                            & $(\sigma^2_1, \sigma^2_2)$ & (0.1, 0.2)       \\ \cline{2-3} 
                            & $\rho^{12}$                & 0.35             \\ \hline
\multirow{3}{*}{Merton}     & $(a_1, a_2)$               & (-0.005, -0.003) \\ \cline{2-3} 
                            & $(b_1,b_2)$                & (0.015, 0.02)    \\ \cline{2-3} 
                            & $(\lambda_1, \lambda_2)$   & (100, 100)       \\ \hline
\end{tabular}
\caption{Parameters used in the simulation for the Geometric Brownian Motion and the Merton Model.}
\label{tab:param1}
\end{table}

\subsection*{Merton Model}

The bivariate Merton model satisfies the following system of SDEs:
\begin{equation} \label{eq:comp:2}
  \frac{dP^i_t}{P^i_{t-}} = \mu_i dt + \sigma_i dW^i_t + dJ^i_t, \ \ \ \ i = 1, 2.
\end{equation}
Here the correlation is $\mathrm{Corr}(dW^1, dW^2) = \rho^{12}$ and the intervals $J^i$ are independent of the $W^i$ with piece-wise constant paths \citep{GLASSERMAN2004}. J is defined as: \begin{equation} \label{eq:comp:3}
  J^i_t = \sum_{j=1}^{N(t)} (Y_j - 1),
\end{equation}
where $N(t)$ is a Poisson process with rate $\lambda_i$, $Y_j \sim LN(a,b)$ i.i.d and also independent of $N(t)$.\footnote{The $t^{-}$ on the LHS of \eqref{eq:comp:2} is used to indicate the C\`{a}dl\`{a}g nature of the process near jumps.} The parameters used in the simulation are given in \Cref{tab:param1}.

\subsection*{Heston Model}

The stochastic volatility models are simulated such that the entire volatility matrix is stochastic. Concretely, the bi-variate Heston model takes the form:
\begin{equation} \label{eq:comp:4}
\begin{aligned}
    &{X}_t = {X}_0 + \int_0^t - \frac{1}{2} \Sigma^{\text{diag}}(s) ds + \int_0^t \sqrt{\Sigma(s)} dZ_s,      \\
    &\Sigma(t) = \Sigma(0) + \int_0^t \left( b + M \Sigma(t) + \Sigma(t) M^\top \right) dt    \\
    &\quad \quad \quad + \sqrt{\Sigma(t)} dB_t H + H dB_t^\top \sqrt{\Sigma(t)},
\end{aligned}
\end{equation}
where $M$ and $H$ are invertible matrices, $\Sigma(0)$ a 2-dimensional positive semi-definite symmetric matrix $S^+_2$, $b - H^2 \in S^+_2$, and $Z$ is a 2-dimensional Brownian motion correlated with the 2x2 matrix of Brownian motions $B$ such that $Z = \sqrt{1 - \rho^\top \rho}W + B \rho$, where $\rho \in [-1, 1]^2$ such that $\rho^\top \rho \leq 1$ and $W$ is a 2-dimensional Brownian motion independent of $B$. The parameters for the simulation are given in \Cref{tab:param2}.

\begin{table}[H]
\centering
\begin{tabular}{|c|l|l|}
\hline
Model                       & Parameter                  & Values   \\  \hline
\multirow{6}{*}{Heston/Bates} & $(X^1_0, X^2_0)$                                                                                                                               & (4.6, 4.6)                                                                                                  \\ \cline{2-3} 
                              & \begin{tabular}[c]{@{}l@{}}$\begin{pmatrix} \Sigma^{11}(0) & \Sigma^{12}(0) \\ \Sigma^{12}(0) & \Sigma^{22}(0) \end{pmatrix}$\end{tabular} & \begin{tabular}[c]{@{}l@{}}$\begin{pmatrix} 0.09 & -0.036 \\ -0.036 & 0.09 \end{pmatrix}$\end{tabular}  \\ \cline{2-3} 
                              & $M$                                                                                                                                            & \begin{tabular}[c]{@{}l@{}}$\begin{pmatrix} -1.6 & -0.2 \\  -0.4 & -1 \end{pmatrix}$\end{tabular} \\ \cline{2-3}  
                              & $\alpha = H^2$                                                                                                                                 & \begin{tabular}[c]{@{}l@{}}$\begin{pmatrix} 0.0725 & 0.06 \\  0.06 & 0.1325 \end{pmatrix}$\end{tabular} \\ \cline{2-3} 
                              & $b$                                                                                                                                            & $3.5\alpha$                                                                                                 \\ \cline{2-3} 
                              & $\rho$                                                                                                                                         & (-0.3, -0.5)                                                                                                \\ \hline
\multirow{5}{*}{Bates}        & $(\lambda_1^X, \lambda_2^X)$                                                                                                                   & (100, 100)                                                                                                  \\ \cline{2-3} 
                              & $(a_1, a_2)$                                                                                                                                   & (-0.005, -0.003)                                                                                            \\ \cline{2-3} 
                              & $(b_1, b_2)$                                                                                                                                   & (0.015, 0.02)                                                                                               \\ \cline{2-3} 
                              & $\lambda^{\Sigma^{11}}$                                                                                                                        & 10                                                                                                          \\ \cline{2-3} 
                              & $\theta$                                                                                                                                       & 0.05                                                                                                        \\ \hline
\end{tabular}
\caption{Parameters used in the simulation for the Heston and the Bates-type Model. The parameters are borrowed from \cite{CT2015}.}
\label{tab:param2}
\end{table}

\subsection*{Bates model}

Here we consider a Bates-type model (henceforth referred to as the Bates model for simplicity) where jumps occur in the log-price $X_t$ and in the volatility process $\Sigma(t)$.\footnote{The jumps in the volatility process only happen for $\Sigma^{11}(t)$ as in \cite{CT2015}.} The 2-dimensional Bates model takes the form:
\begin{equation} \label{eq:comp:5}
\begin{aligned}
    &{X}_t = {X}_0 + \int_0^t b_s ds + \int_0^t \sqrt{\Sigma(s^-)} dZ_s   \\
    &\quad + \int_{0}^{t} \int_{\mathbb{R}^{2}} \xi \mu^{X}(d \xi, d s),   \\  
    &\Sigma(t) = \Sigma(0) + \int_0^t \left( b + M \Sigma(t) + \Sigma(t) M^\top \right) dt    \\
    &\quad + \sqrt{\Sigma(t)} dB_t H + H dB_t^\top \sqrt{\Sigma(t)} + \int_{0}^{t} \int_{\mathbb{R}^{2}} \xi \mu^{\Sigma}(d \xi, d s).
\end{aligned}
\end{equation}
The Brownian motions $Z$ and $B$, and the parameters $b$, $H$ and $M$ are defined as before in the Heston model. Here $\mu^{X}(d \xi, d t)$ is the random measure associated with the jumps of $X$. The jumps are Gaussian with mean $a_i$, standard deviation $b_i$, and the rate of jumps is $\lambda_i^X$. Here $\mu^{\Sigma}(d \xi, d t)$ is the random measure associated with the jumps of $\Sigma$. The jumps are exponential with parameter $\theta$ and the rate of jumps is $\lambda^{\Sigma^{11}}$. Lastly, the drift of $X$ is given by $b_{s,i} = -\frac{1}{2} \Sigma^{ii}(s) - \lambda^{X}_i\left(e^{a_{i}-\frac{1}{2} b_{i}^{2}}-1\right)$. The parameters for the simulation are given in \Cref{tab:param2}.

\subsection{Synchronous case}\label{subsec:comparisonSyn}

Let us consider the impact of jumps against no jumps for the two spot volatility estimators when the observations are observed on an equidistant synchronous grid ($n_1 = n_2 = n$). The number of grid points is set to $n = 28,800$ which corresponds to 1-second data for a trading day of 8 hours. The four aforementioned stochastic models are simulated using the parameters given in the tables.

\begin{figure*}[p]
    \centering
    \subfloat[GBM $\Sigma^{11}(t)$, N = Nyq., M = 100]{\label{fig:SynInst:a}\includegraphics[width=0.33\textwidth]{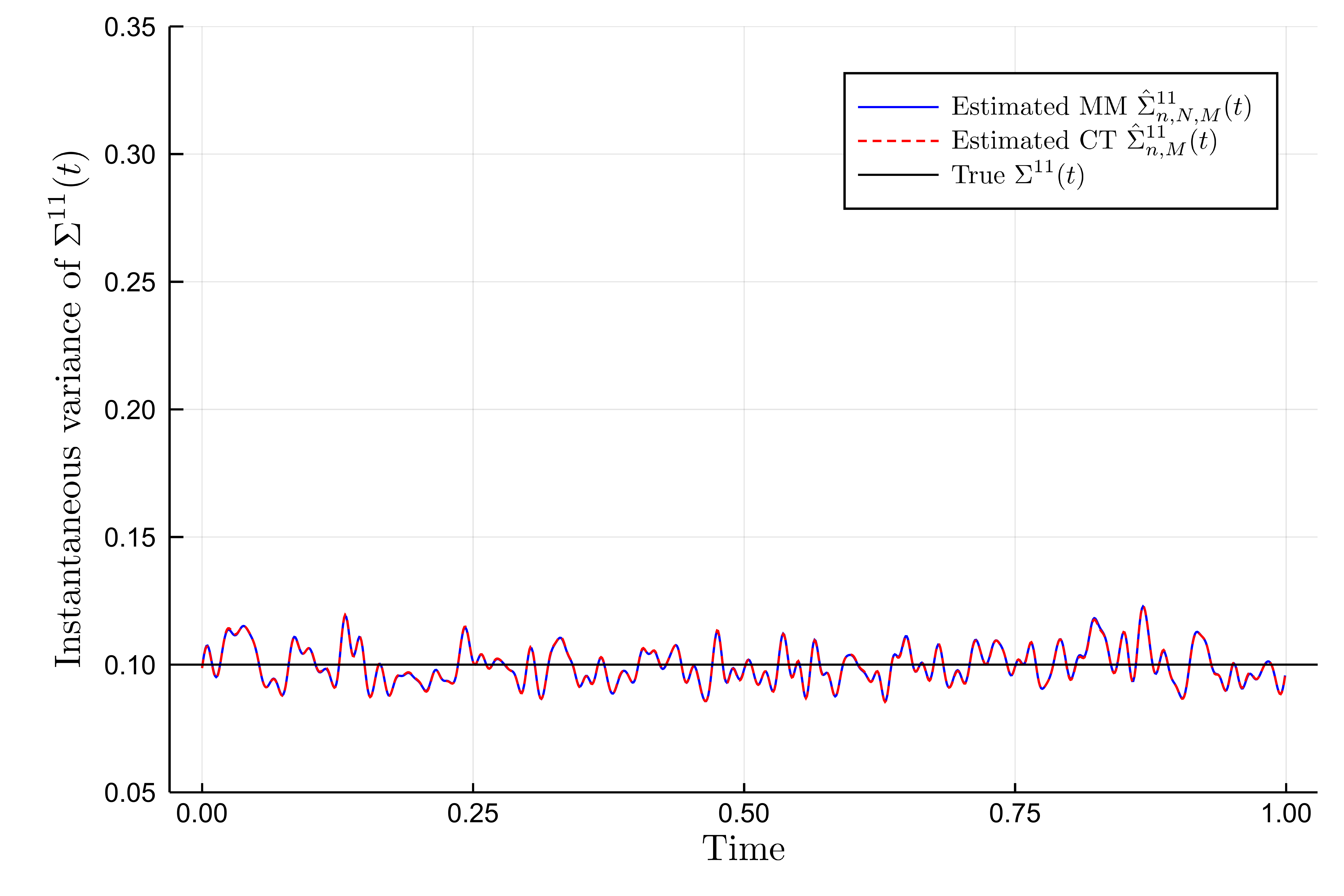}}
    \subfloat[GBM $\Sigma^{22}(t)$, N = Nyq., M = 100]{\label{fig:SynInst:b}\includegraphics[width=0.33\textwidth]{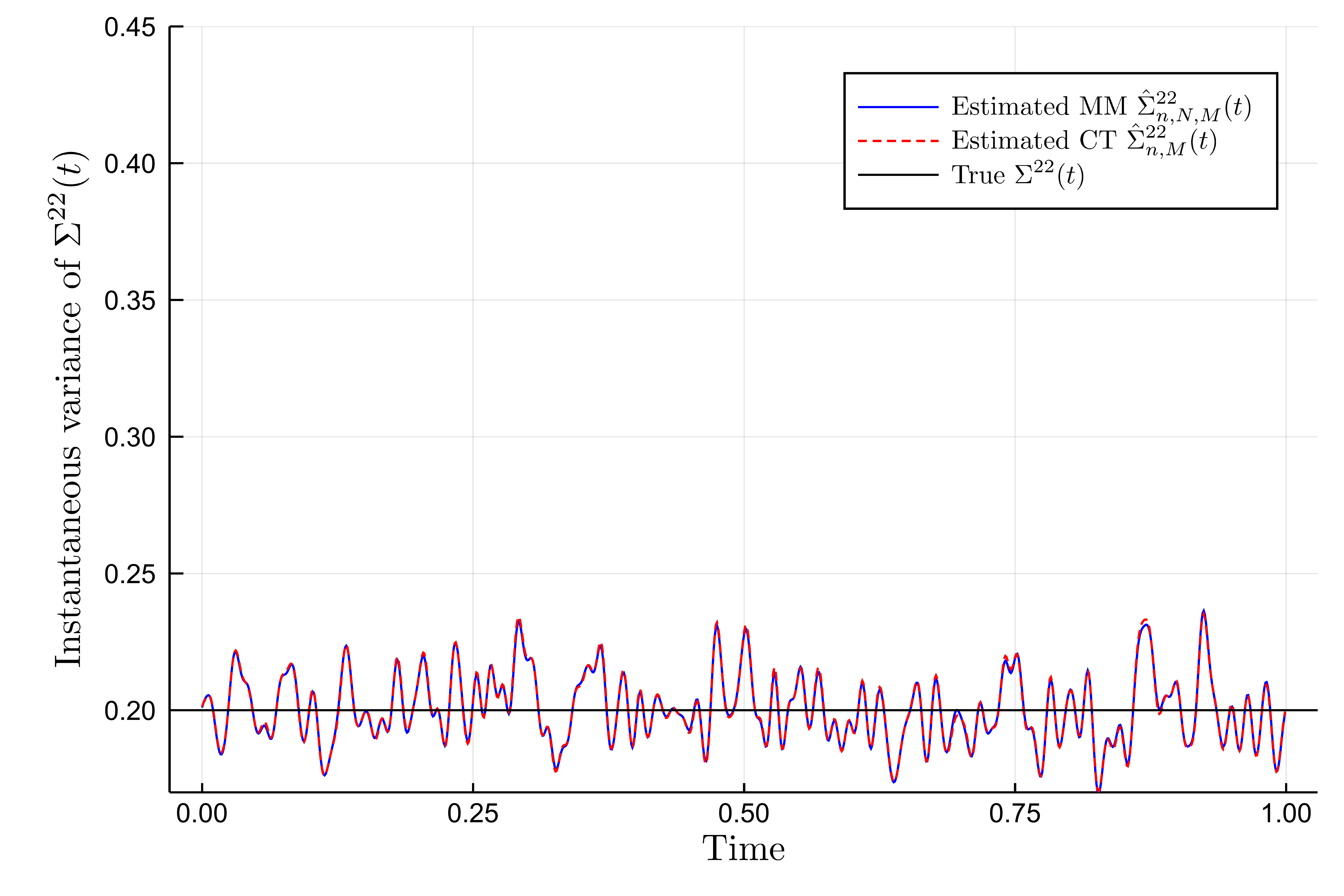}}    
    \subfloat[GBM $\Sigma^{12}(t)$, N = Nyq., M = 100]{\label{fig:SynInst:c}\includegraphics[width=0.33\textwidth]{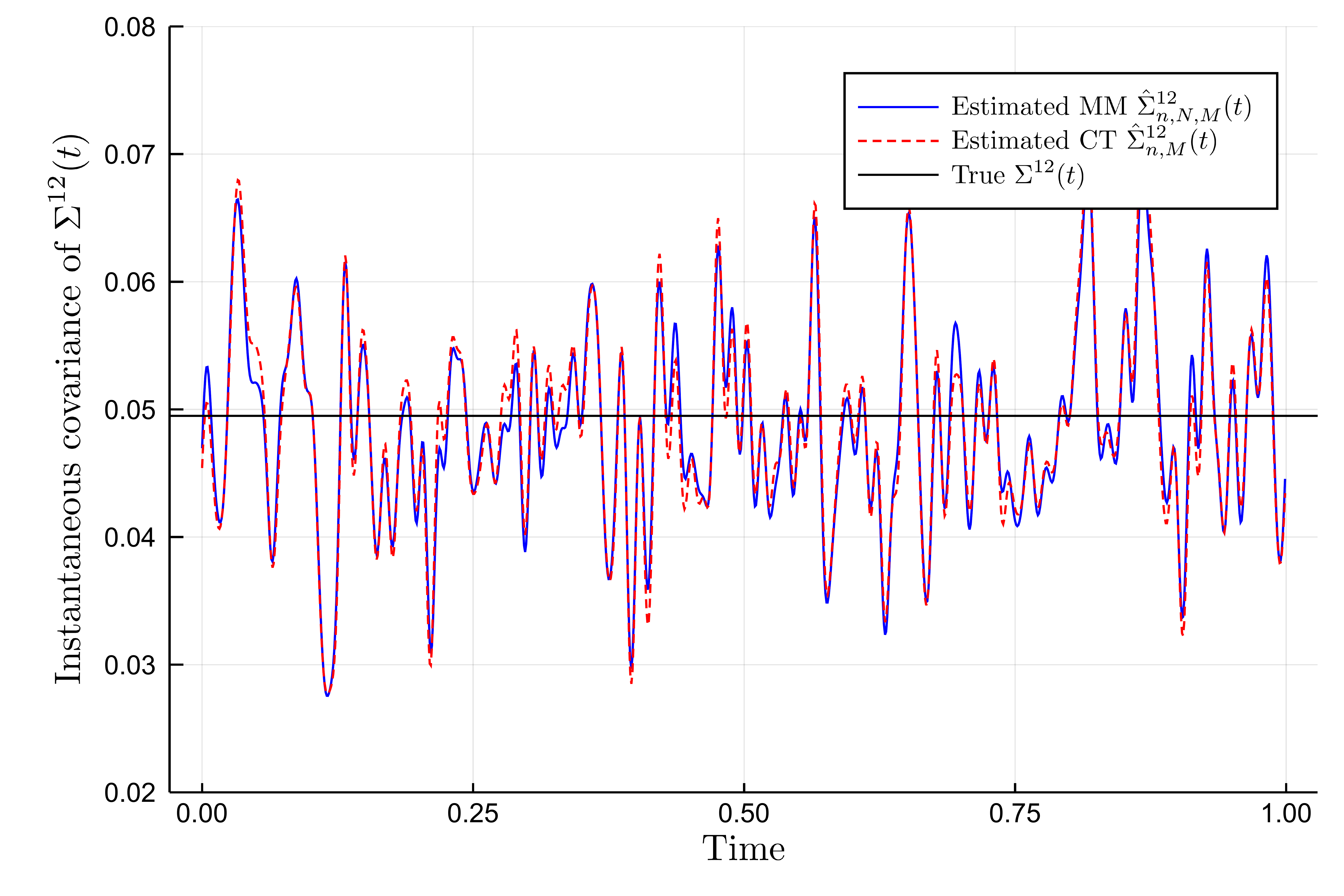}}  \\
    \subfloat[Merton Model $\Sigma^{11}(t)$, N = Nyq., M = 100]{\label{fig:SynInst:d}\includegraphics[width=0.33\textwidth]{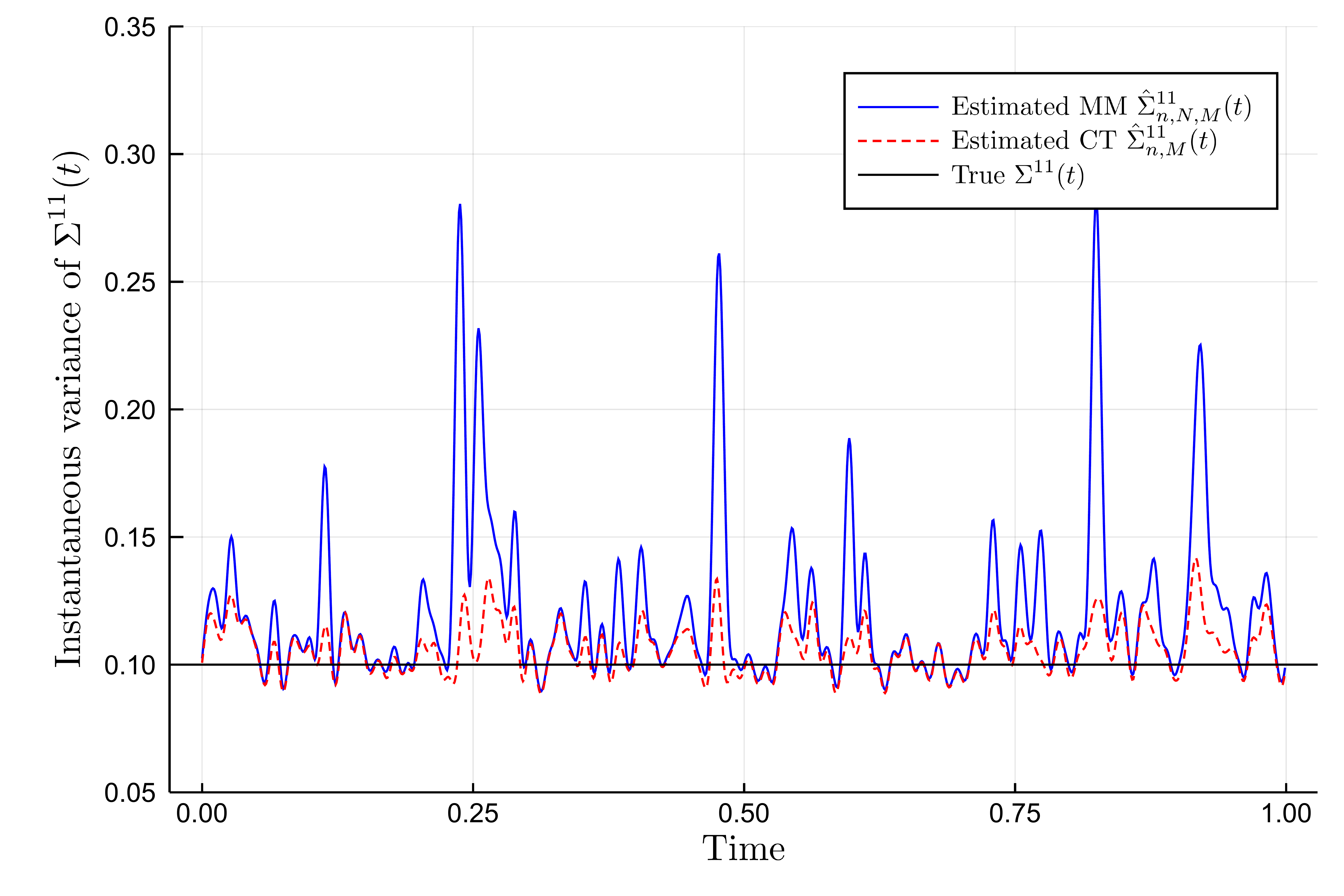}}
    \subfloat[Merton Model $\Sigma^{22}(t)$, N = Nyq., M = 100]{\label{fig:SynInst:e}\includegraphics[width=0.33\textwidth]{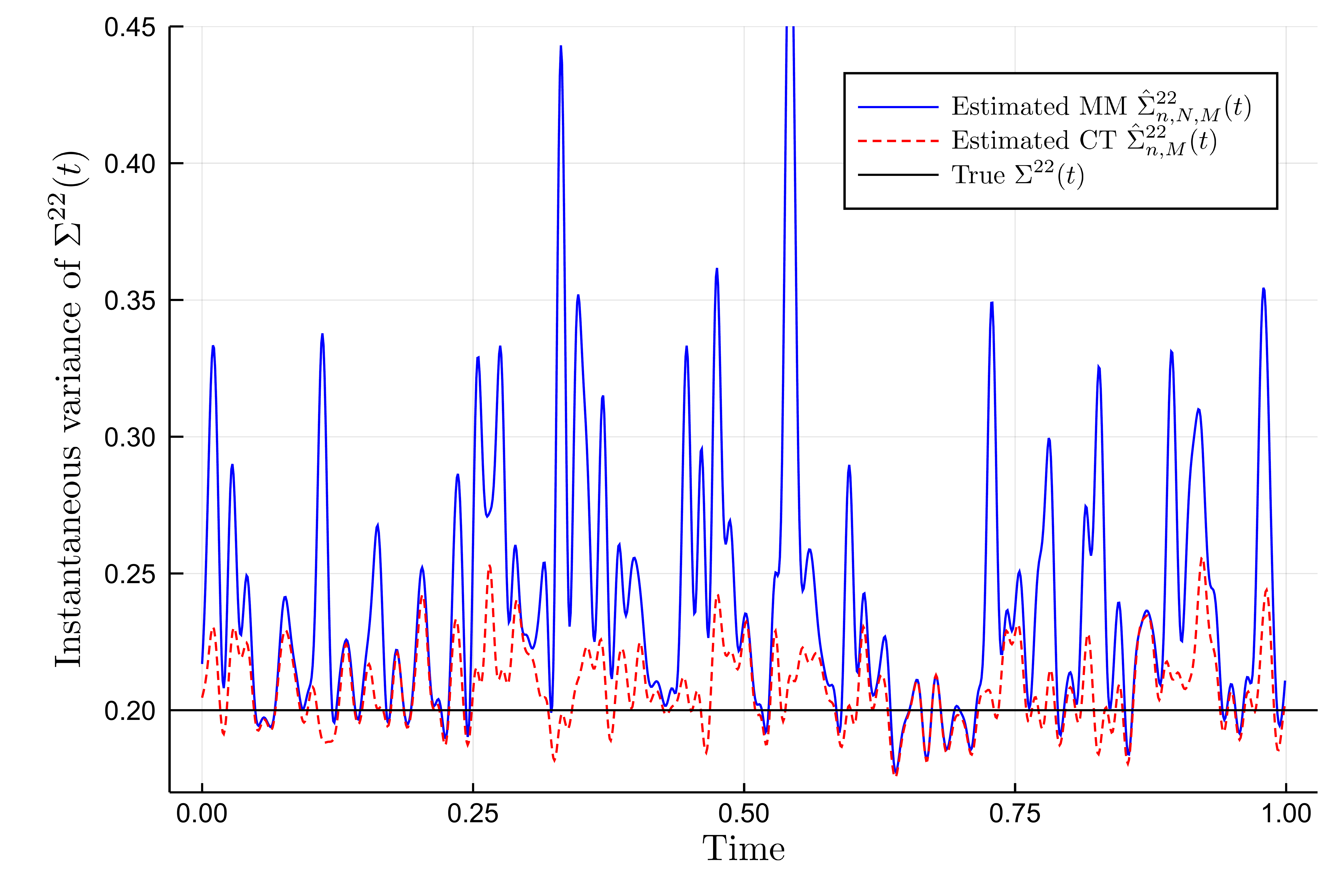}}    
    \subfloat[Merton Model $\Sigma^{12}(t)$, N = Nyq., M = 100]{\label{fig:SynInst:f}\includegraphics[width=0.33\textwidth]{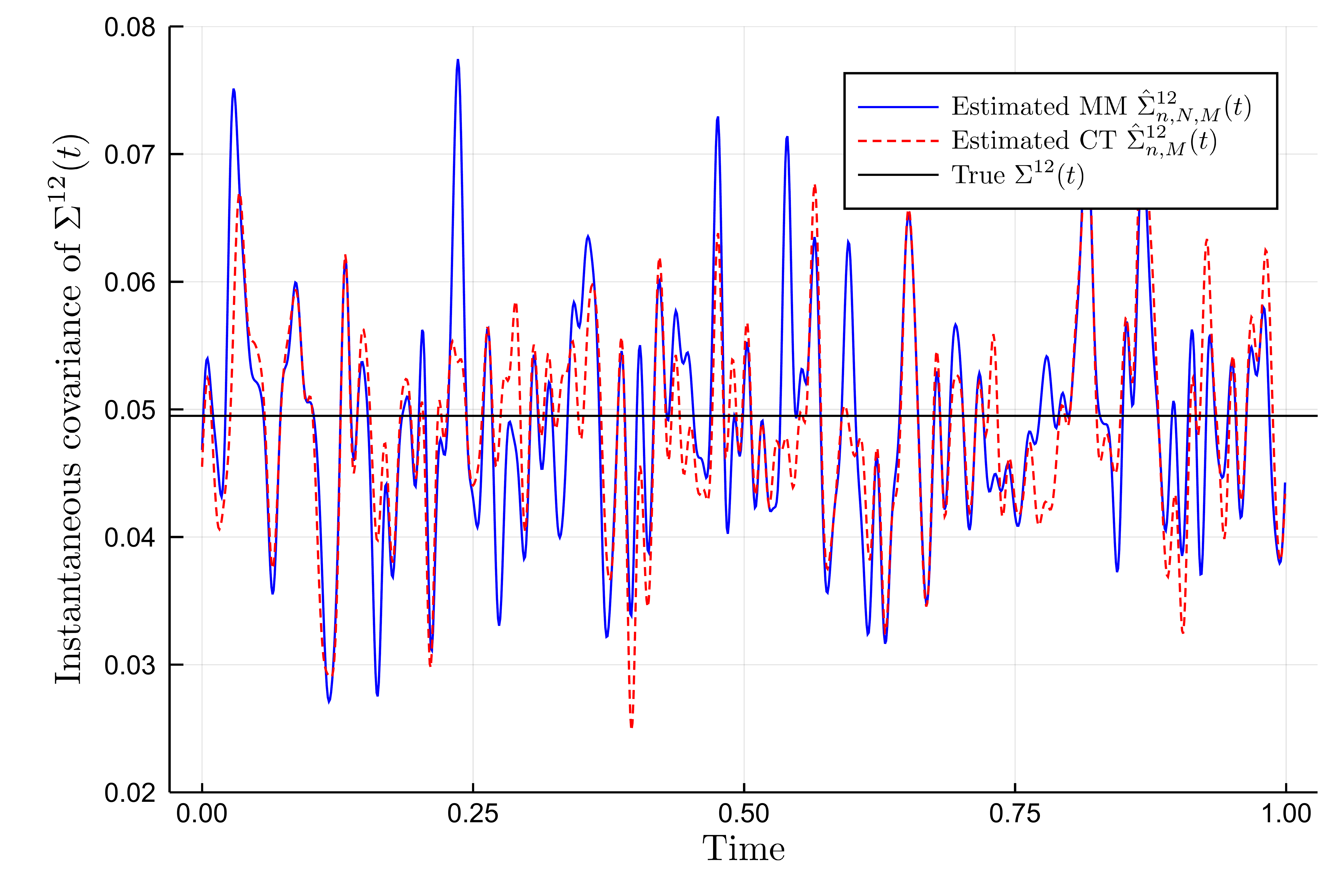}}  \\
    \subfloat[Heston Model $\Sigma^{11}(t)$, N = Nyq., M = 100]{\label{fig:SynInst:g}\includegraphics[width=0.33\textwidth]{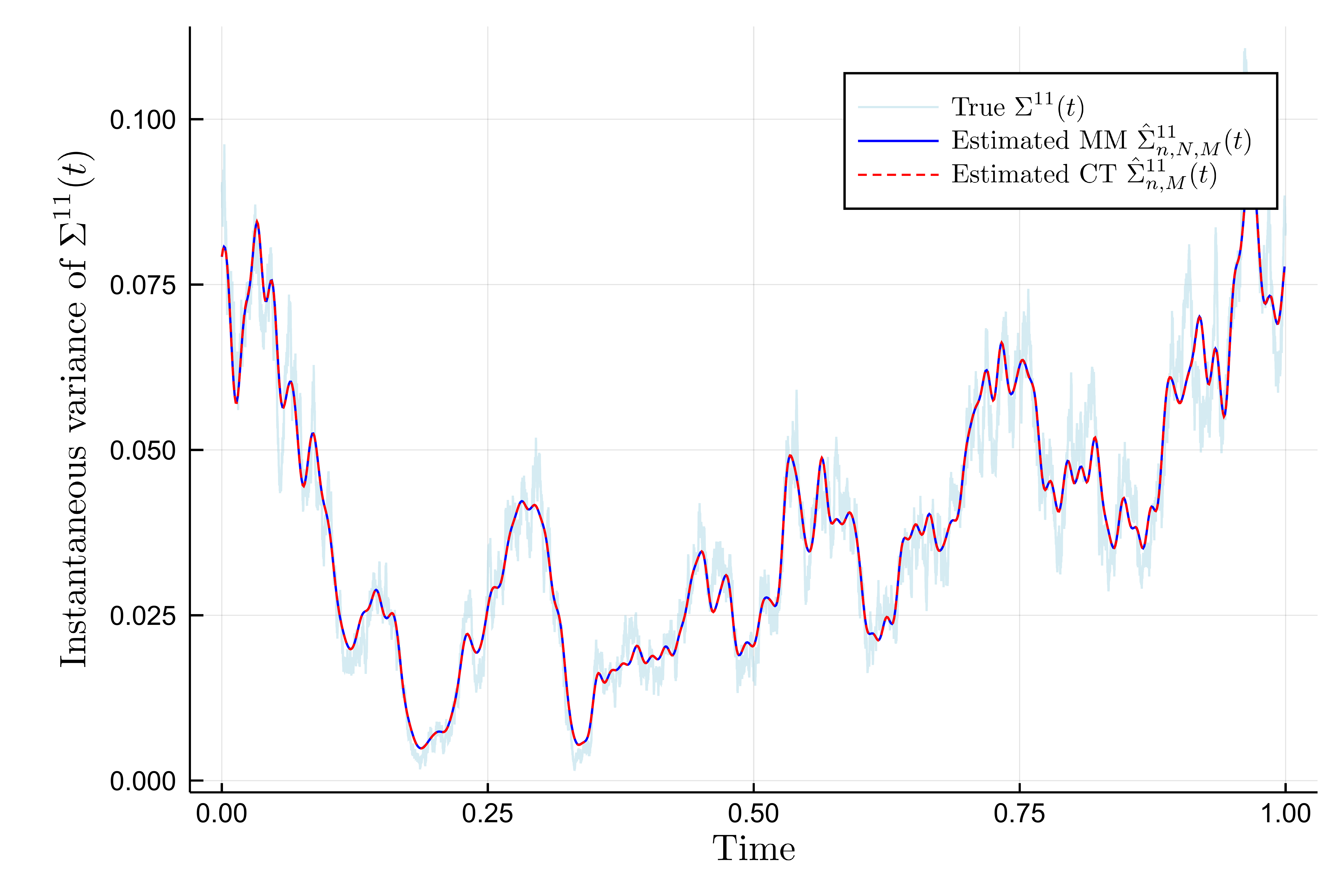}}
    \subfloat[Heston Model $\Sigma^{22}(t)$, N = Nyq., M = 100]{\label{fig:SynInst:h}\includegraphics[width=0.33\textwidth]{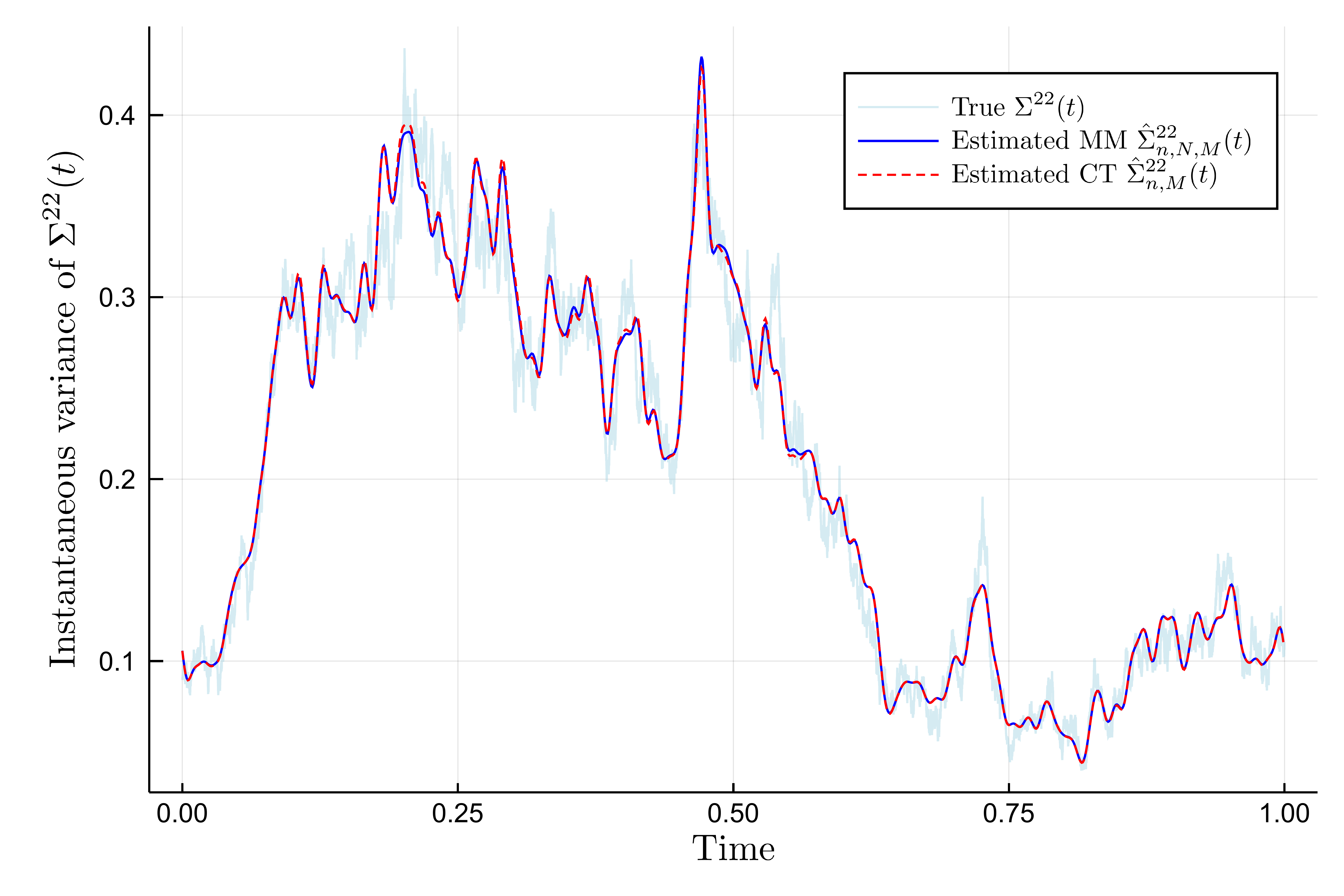}}    
    \subfloat[Heston Model $\Sigma^{12}(t)$, N = Nyq., M = 100]{\label{fig:SynInst:i}\includegraphics[width=0.33\textwidth]{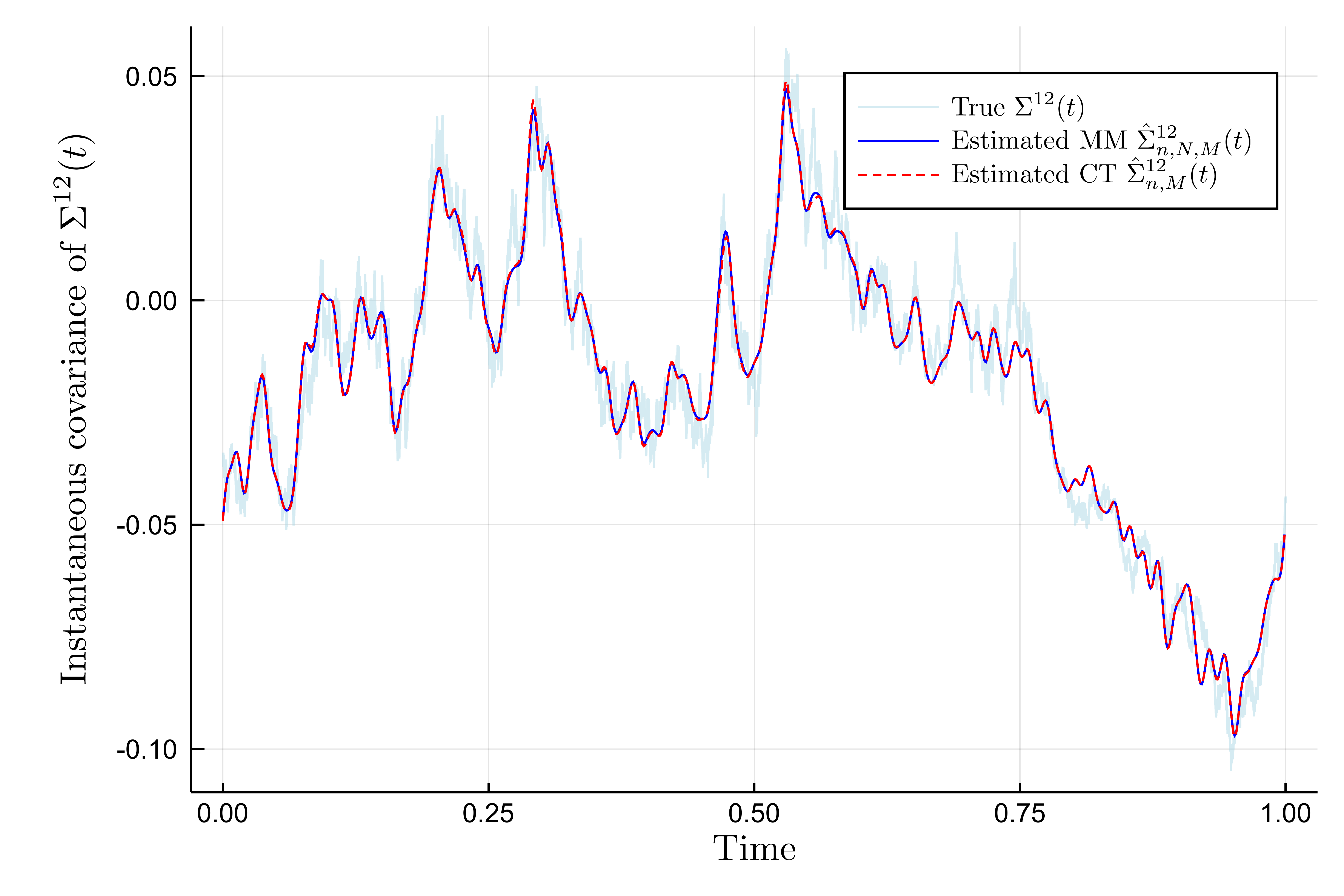}}  \\
    \subfloat[Bates Model $\Sigma^{11}(t)$, N = Nyq., M = 100]{\label{fig:SynInst:j}\includegraphics[width=0.33\textwidth]{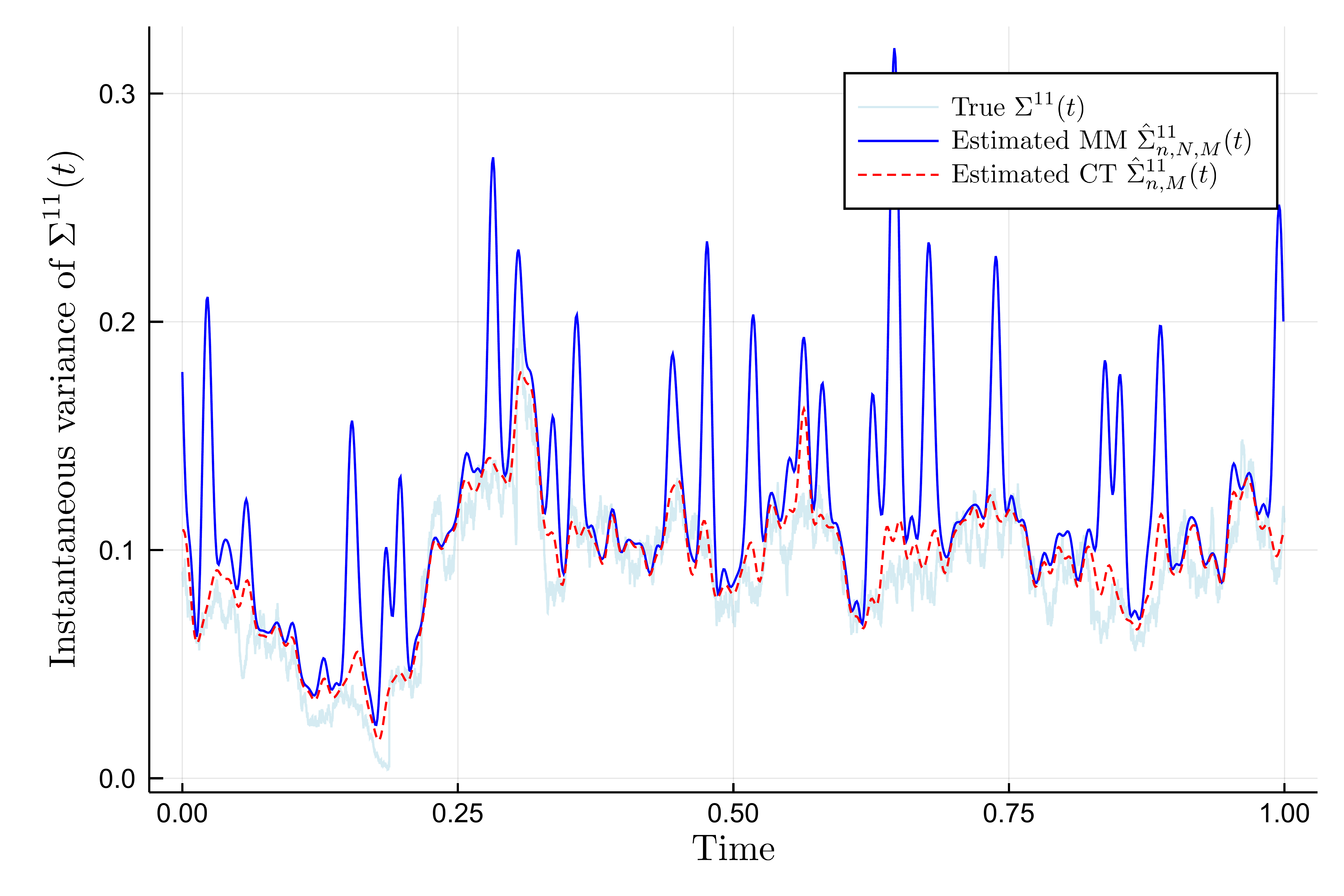}}
    \subfloat[Bates Model $\Sigma^{22}(t)$, N = Nyq., M = 100]{\label{fig:SynInst:k}\includegraphics[width=0.33\textwidth]{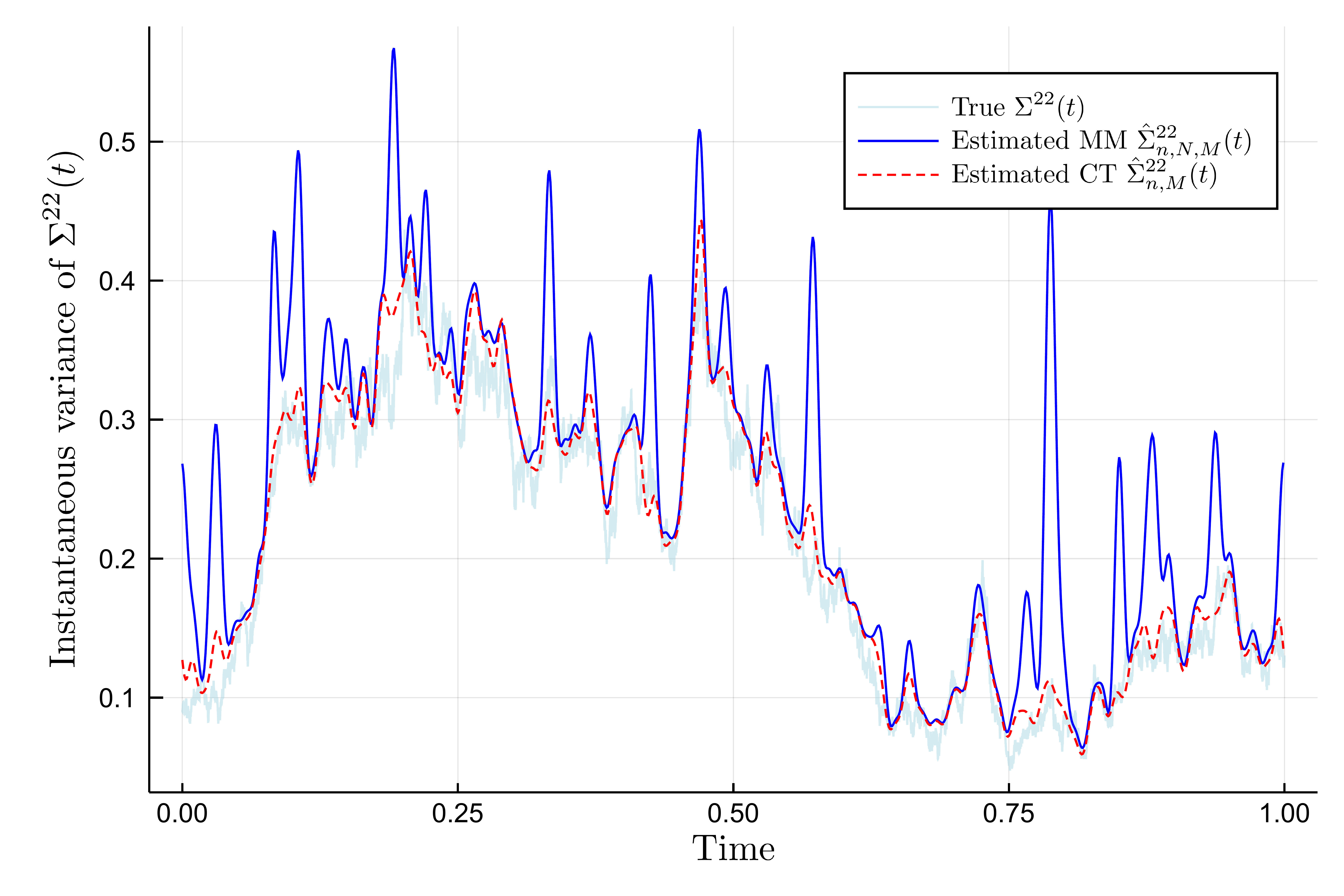}}    
    \subfloat[Bates Model $\Sigma^{12}(t)$, N = Nyq., M = 100]{\label{fig:SynInst:l}\includegraphics[width=0.33\textwidth]{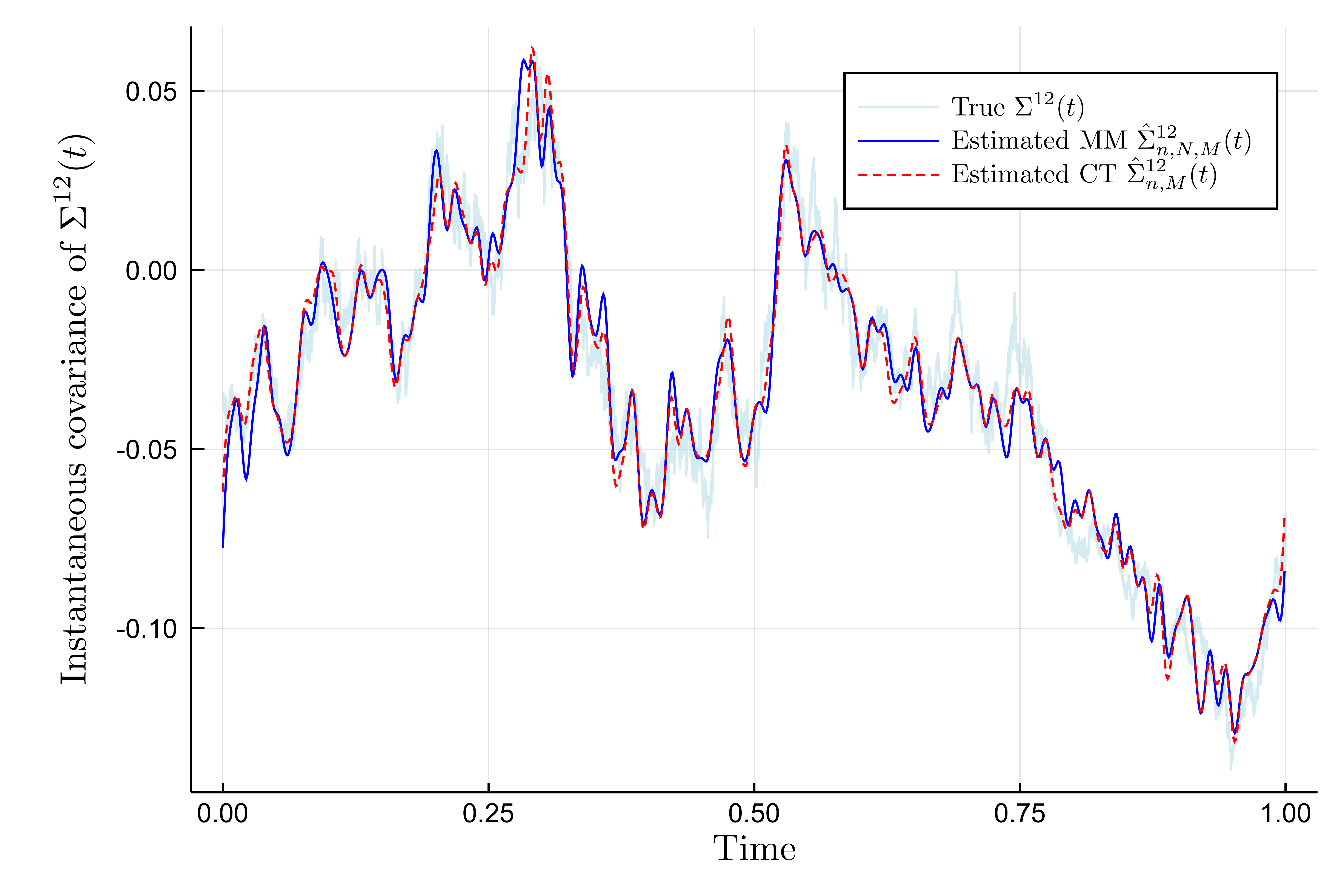}}
    \caption{Here we compare the Malliavin-Mancino (blue line with label ``Estimated MM'') and Cuchiero-Teichmann (red dashes with label ``Estimated CT'') spot volatility/co-volatility estimator for stochastic models with constant volatility against stochastic volatility, and jumps against no jumps. The models are simulated with $n=28,800$ synchronous grid points. The spot estimates are compared against the true underlying instantaneous volatility matrix (black and light-blue lines with label ``True''). We see that for the models with no jumps, both estimators recover the instantaneous volatility matrix with high accuracy. For the models with jumps, the Malliavin-Mancino volatility estimates have spikes in volatility caused by jumps while the Cuchiero-Teichmann volatility estimates are not as severely affected. In the case of the co-volatility estimates, both estimators are not affected by jumps and recover the underlying co-volatility.}
\label{fig:SynInst}
\end{figure*}

\Cref{fig:SynInst} compares the true underlying volatility (black and light-blue lines) against the Malliavin-Mancino (blue lines) and Cuchiero-Teichmann (red dashes) spot volatility estimates. The rows of the figures are: GBM, Merton, Heston, and Bates model from first to last. The columns of the figures are: $\Sigma^{11}(t)$, $\Sigma^{22}(t)$, and $\Sigma^{12}(t)$ from first to last. The cutting frequencies used are $N$ as the Nyquist frequency for the Malliavin-Mancino estimator, and $M = 100$ for both the Malliavin-Mancino and Cuchiero-Teichmann estimator. For the GBM and Heston model, we see that both the Malliavin-Mancino and Cuchiero-Teichmann estimators recover the entire volatility matrix with high fidelity. For the Merton and Bates model, we see that the volatility ($\Sigma^{ii}(t)$, $i=1,2$) estimates using the Malliavin-Mancino estimator presents regions with spikes in volatility when there are jumps. On the other hand, the Cuchiero-Teichmann estimator does not present large spikes in volatility because it is robust to jumps. However, the effect of jumps can still be seen in the Cuchiero-Teichmann estimator as the volatility becomes slightly higher than in the case with no jumps. Interestingly, the co-volatility ($\Sigma^{12}(t)$) estimates are not severely altered by jumps for both the Malliavin-Mancino and Cuchiero-Teichmann estimators. Jumps result in a negative bias in the integrated correlation estimate \citep{PCRBTG2019}, here we see that the negative bias is a result of volatility spikes which results in a larger normalisation factor for the correlation and ultimately leads to a downward bias.

\subsection{Asynchronous case}\label{subsec:comparisonAsyn}

\begin{figure*}[p]
    \centering
    \subfloat[GBM $\Sigma^{11}(t)$]{\label{fig:AsynInst:a}\includegraphics[width=0.33\textwidth]{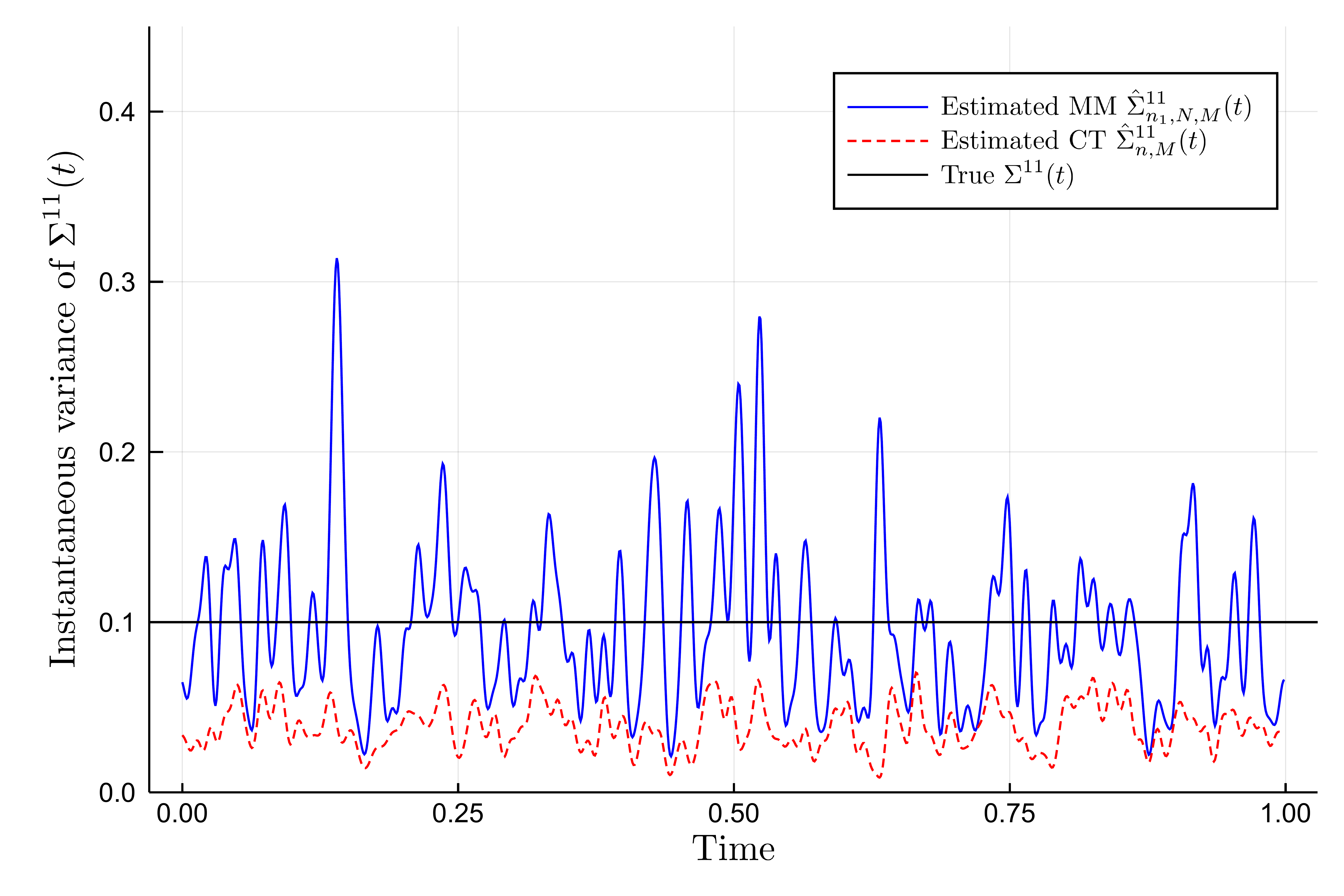}}
    \subfloat[GBM $\Sigma^{22}(t)$]{\label{fig:AsynInst:b}\includegraphics[width=0.33\textwidth]{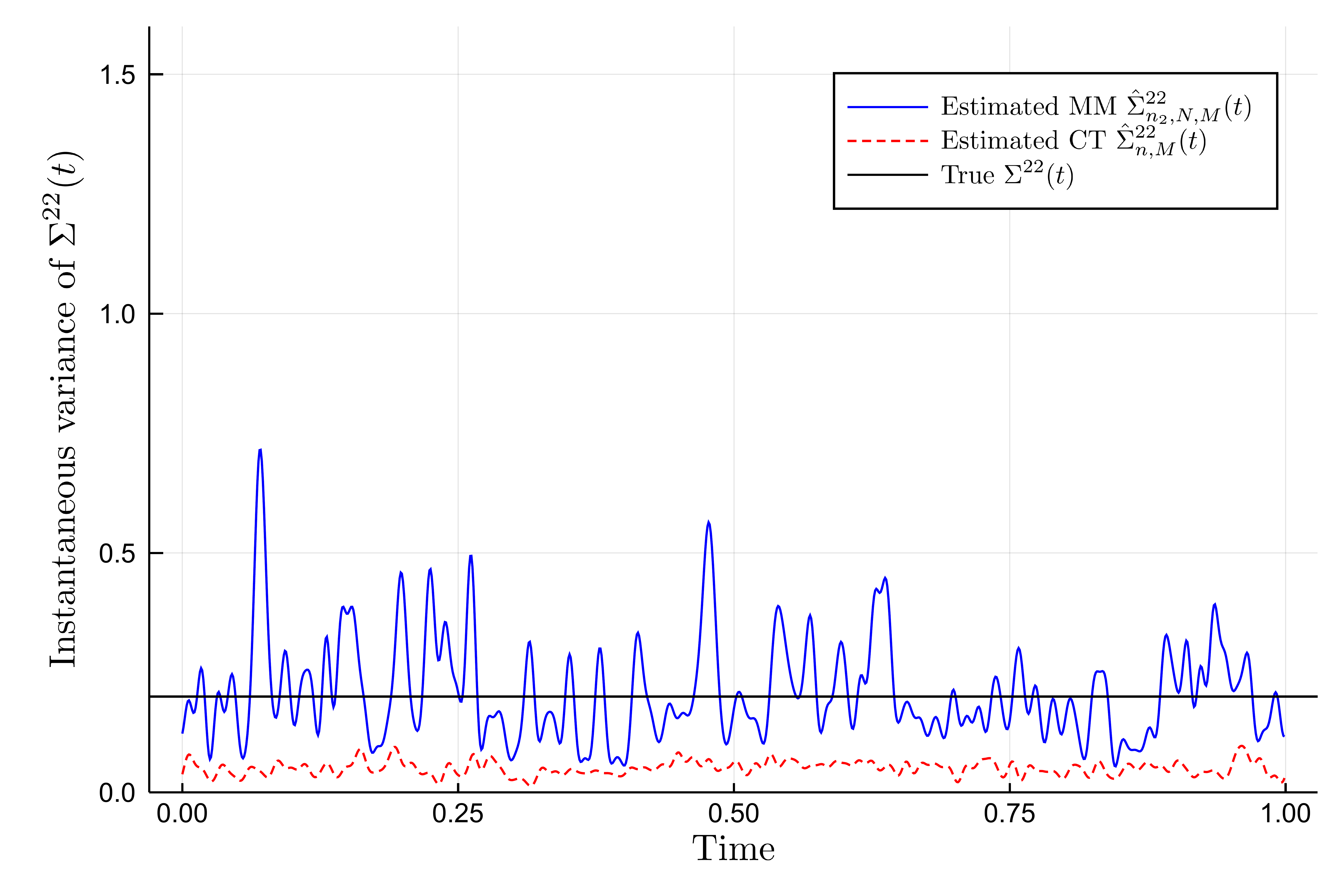}}    
    \subfloat[GBM $\Sigma^{12}(t)$]{\label{fig:AsynInst:c}\includegraphics[width=0.33\textwidth]{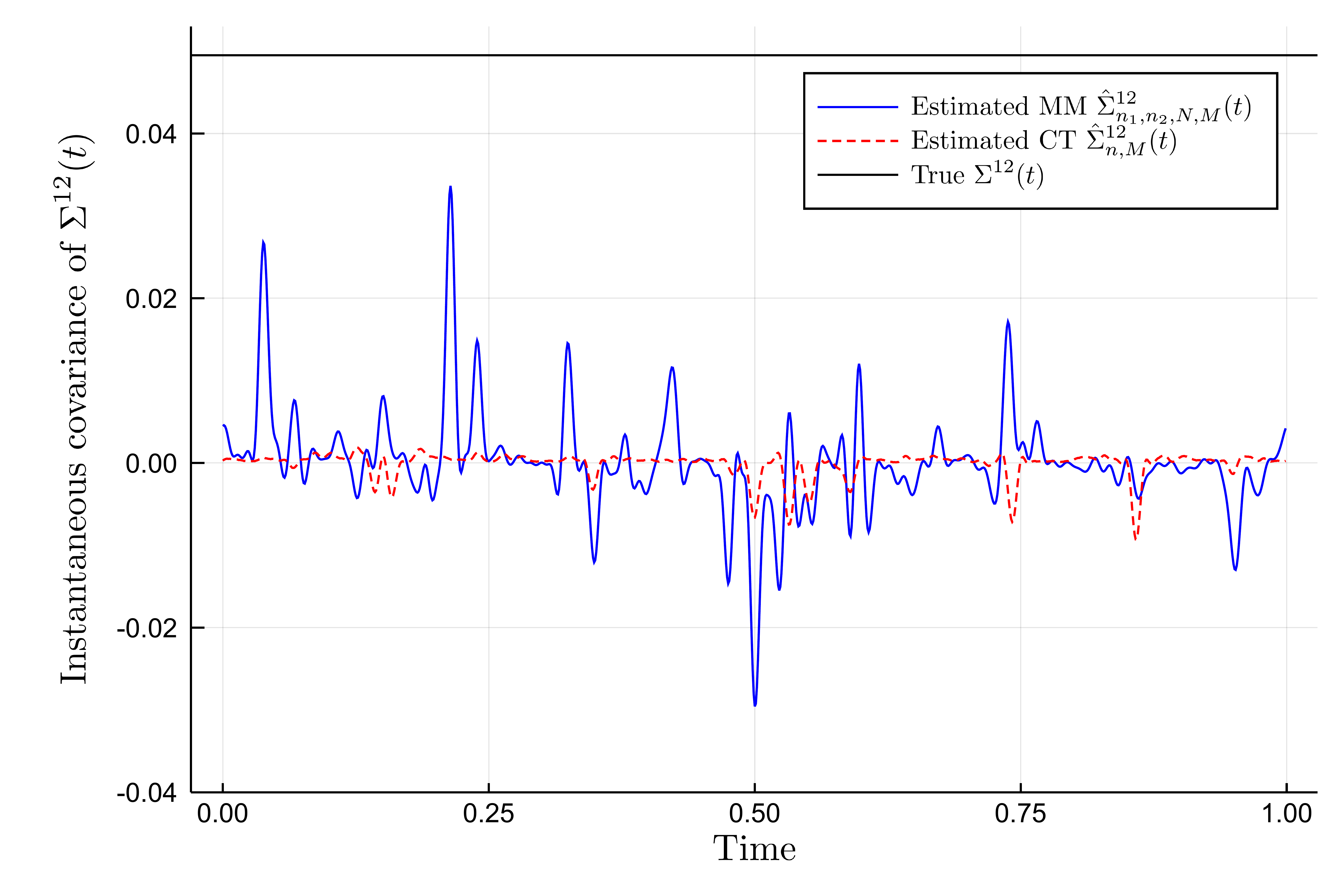}}  \\
    \subfloat[Merton Model $\Sigma^{11}(t)$]{\label{fig:AsynInst:d}\includegraphics[width=0.33\textwidth]{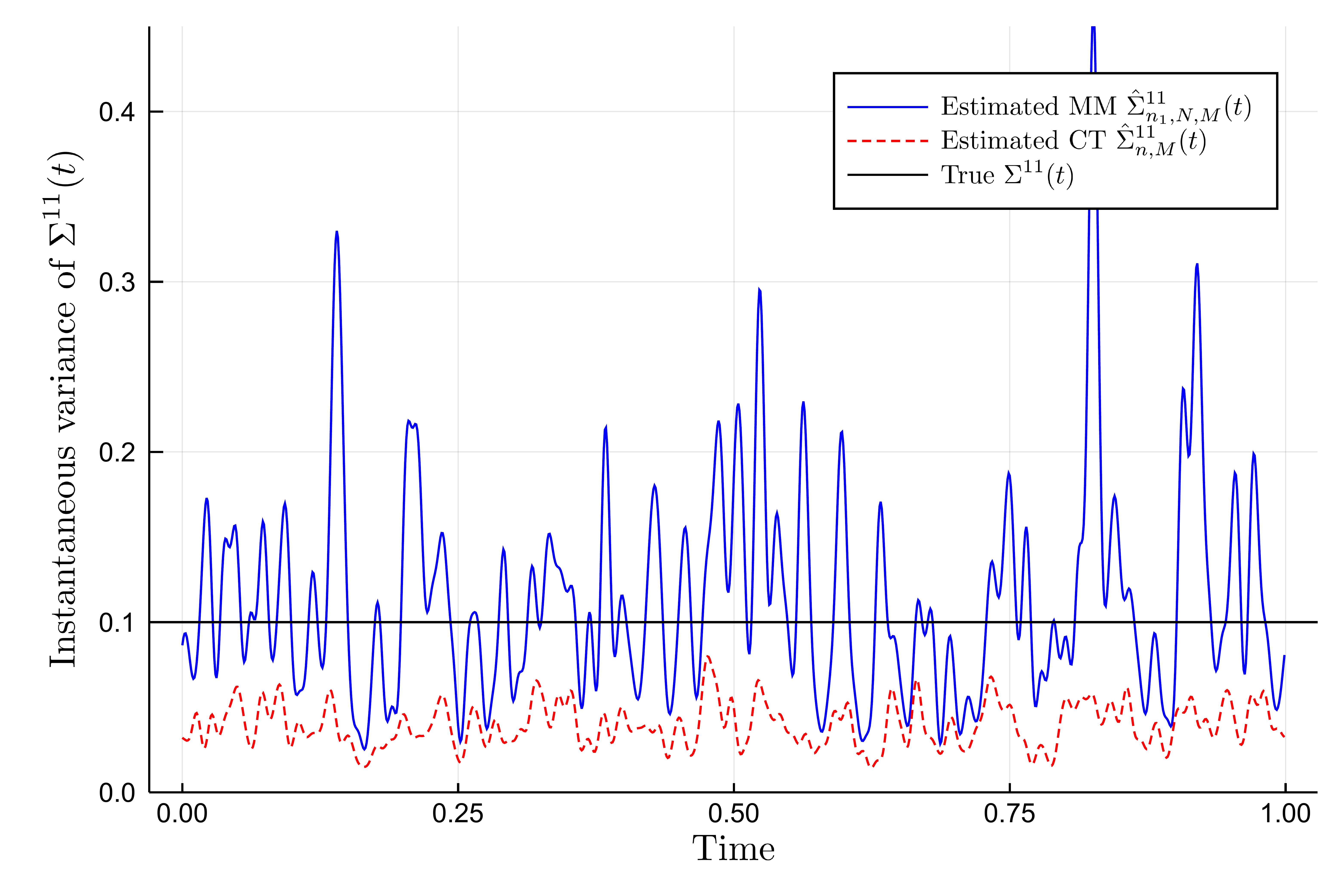}}
    \subfloat[Merton Model $\Sigma^{22}(t)$]{\label{fig:AsynInst:e}\includegraphics[width=0.33\textwidth]{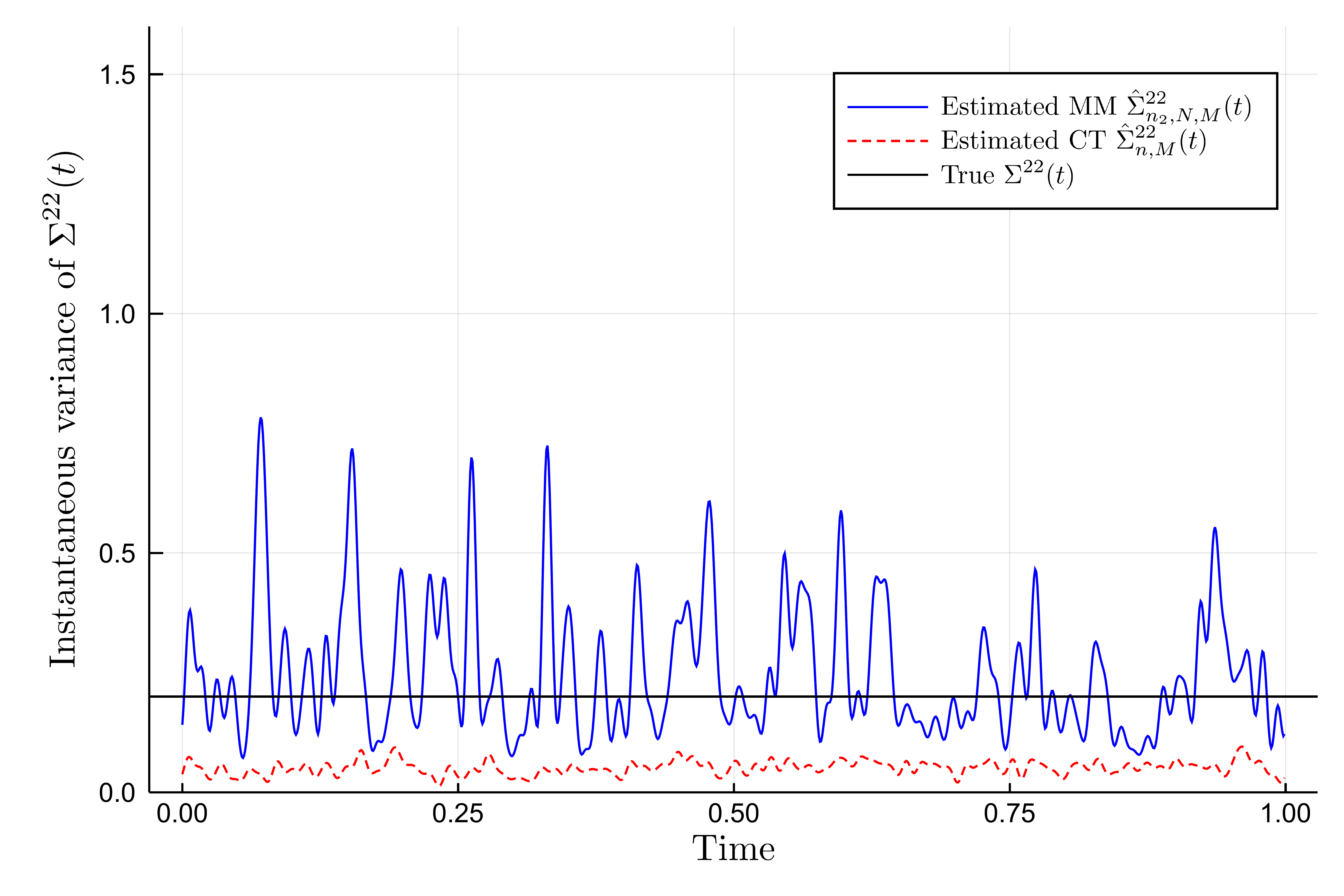}}    
    \subfloat[Merton Model $\Sigma^{12}(t)$]{\label{fig:AsynInst:f}\includegraphics[width=0.33\textwidth]{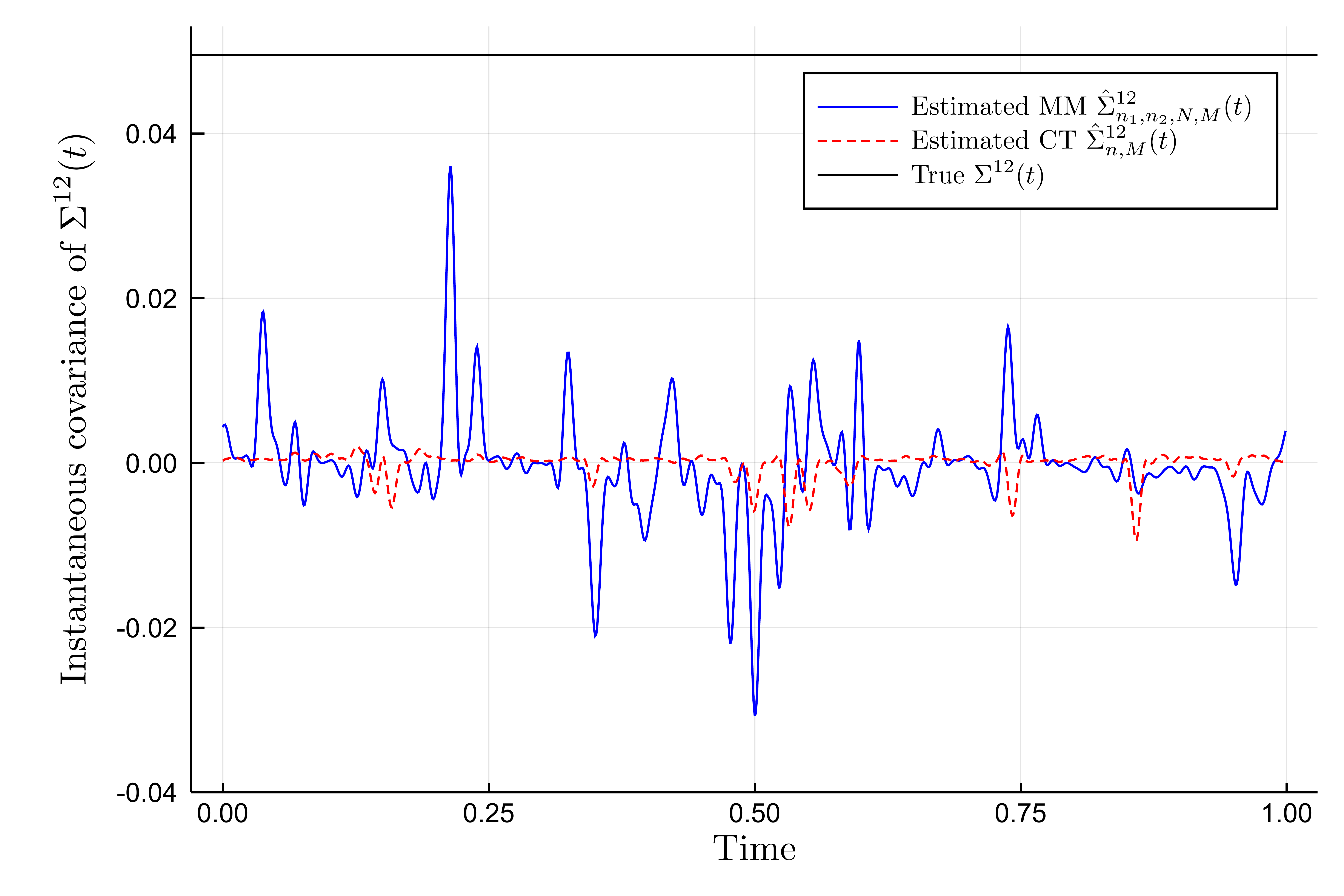}}  \\
    \subfloat[Heston Model $\Sigma^{11}(t)$]{\label{fig:AsynInst:g}\includegraphics[width=0.33\textwidth]{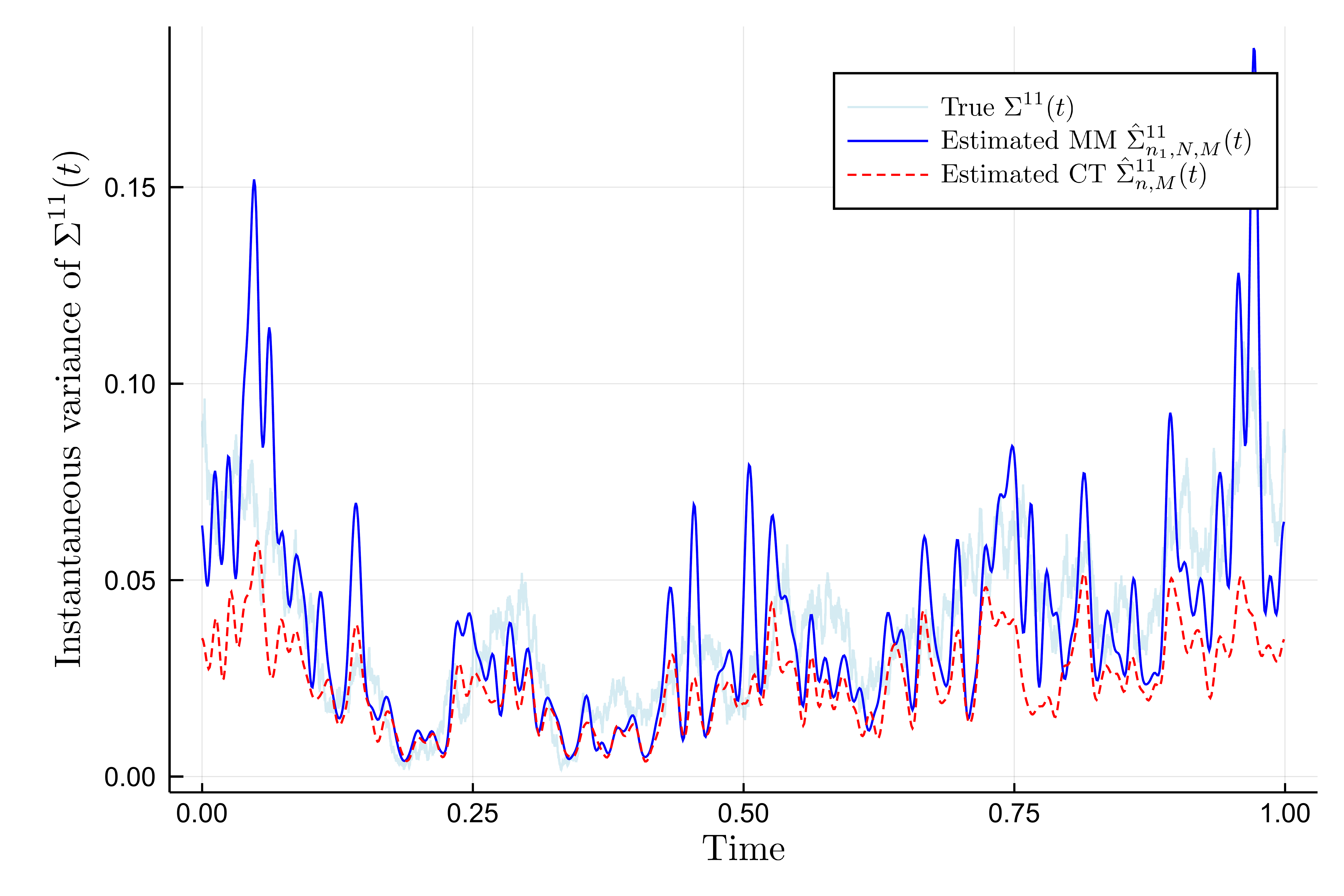}}
    \subfloat[Heston Model $\Sigma^{22}(t)$]{\label{fig:AsynInst:h}\includegraphics[width=0.33\textwidth]{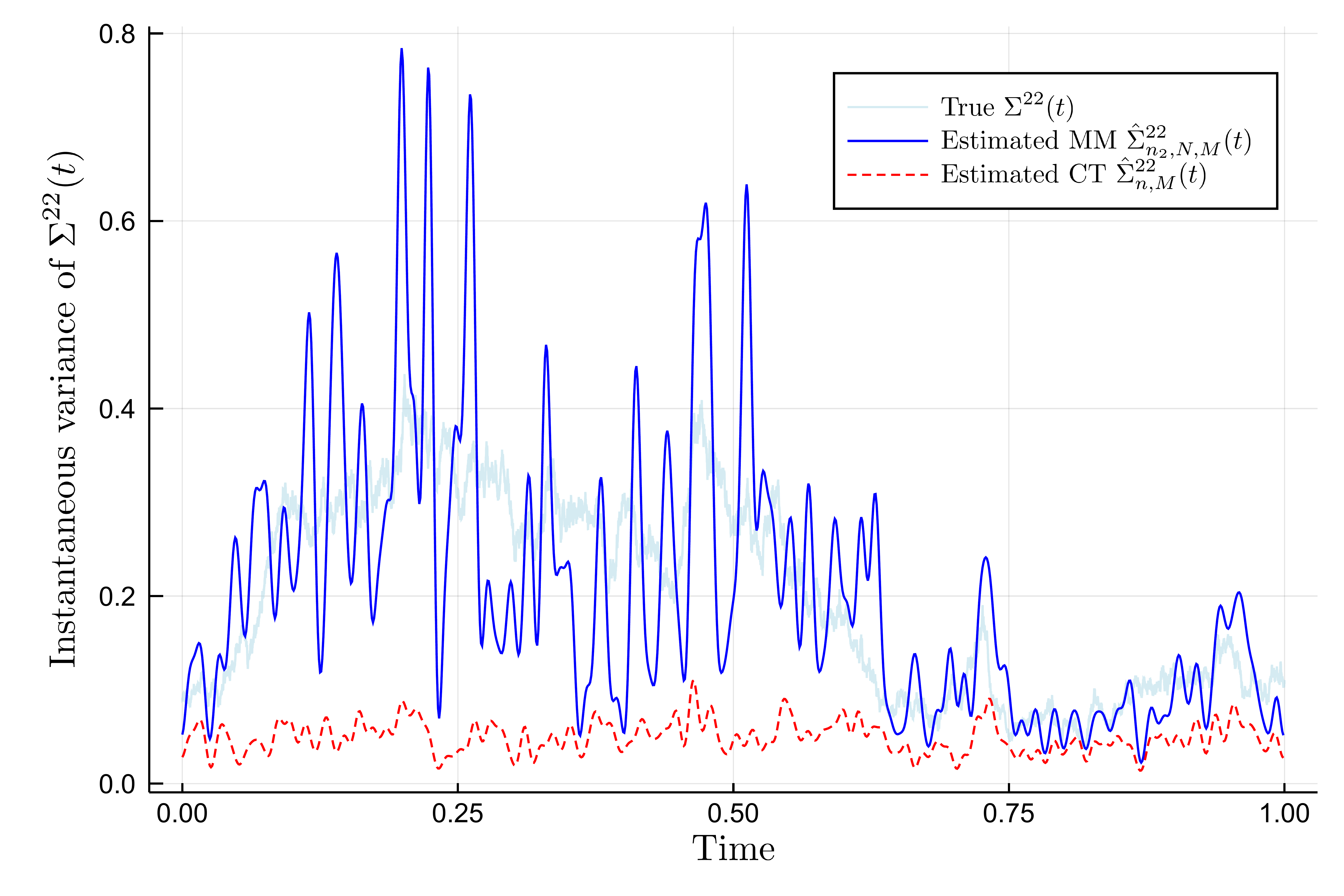}}    
    \subfloat[Heston Model $\Sigma^{12}(t)$]{\label{fig:AsynInst:i}\includegraphics[width=0.33\textwidth]{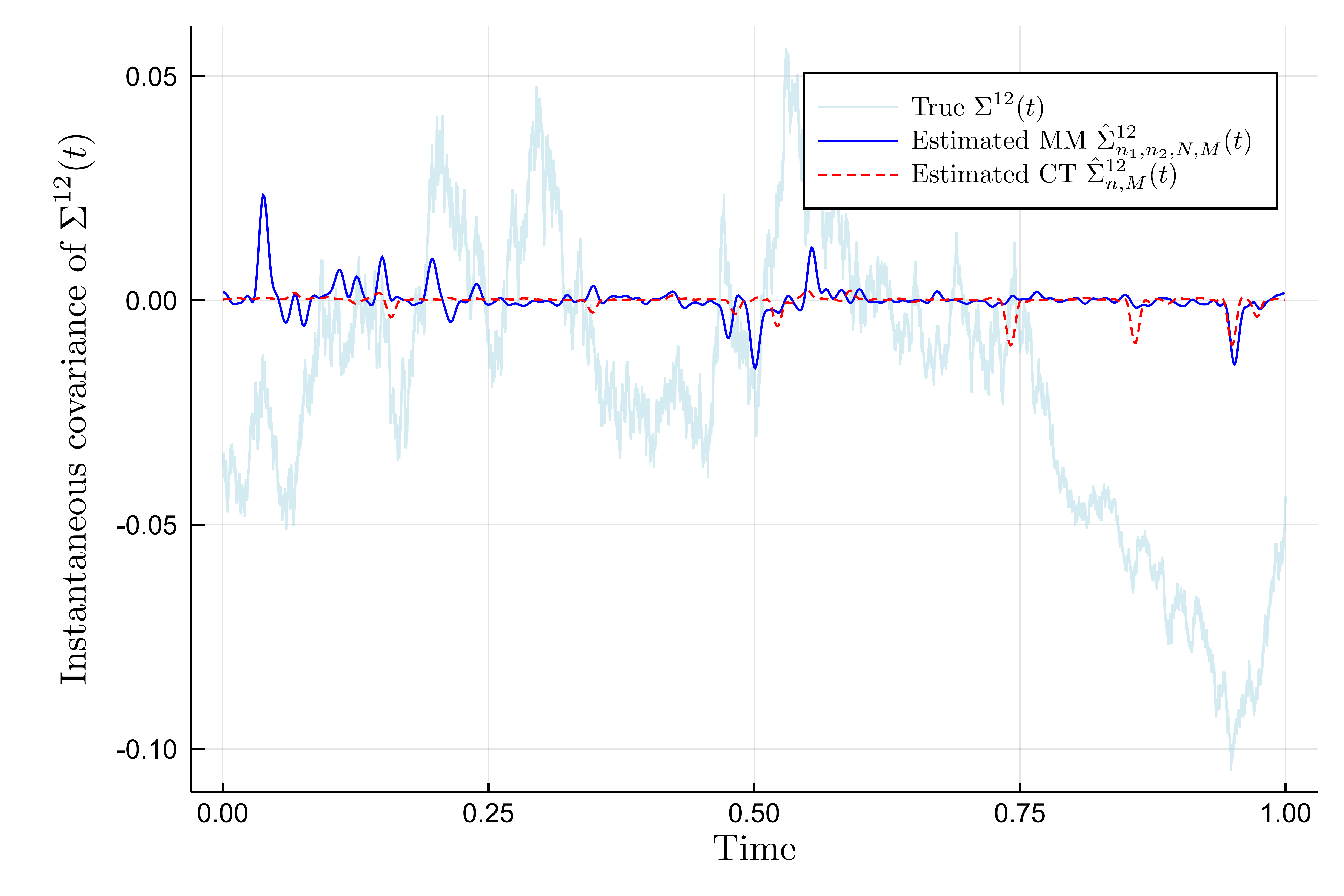}}  \\
    \subfloat[Bates Model $\Sigma^{11}(t)$]{\label{fig:AsynInst:j}\includegraphics[width=0.33\textwidth]{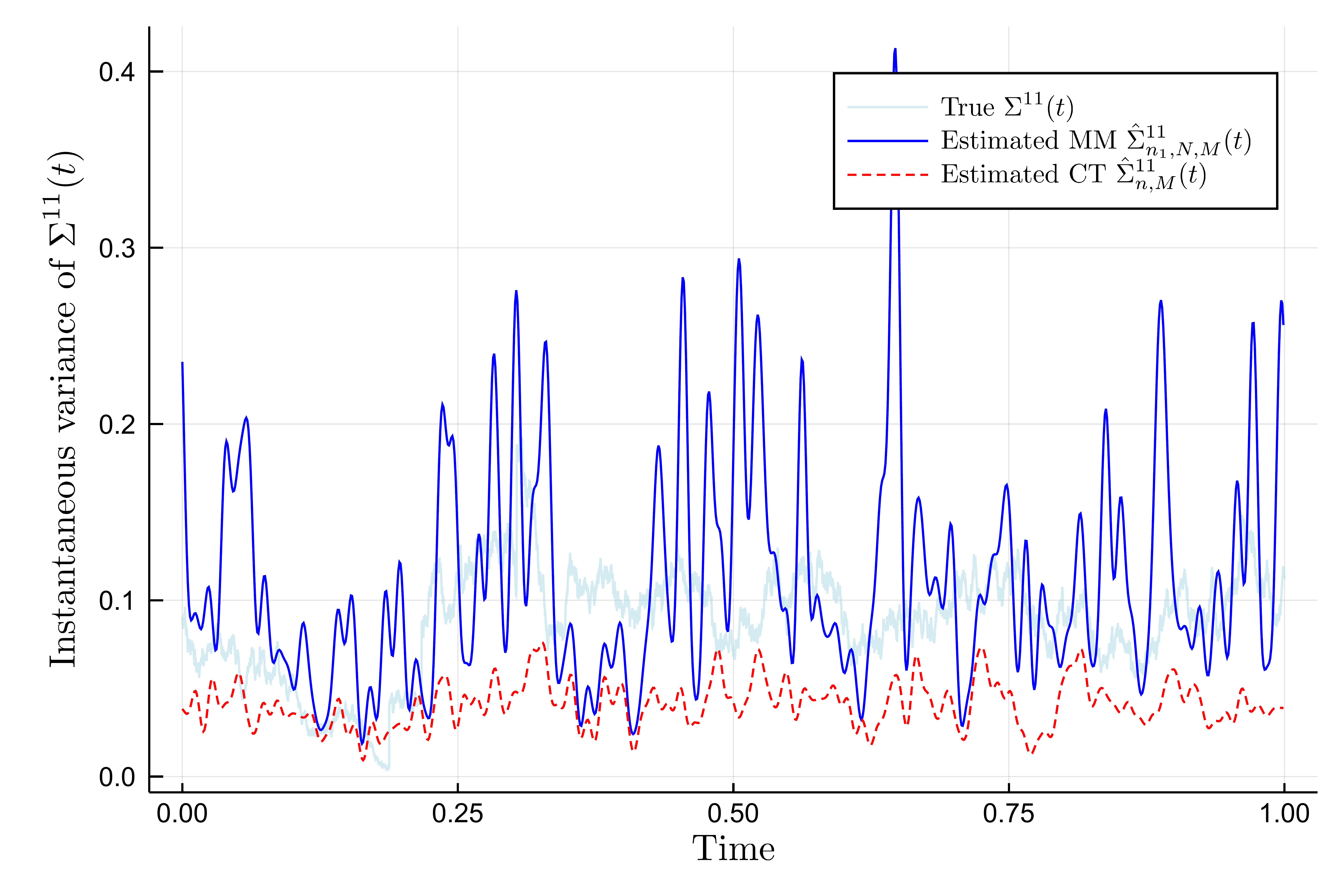}}
    \subfloat[Bates Model $\Sigma^{22}(t)$]{\label{fig:AsynInst:k}\includegraphics[width=0.33\textwidth]{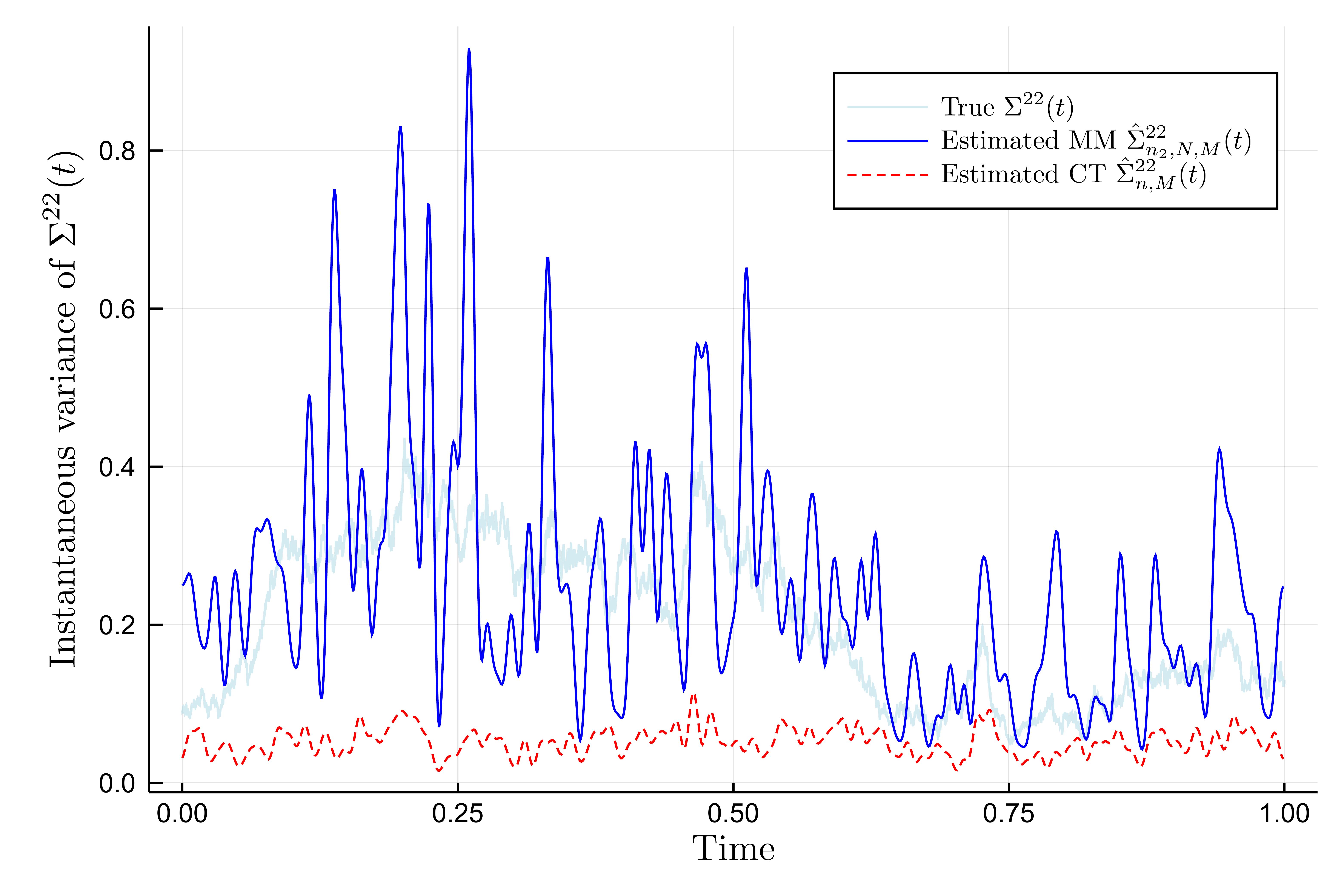}}    
    \subfloat[Bates Model $\Sigma^{12}(t)$]{\label{fig:AsynInst:l}\includegraphics[width=0.33\textwidth]{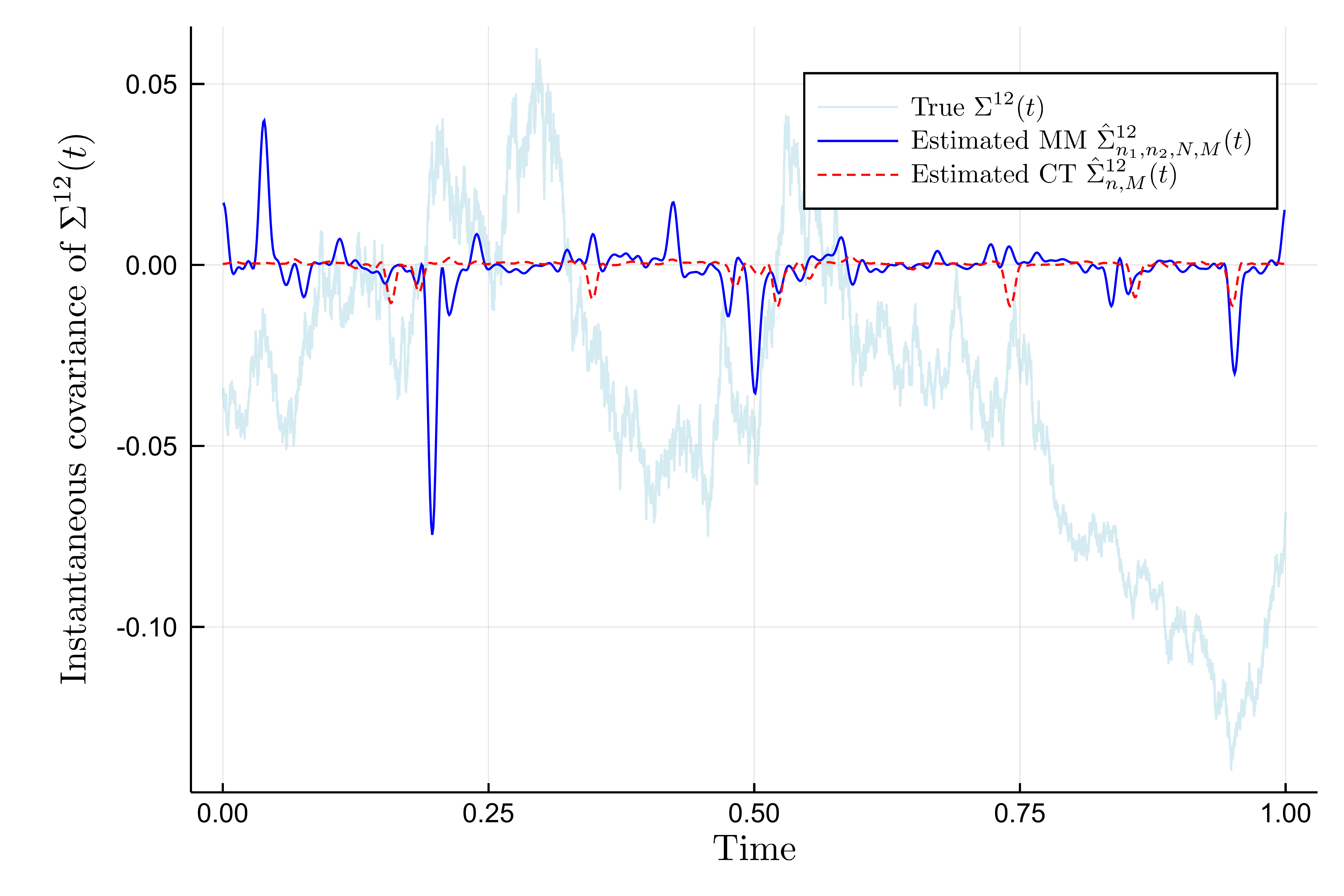}}
    \caption{Here we compare the Malliavin-Mancino (blue line with label ``Estimated MM'') and Cuchiero-Teichmann (red dashes with label ``Estimated CT'') spot volatility/co-volatility estimator for stochastic models with constant volatility against stochastic volatility, and jumps against no jumps under the influence of asynchrony. Asynchrony is introduced by sampling each synchronous grid with $n=28,800$ grid points using an exponential inter-arrival with mean 30, yielding $n_i \approx n / \lambda_i$. A time-scale of $\Delta t = 1$ second is investigated. The spot estimates are compared against the true underlying instantaneous volatility matrix (black and light-blue lines with label ``True''). We see that the Cuchiero-Teichmann estimator under-estimates the volatilities due to the multiple zero returns caused by previous tick interpolation, while the Malliavin-Mancino estimator recovers the volatility estimates but the estimates are more volatile as a result of asynchrony. The effect of jumps increasing the volatility of the estimates are further confounded with the effects of asynchrony in the Malliavin-Mancino estimator. Both estimators have co-volatility estimates around zero due to the Epps effect arising from asynchrony.}
\label{fig:AsynInst}
\end{figure*}

Let us investigate the additional impact of asynchrony on top of jumps against no jumps for the spot volatility estimators. Asynchrony is introduced using an arrival time representation, where the inter-arrival time between trades follow an exponential distribution with parameter $\lambda$ (also known as Poissonian sampling). Thus the mean inter-arrival time is given as $1/\lambda$.

The Cuchiero-Teichmann estimator does not have the ability to deal with asynchronous observations, so we need a method to synchronise the data at an appropriate time-scale. We will use the previous tick interpolation to impute the asynchronous observations onto a new uniform grid. To this end, let $U^i = \{ t_k^i \}_{k \in \mathbb{Z}}$, be the set of asynchronous arrival times observed between $\left[0, T\right]$ for asset $i$. The synchronised process is then given by $\tilde{X}^i_t = X^i_{\gamma_i(t)}$, where $\gamma_i(t) = \max \{t_k^i : t_k^i \leq t \}$ for $t \in \left[0,T\right]$. The resulting synchronised process is piece-wise constant with jumps at $t_k^i \in U^i$. The new uniform grid has width $\Delta t$, which will be the time-scale of investigation.

The Malliavin-Mancino estimator deals with asynchrony by performing the operations in the frequency domain where asynchrony is not an issue anymore. Moreover, the estimator has an implicit method to investigate various time-scales through an appropriate choice of $N$. This is discussed in \cite{PCEPTG2020a}. The link between $\Delta t$ and $N$ is given by:
\begin{equation} \label{eq:comp:6}
    N =  \left\lfloor \frac{1}{2} \left( \frac{T}{\Delta t} - 1 \right) \right\rfloor,
\end{equation}
where $T$ is the entire interval of investigation. $T$ should be measured in the same units as $\Delta t$, which is seconds in our case. This means $T=28,800$ seconds for an 8 hour trading day. Note that the conversion may not be exact because $N$ is an integer.

We simulate $n = 28,800$ synchronous grid points for the above processes. These are then each sampled with an exponential with a mean of $1/\lambda_i = 30$ seconds. Thus $n_i \approx T/\lambda_i$ asynchronous observations for the Malliavin-Mancino estimator. We pick $N$ such that $\Delta t = 1$ second is investigated. For the Cuchiero-Teichmann estimator, a new synchronised grid $\tilde{X}^i_t$ is constructed using the previous tick interpolation with $\Delta t = 1$ second. The reconstruction frequency $M$ is chosen to again be 100 so that comparisons can be made with \Cref{fig:SynInst}.

\Cref{fig:AsynInst} compares the true underlying volatility (black and light-blue lines) against the Malliavin-Mancino (blue lines) and Cuchiero-Teichmann (red dashes) spot volatility estimates under the presence of asynchrony. The rows of the figures are: GBM, Merton, Heston, and Bates model from first to last. The columns of the figures are: $\Sigma^{11}(t)$, $\Sigma^{22}(t)$, and $\Sigma^{12}(t)$ from first to last. Before in \Cref{fig:SynInst}, the Cuchiero-Teichmann could recover the entire volatility matrix with high fidelity for the models with jumps and no jumps. Here we now see that the volatility ($\Sigma^{ii}(t)$, $i=1,2$) is under-estimated. This is because of the previous tick interpolation, since $\Delta t = 1$ the synchronised process results in many zero returns and results in lower volatility estimates. On the other hand, here the Malliavin-Mancino estimator can recover the volatility. However, it seems that asynchrony makes the Malliavin-Mancino volatility estimates significantly more volatile with regions of large deviation from the true volatility. Moreover, the spikes in volatility from jumps are confounded with the increased volatility in the estimates caused by asynchrony. For the case of the co-volatility, both spot volatility estimators have estimates around zero. For the Malliavin-Mancino estimator there are spikes in volatility around zero and the effect is enhanced where there are jumps. The reason behind the decay of the covariance estimate is due to the Epps effect, which is the decay of correlations as the sampling interval $\Delta t$ decreases (See \cite{EPPS1979}). 

We will investigate the impact of the Epps effect arising from asynchrony for the instantaneous correlation in \Cref{subsec:cutfreqN,subsec:cutfreq_OT}. Here we have demonstrated the effect of small $\Delta t$ (large $N$) on the individual volatility components. The results are consistent with the case of the integrated volatility. \cite{MRS2017} performed a bias-MSE analysis for the Malliavin-Mancino integrated covariance estimates under the conditions of asynchrony. They found that the integrated volatility presents little bias for any choice of $N$, while larger $N$ results in a larger bias for the case of the integrated co-volatility which we have demonstrated in the case of the instantaneous volatility.

% The impact of the Epps effect arising from asynchrony on the integrated covariance has been investigated in depth (See \cite{RENO2001, PI2007, MMZ2011, MSG2011, TK2007, TK2009, PCEPTG2020b}). For the case of the integrated covariance, it is understood that in order to avoid the Epps effect caused by asynchrony $N$ must be small \citep{RENO2001}. \cite{MRS2017} perform as bias-MSE analysis for the Malliavin-Mancino integrated covariance estimates under the conditions of asynchrony. They found that the integrated volatility presents little bias for any choice of $N$, while larger $N$ results in a larger bias for the case of the integrated co-volatility. The models in \Cref{fig:AsynInst} with no jumps demonstrate this for the case of instantaneous volatility/co-volatility with large $N$. \Cref{subsec:cutfreqN,subsec:cutfreq_OT} further investigates the impact of the Epps effect arising from asynchrony for the instantaneous correlation.

\section{Cutting frequencies}\label{sec:cutfreq}

In \Cref{sec:comparison}, the cutting frequencies were chosen somewhat arbitrarily. However, the impact of the cutting frequencies $N$ and $M$ play a pivotal role in the Fourier spot volatility estimates. For the Malliavin-Mancino estimator, $N$ and $M$ affects the asymptotic properties and convergence rates of the estimators \citep{MR2015, MRS2017, Chen2019}. The level of averaging $N$ determines the time-scale for which to estimate \cref{eq:instantMM:3}, which controls for the impact of the Epps effect caused by asynchrony. The reconstruction frequency $M$ determines the accuracy of the approximation, the time-scale for the reconstruction of the volatility matrix, and also affects the volatility of volatility estimation either through the path constructed \citep{CT2015}, or by the number of Fourier modes used in the estimation \citep{SCM2015}.

Several choices for $N$ and $M$ have been put forward in the literature. \cite{MRS2017} suggest to use $N = n/2$ and $M = \frac{1}{2 \pi} \frac{1}{8} \sqrt{n} \log n$ for the synchronous case; $N = 0.85 n^{3/4}$ and $M = \frac{1}{2 \pi} \frac{1}{8} \sqrt{n^{3/4}} \log n^{3/4}$ for a special case of asynchrony where $n=n_i=n_j$ and one of the process is observed on non-equidistant grid points and the other on a synchronous grid. Their choice comes from satisfying asymptotic properties and simulation experiments. \cite{Chen2019} suggests $N \leq \lfloor \underline{n} / 2 \rfloor - M$ for the synchronous case, and $N = o( \underline{n}^{4/5} )$ for the asynchronous case where $\underline{n} = \min_i n_i$. \cite{Chen2019} places the additional restriction that $N+M$ must be less than equal the Nyquist frequency (of the price process). This is because ultimately \cref{eq:instantMM:3,eq:instantMM:4} are estimated from the Fourier coefficients of the price process \cref{eq:instantMM:2}. Thus the condition is placed to avoid aliasing. \cite{Chen2019} motivates these choices through convergence and asymptotic properties. \cite{MI2010} use the Nyquist frequencies $N = n/2$ and $M = N/2$. However, they use a modified Fej\'{e}r kernel from \cite{MT2006} containing a positive parameter $\delta$ which allows them to tune the process to the desired time-scale. They adopt a different approach by choosing $\delta$ to minimise the MSE in an attempt to determine if there exists an optimal time-scales for reconstructing the instantaneous volatility matrix. Here we visualise the impact of $M$ and $N$ in order to design a pragmatic approach to pick $N$ and $M$.

\subsection{Impact of M}\label{subsec:cutfreqM}

\begin{figure*}[p]
    \centering
    \subfloat[MM Surface plot, $N$ = Nyquist]{\label{fig:CF_M:a}\includegraphics[width=0.5\textwidth]{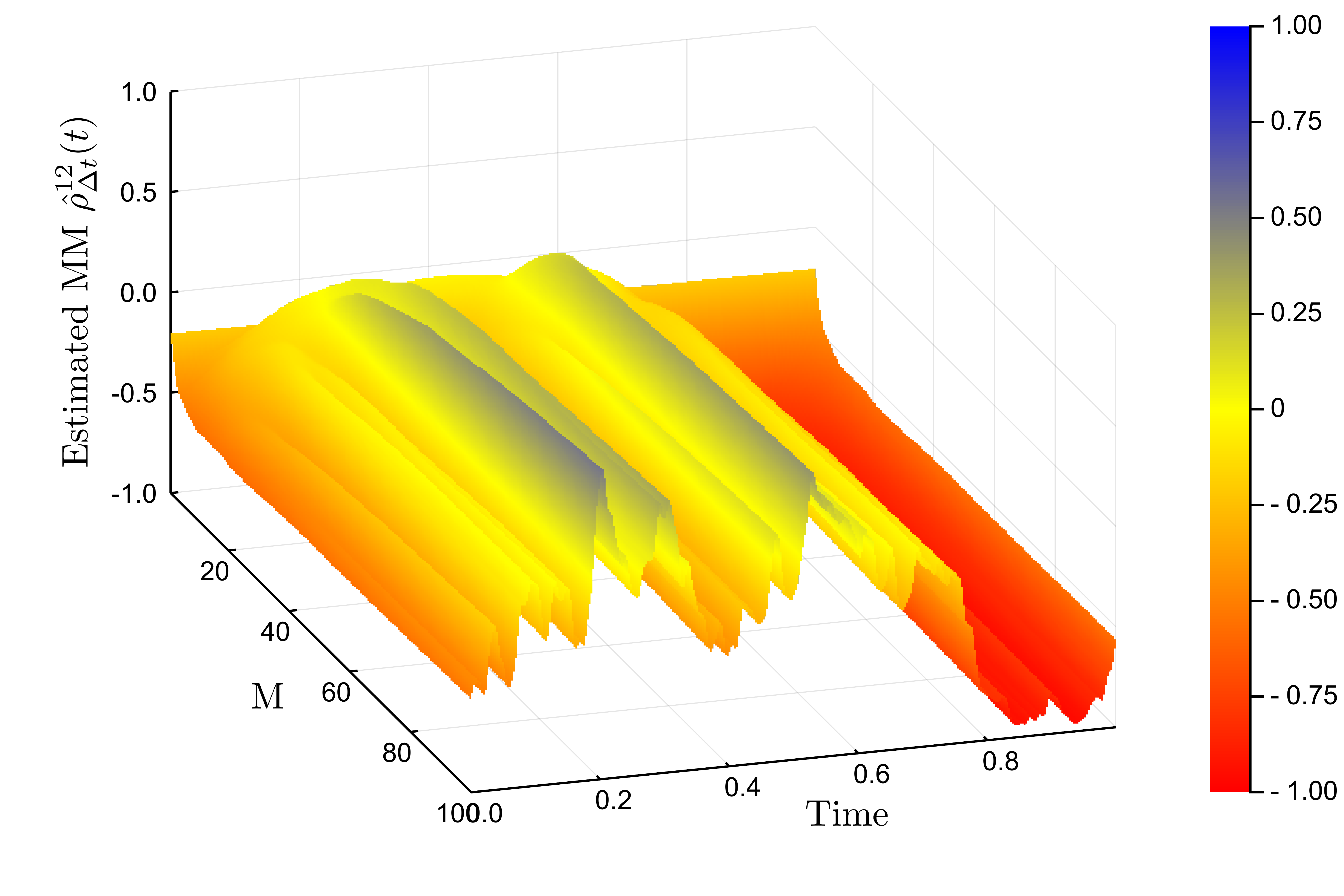}}  
    \subfloat[MM Contour plot, $N$ = Nyquist]{\label{fig:CF_M:b}\includegraphics[width=0.5\textwidth]{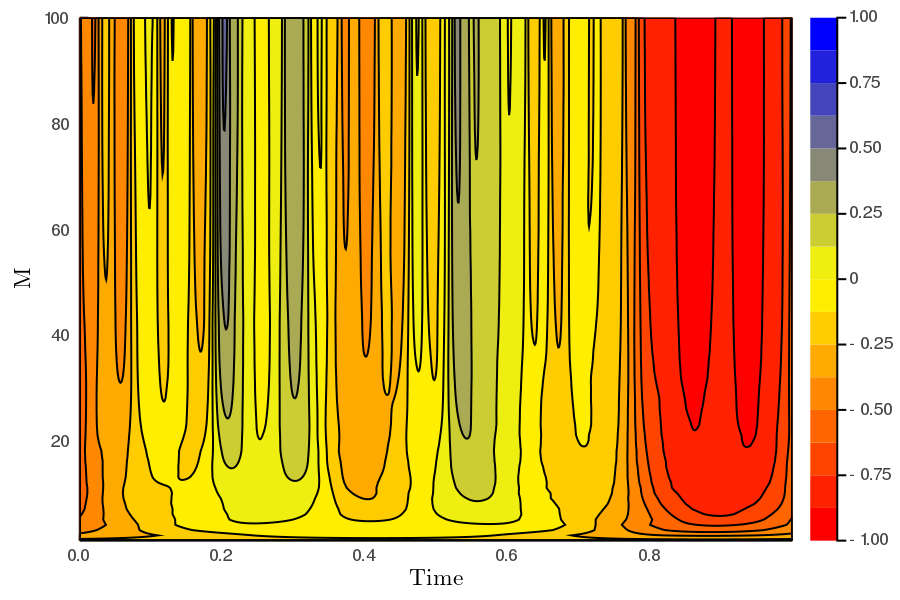}}    \\
    \subfloat[CT Surface plot, $N$ = Nyquist]{\label{fig:CF_M:c}\includegraphics[width=0.5\textwidth]{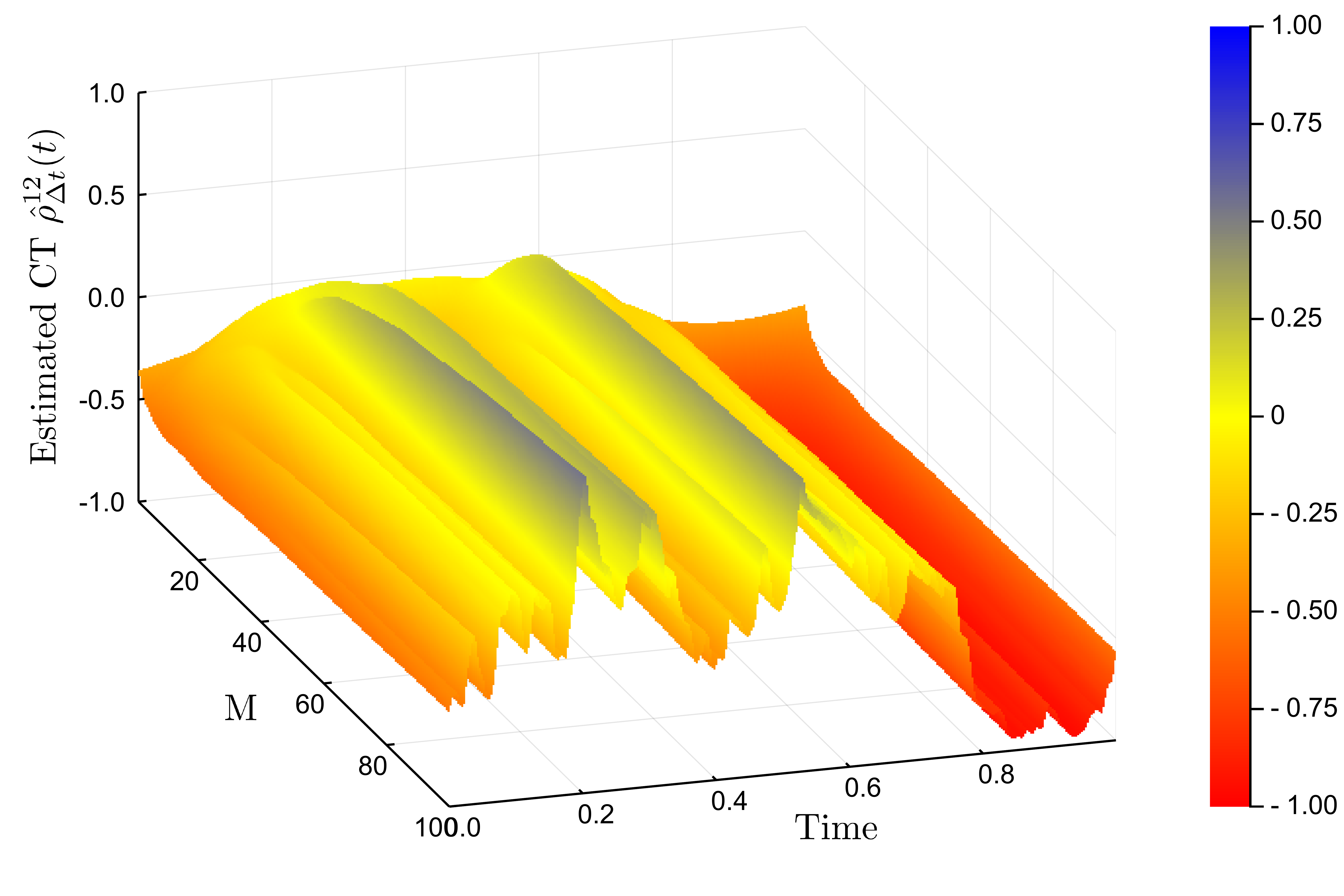}}  
    \subfloat[CT Contour plot, $N$ = Nyquist]{\label{fig:CF_M:d}\includegraphics[width=0.5\textwidth]{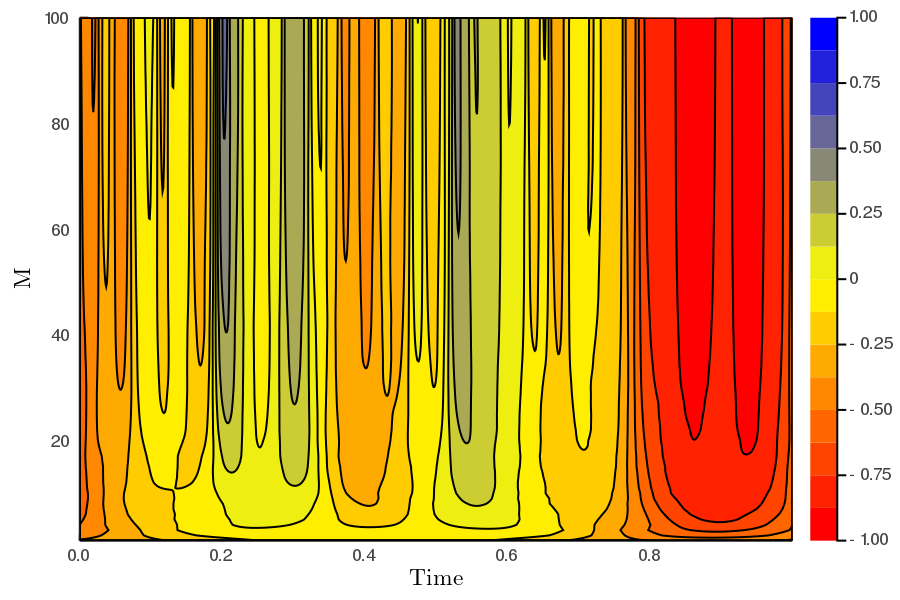}}    \\
    \subfloat[Synchronous, $N$ = Nyquist, $M$ = 100]{\label{fig:CF_M:e}\includegraphics[width=\textwidth, height = 5cm]{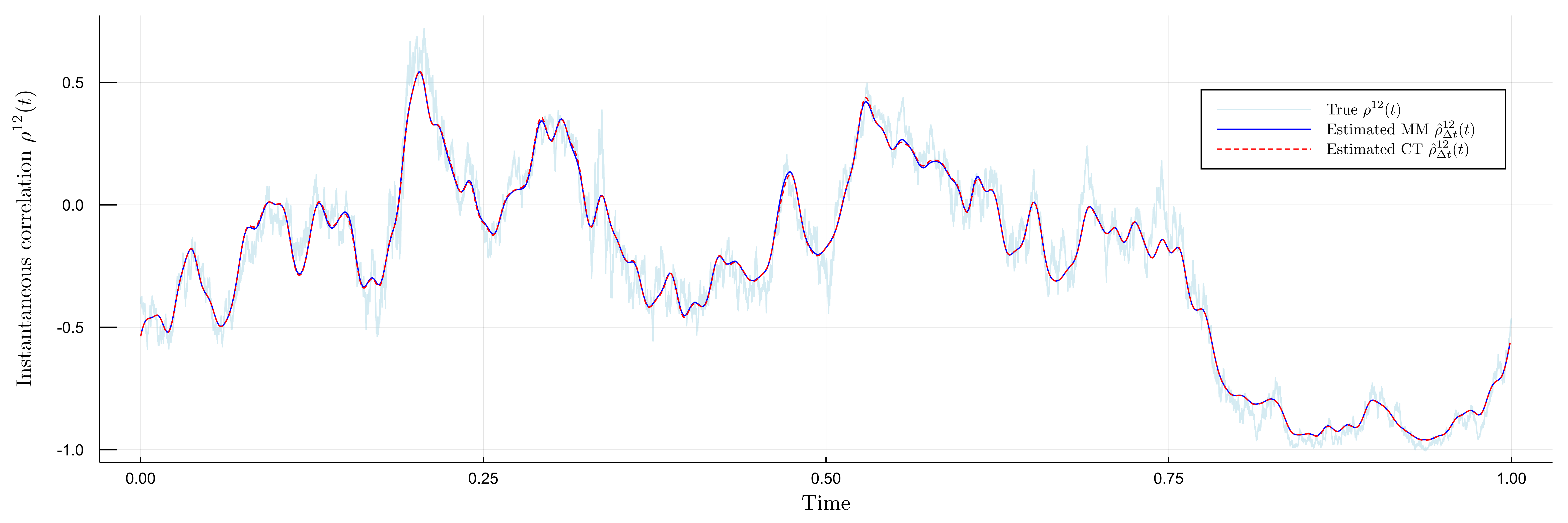}}
    \caption{Here we investigate the impact of $M$ ranging from 1 to 100 on the instantaneous correlation for the synchronous case of a Heston model with parameters in \Cref{tab:param2}. We simulate $n = 28,800$ synchronous grid points. The first row of figures is the Malliavin-Mancino estimates visualised using a surface and contour plot; the second row is that of the Cuchiero-Teichmann estimates. The figure in the last row is a comparison between the Malliavin-Mancino (blue line with label ``Estimated MM'') and the Cuchiero-Teichmann (red dashes with label ``Estimated CT'') estimates against the true instantaneous correlation (light-blue line with label ``True'') for $M=100$. All the figures use the Nyquist frequency for $N$. We see that as $M$ increases the additional harmonics allow us to achieve a better approximation.}
\label{fig:CF_M}
\end{figure*}

The cutting frequency $M$ is the number of harmonics used in the reconstruction of the spot volatility. Let us consider an experiment to demonstrate how this works. We simulate $n = 28,800$ grid points using a Heston model with parameters in \Cref{tab:param2}. First we consider the synchronous case and visualise the impact of various choices of $M$ has on the instantaneous correlation estimate $\hat{\rho}_{\Delta t}^{12}(t)$ in \Cref{fig:CF_M}.

The first row of \Cref{fig:CF_M} are the Malliavin-Mancino spot estimates for $M$ ranging from 1 to 100 plotted as a surface plot and a contour plot; analogous to the first row, the second row are the Cuchiero-Teichmann spot estimates. The last row is the Malliavin-Mancino (blue line) and Cuchiero-Teichmann (red dashes) spot estimates compared against the ground truth (light-blue line) for $M = 100$. Here we follow \cite{MRS2017} and $N$ is chosen to be the Nyquist frequency. From the surface and contour plots, we see how the harmonics build up to achieve a better approximation by allowing the estimates to reach higher peaks and lower troughs. Moreover, we see that when $M$ is too small the spot estimates cannot recover the finer details. However, we note that when $M$ is too large the spot estimates presents a rapid zigzagging behaviour. How one achieves this subtle balance is still an open question relating to if there exist an optimal time-scale to reconstruct the spot volatility estimates \citep{MRS2017}. It must be noted that the choice of $M = N/2$ in \cite{MI2010} does not present the rapid zigzagging behaviour due to the smoothing parameter $\delta$ in the modified Fej\'{e}r kernel.

\begin{figure*}
    \centering
    \subfloat[MM Surface plot, $N=$ Nyquist]{\label{fig:OT_Syn:a}\includegraphics[width=0.33\textwidth]{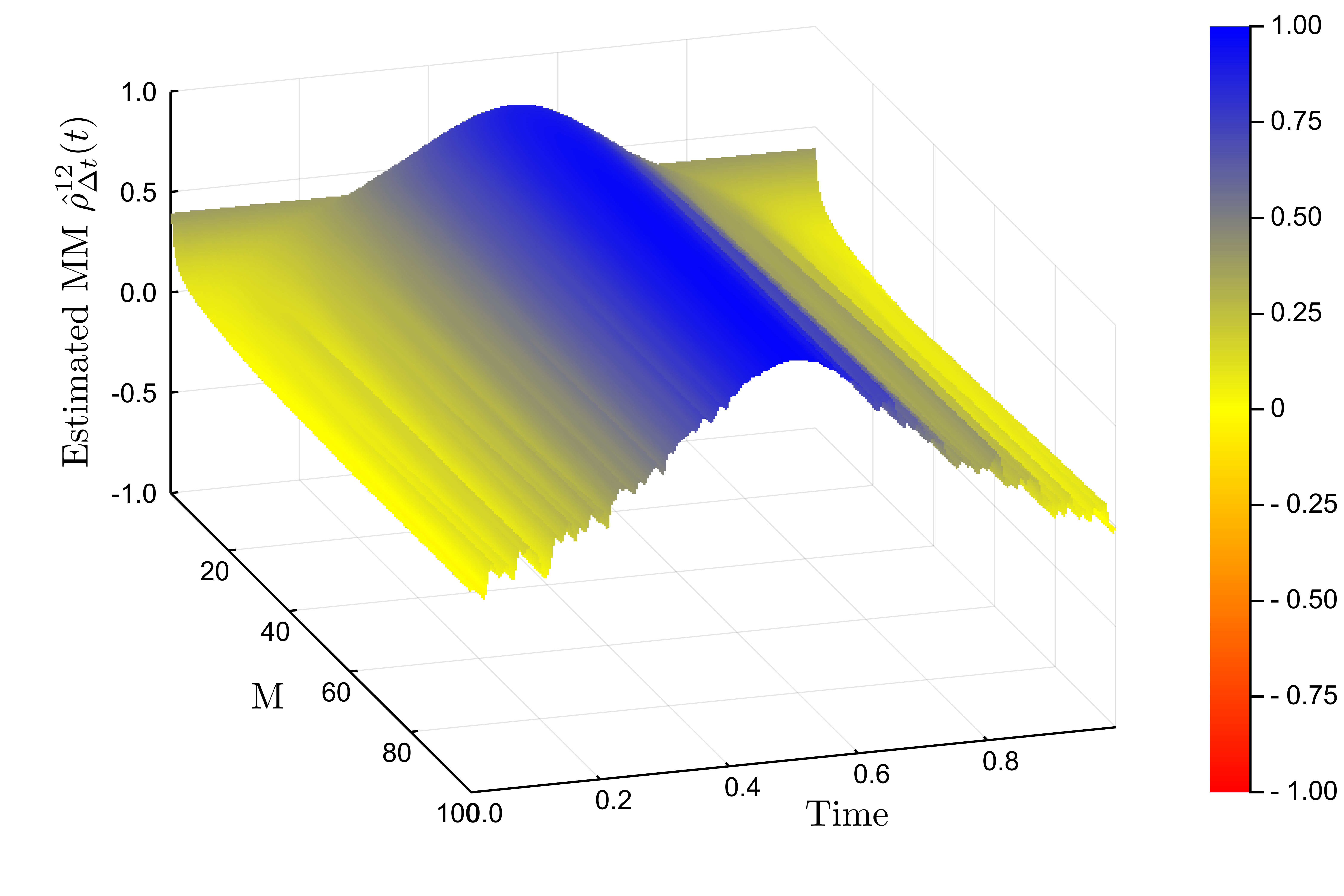}}  
    \subfloat[MM Contour plot, $N=$ Nyquist]{\label{fig:OT_Syn:b}\includegraphics[width=0.33\textwidth]{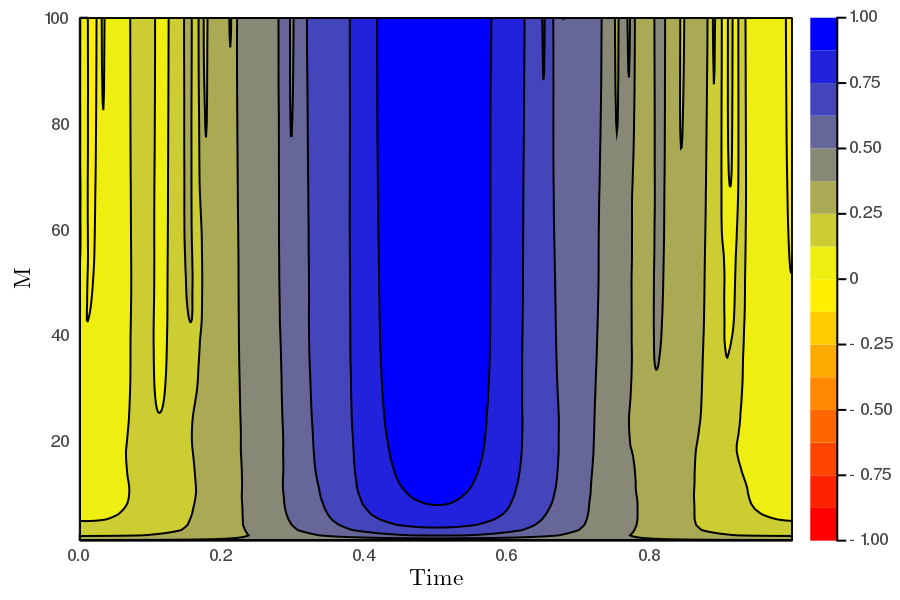}}
    \subfloat[MM, $N=$ Nyquist, $M = 20$]{\label{fig:OT_Syn:c}\includegraphics[width=0.33\textwidth]{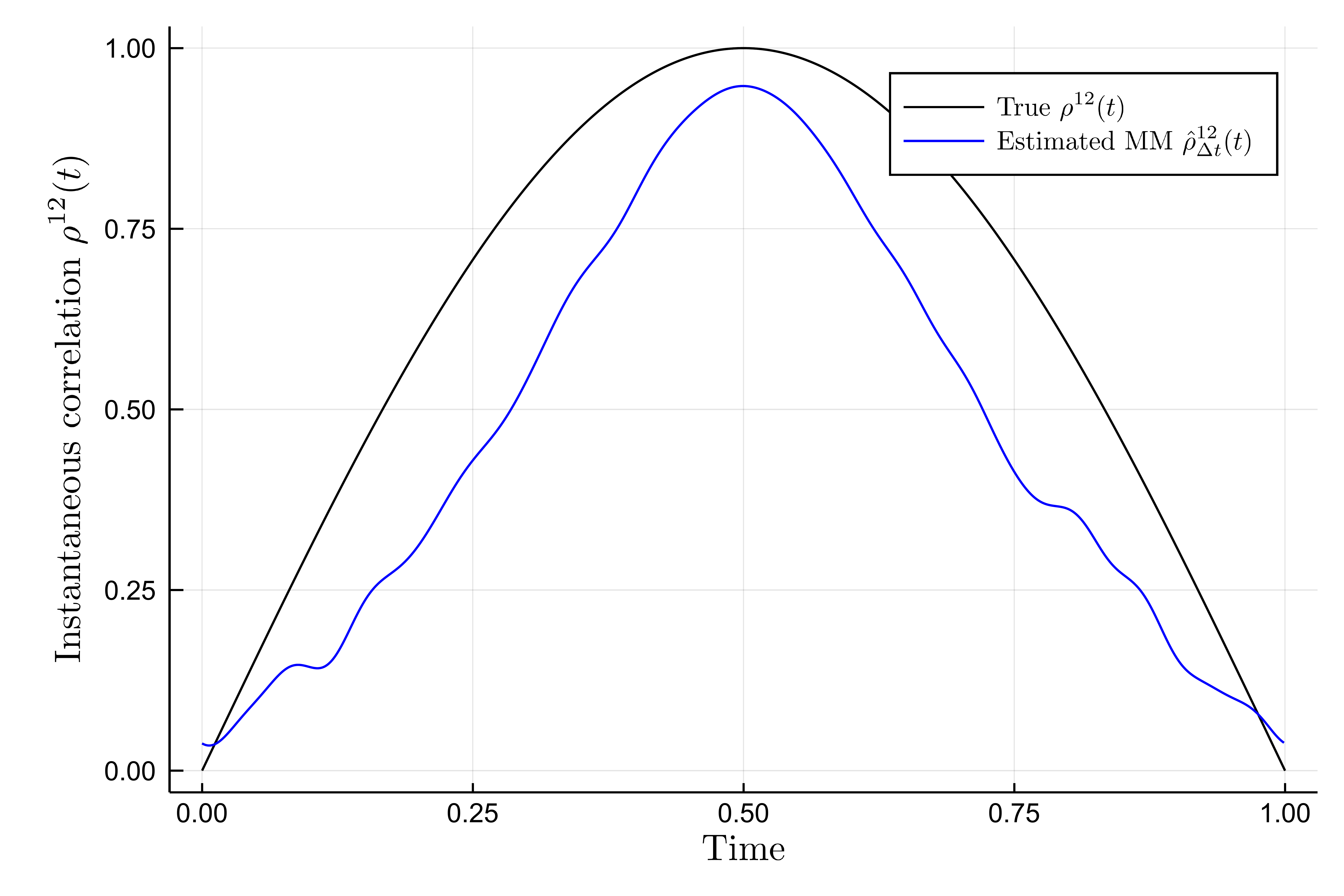}}
    \caption{Here we investigate the ground truth for the simple diffusion model with deterministic correlation in \cref{eq:CF:1,eq:CF:2}. We simulate $n=28,800$ synchronous grid points. \Cref{fig:OT_Syn:a,fig:OT_Syn:b} are the surface and contour plot respectively, investigating the impact of different values for $M$ when $N$ is the Nyquist frequency. \Cref{fig:OT_Syn:c} compares the Malliavin-Mancino spot correlation (blue line with label ``Estimated MM'') using $M=20$ and $N=$ Nyquist against the true instantaneous correlation (black line with label ``True''). We see that for the very simple instantaneous correlation, larger $M$ presents more zigzagging behaviour while a small $M$ is adequate.}
\label{fig:OT_Syn}
\end{figure*}

Based on the preliminary investigation here, having a one size fit all choice for $M$ does not seem to be feasible. This is because a good choice of $M$ depends on the composition of the true spot parameter of interest. For example, the instantaneous stochastic correlation in the Heston model takes on a more complex form and therefore requires more harmonics $M$ to provide a good approximation. If we use the synchronous choice of $M$ from \cite{MRS2017} we obtain $M = 34$ for this experiment, which from \Cref{fig:CF_M} does not provide a good approximation. On the other hand, if the instantaneous correlation takes on a simple shape such as \Cref{fig:OT_Syn} then few harmonics are already enough to obtain a good approximation. Therefore, additional harmonics adds redundant information which results in an unsatisfactory approximation \citep{MI2010}. The conundrum presented here is based on the synchronous case, before adding the additional complication of asynchrony. With the added impact of asynchrony, one needs to first pick an appropriate $N$ to avoid the Epps effect before deciding on an appropriate choice of $M$.

\begin{figure*}[p]
    \centering
    \subfloat[MM, $M$=10]{\label{fig:CF_N:a}\includegraphics[width=0.33\textwidth]{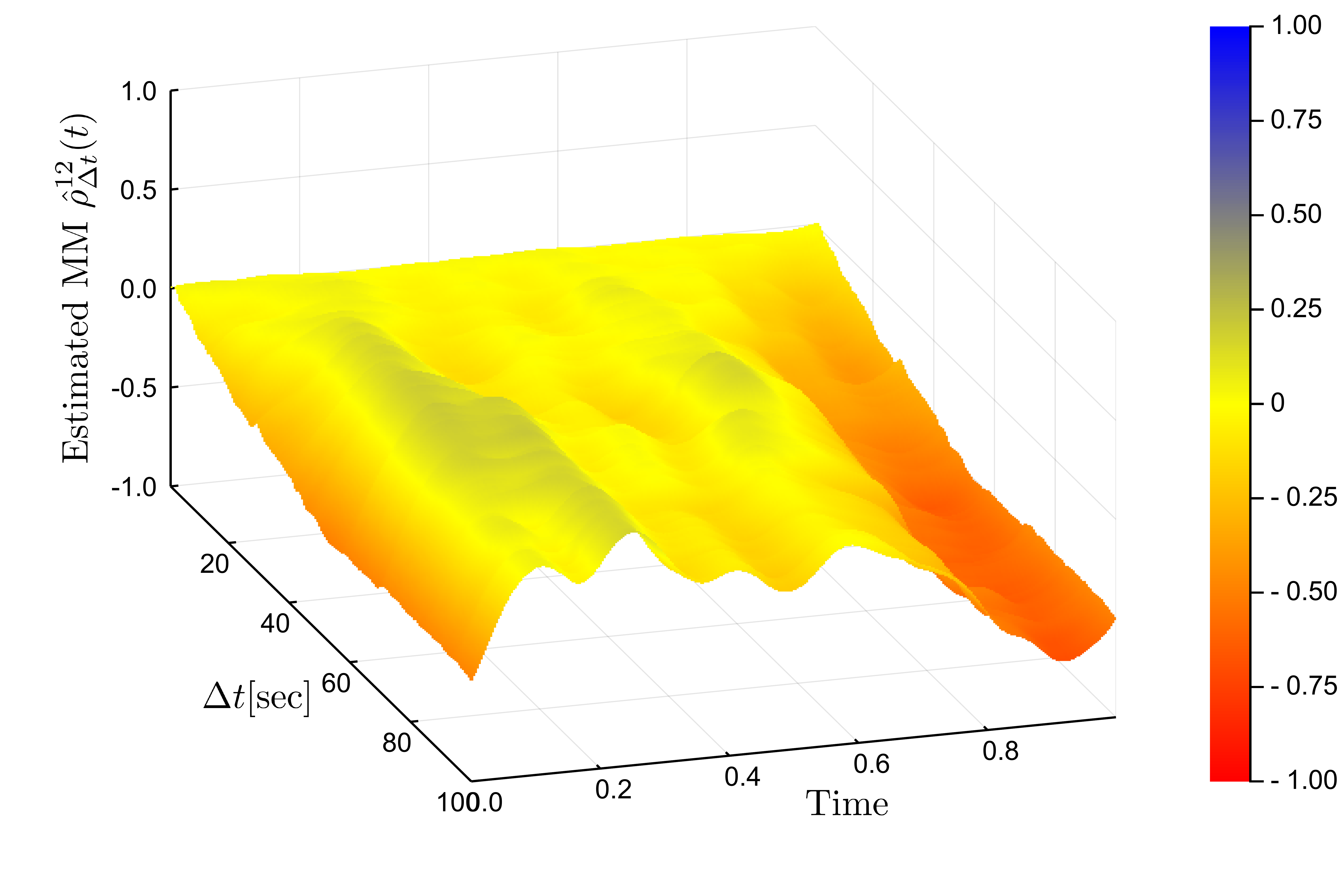}}  
    \subfloat[MM, $M$=20]{\label{fig:CF_N:b}\includegraphics[width=0.33\textwidth]{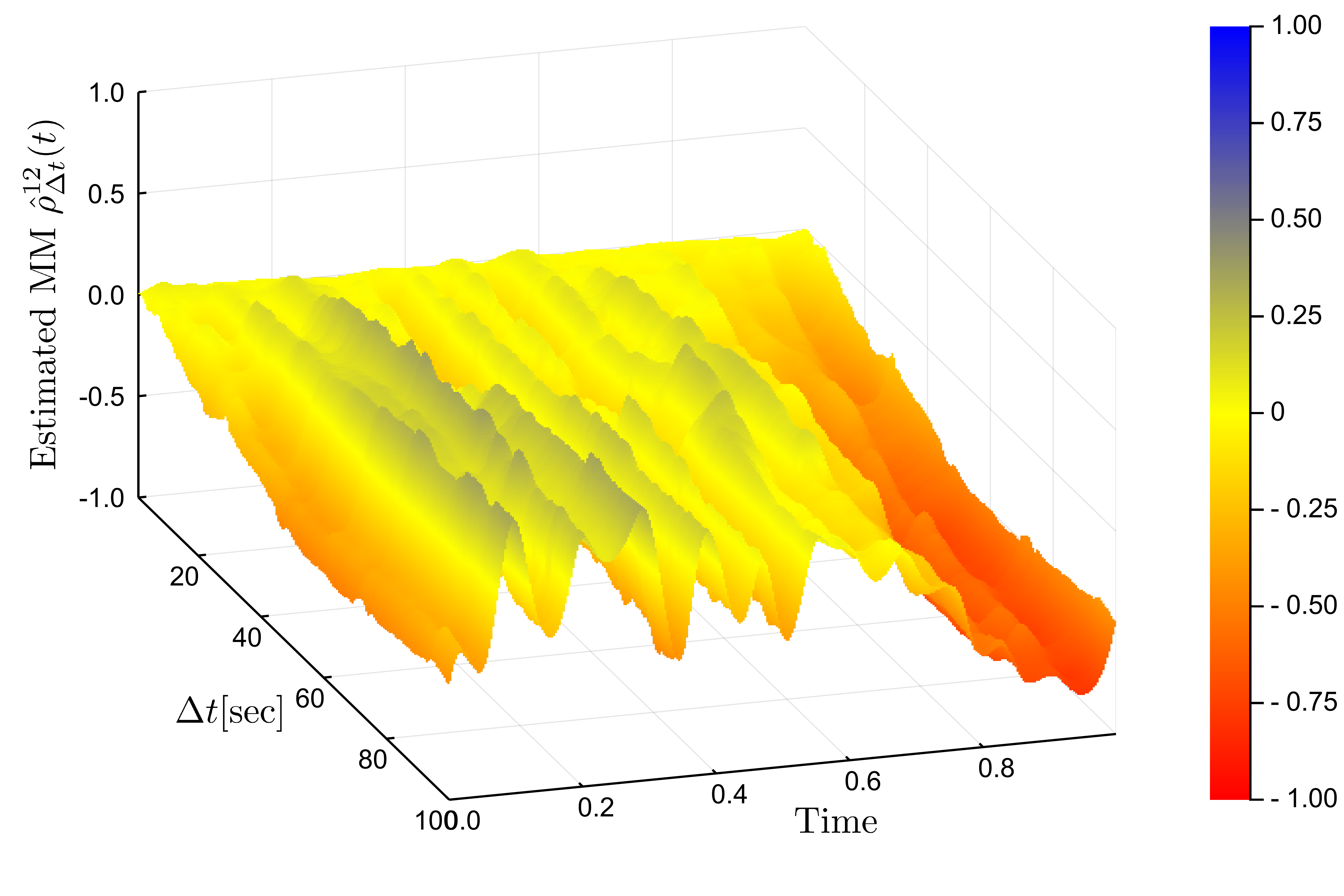}}
    \subfloat[MM, $M$=50]{\label{fig:CF_N:c}\includegraphics[width=0.33\textwidth]{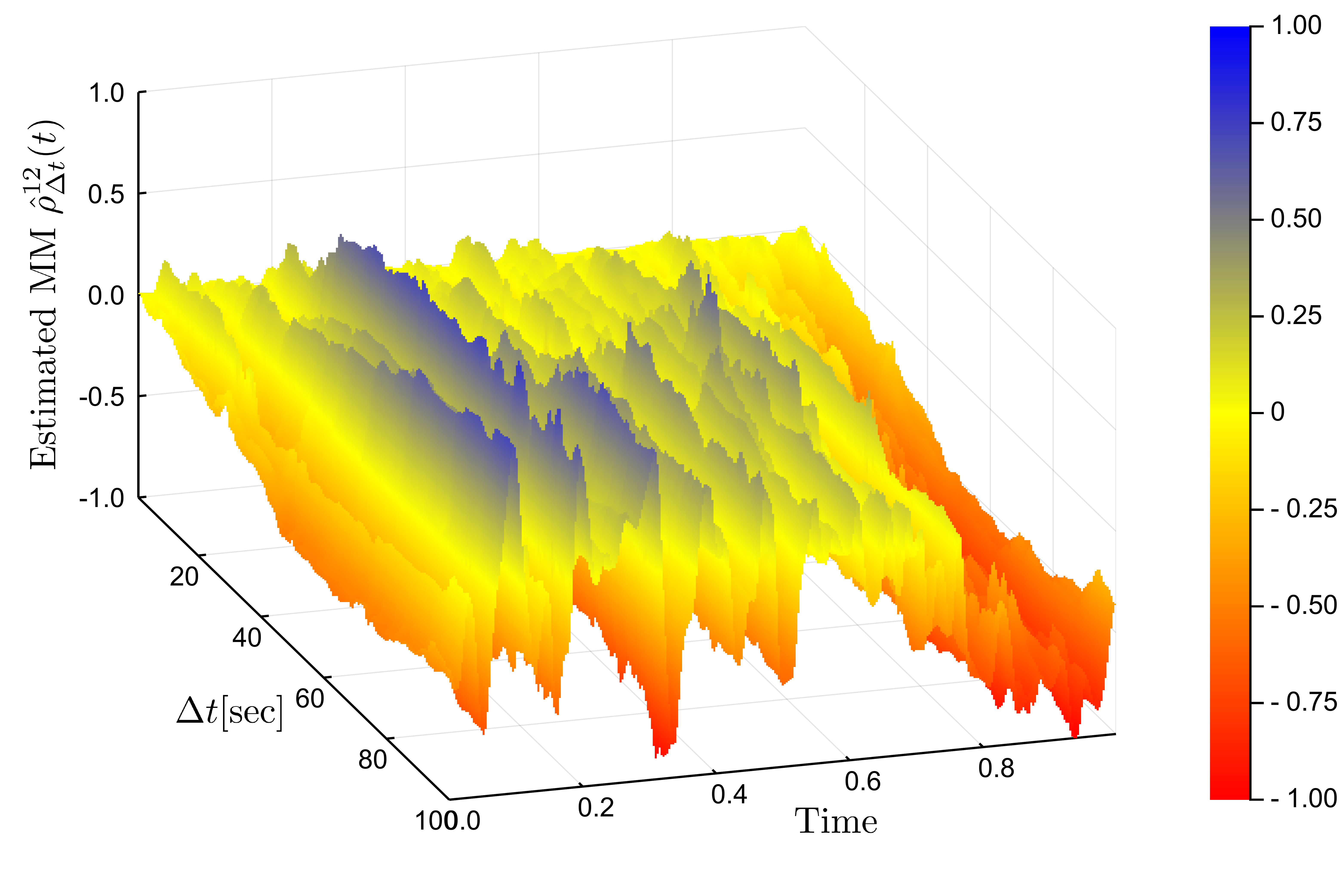}}   \\
    \subfloat[MM, $M$=10]{\label{fig:CF_N:d}\includegraphics[width=0.33\textwidth]{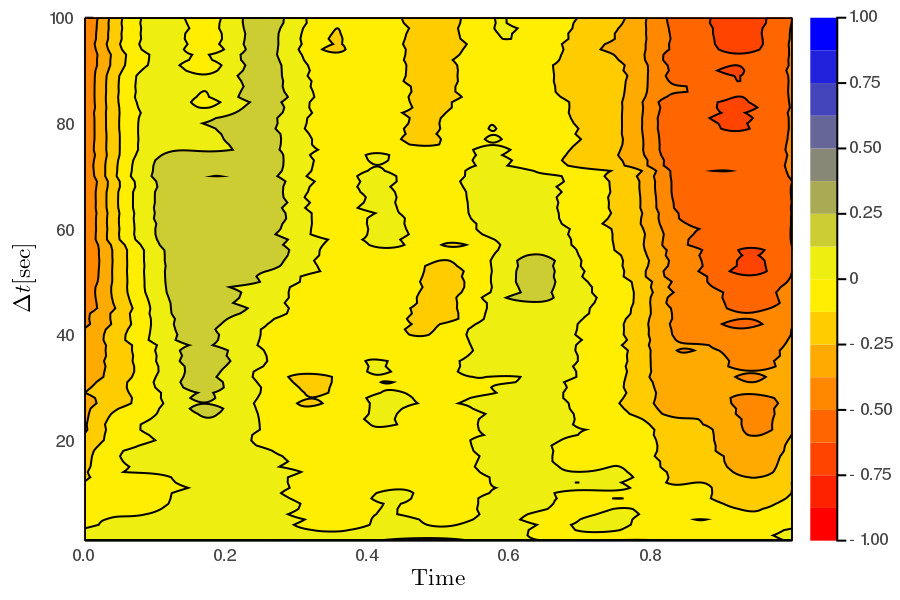}}  
    \subfloat[MM, $M$=20]{\label{fig:CF_N:e}\includegraphics[width=0.33\textwidth]{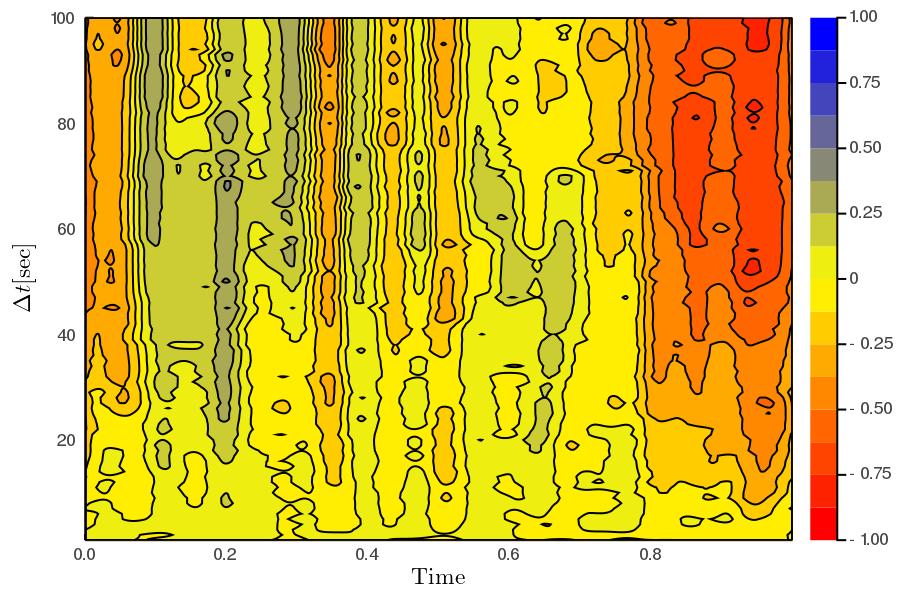}}
    \subfloat[MM, $M$=50]{\label{fig:CF_N:f}\includegraphics[width=0.33\textwidth]{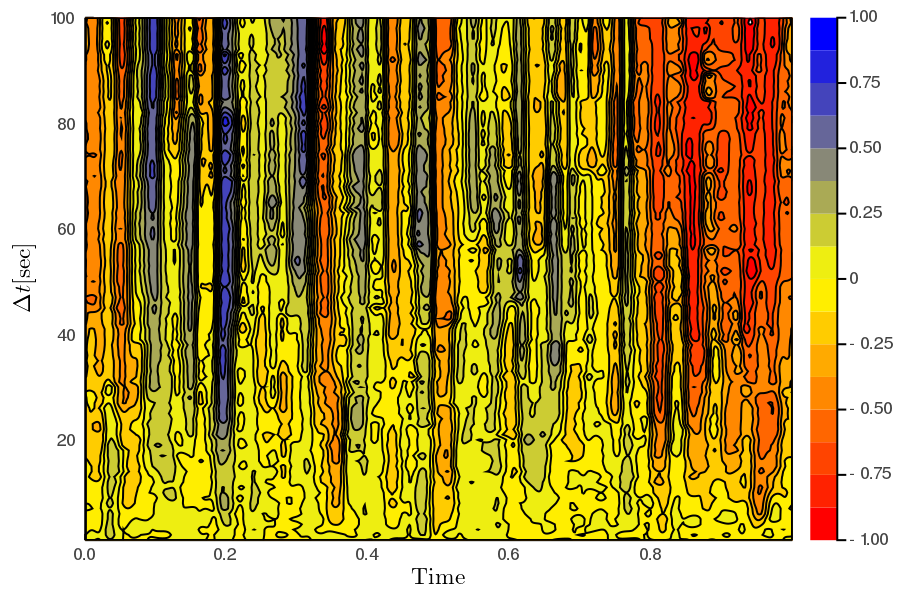}}   \\
    \subfloat[CT, $M$=10]{\label{fig:CF_N:g}\includegraphics[width=0.33\textwidth]{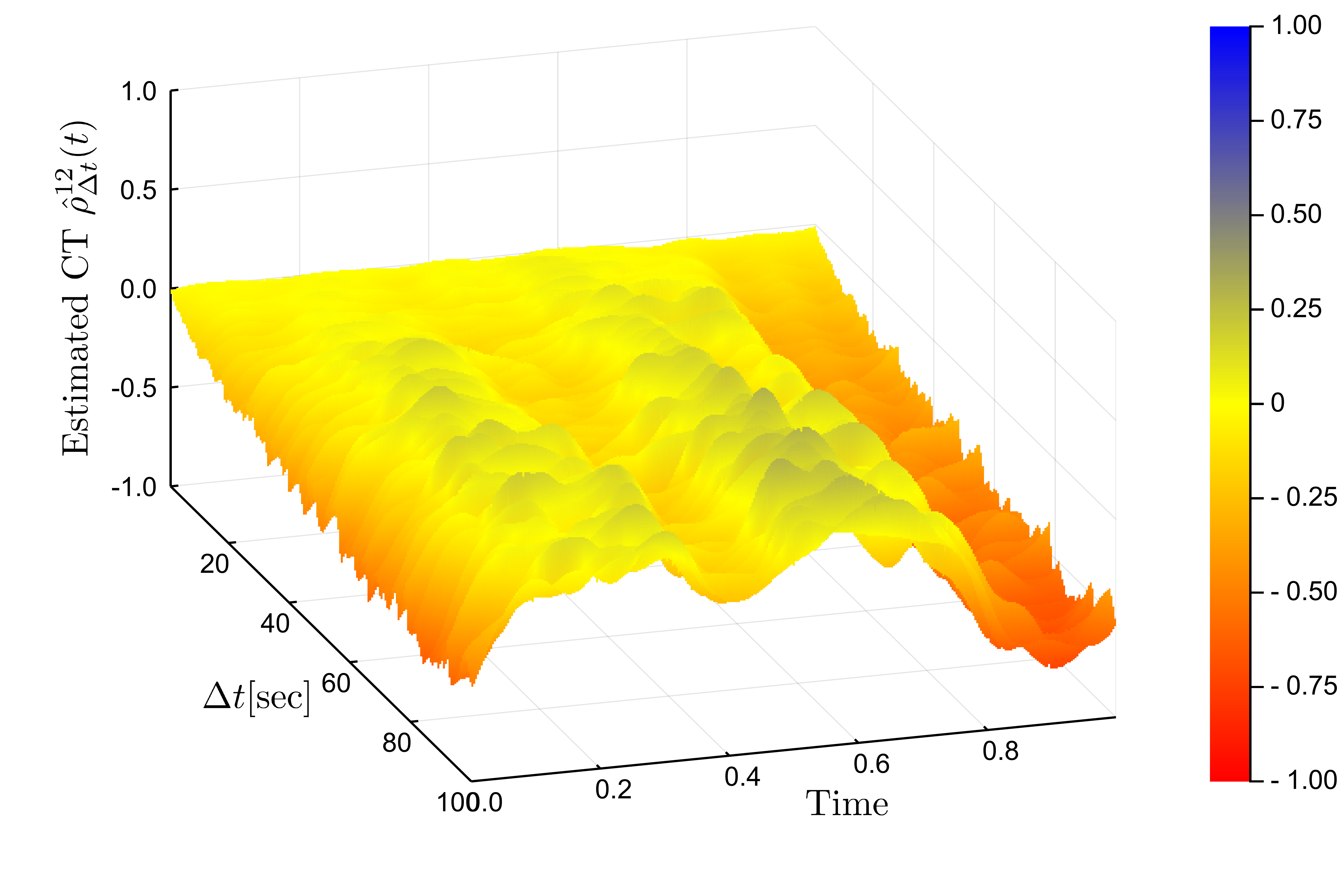}}  
    \subfloat[CT, $M$=20]{\label{fig:CF_N:h}\includegraphics[width=0.33\textwidth]{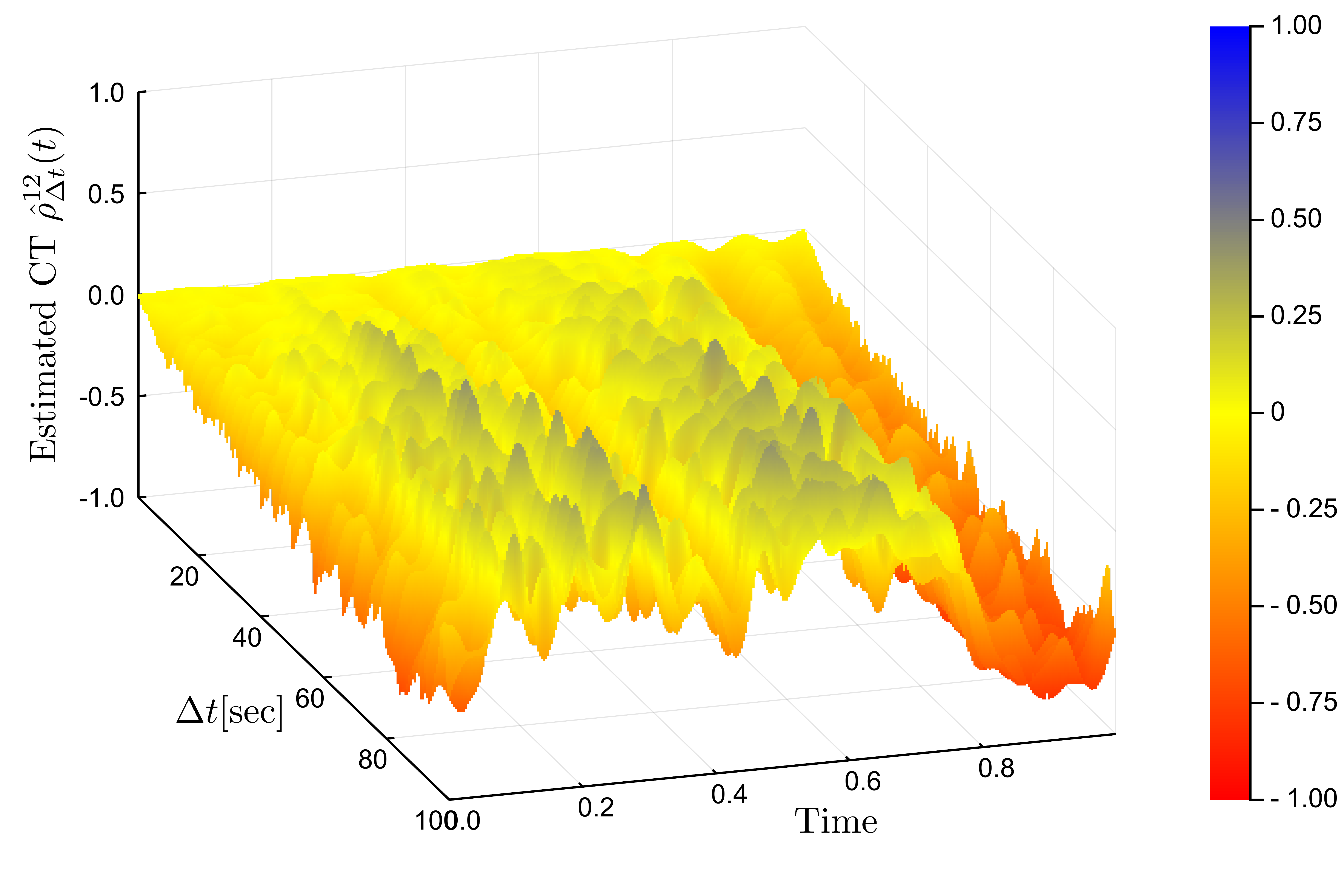}}
    \subfloat[CT, $M$=50]{\label{fig:CF_N:i}\includegraphics[width=0.33\textwidth]{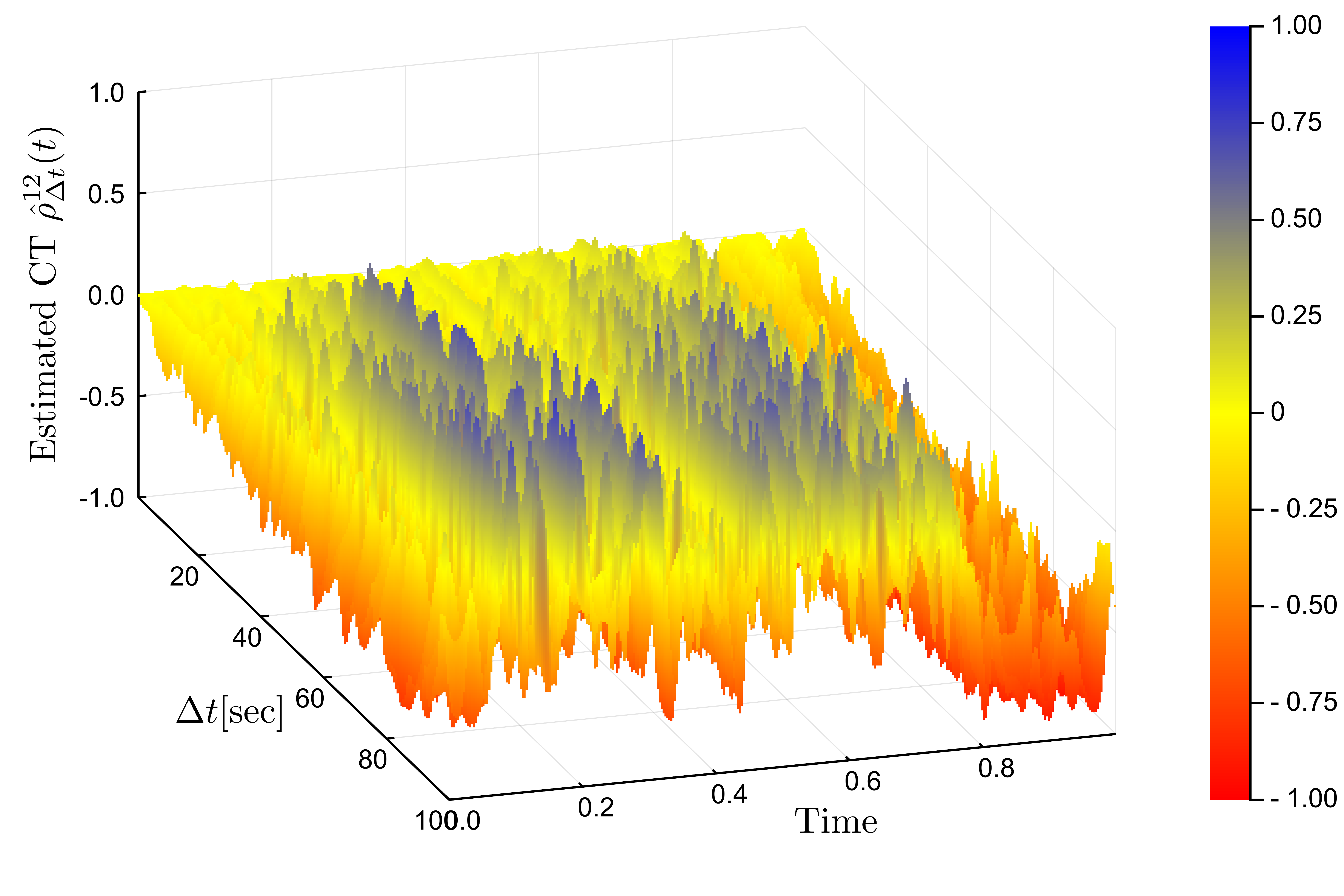}}   \\
    \subfloat[CT, $M$=10]{\label{fig:CF_N:j}\includegraphics[width=0.33\textwidth]{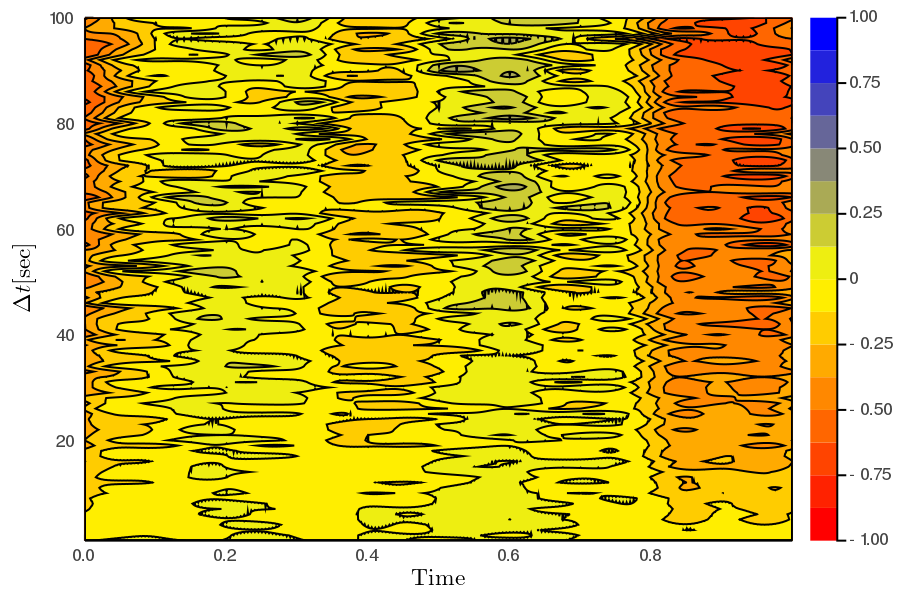}}  
    \subfloat[CT, $M$=20]{\label{fig:CF_N:k}\includegraphics[width=0.33\textwidth]{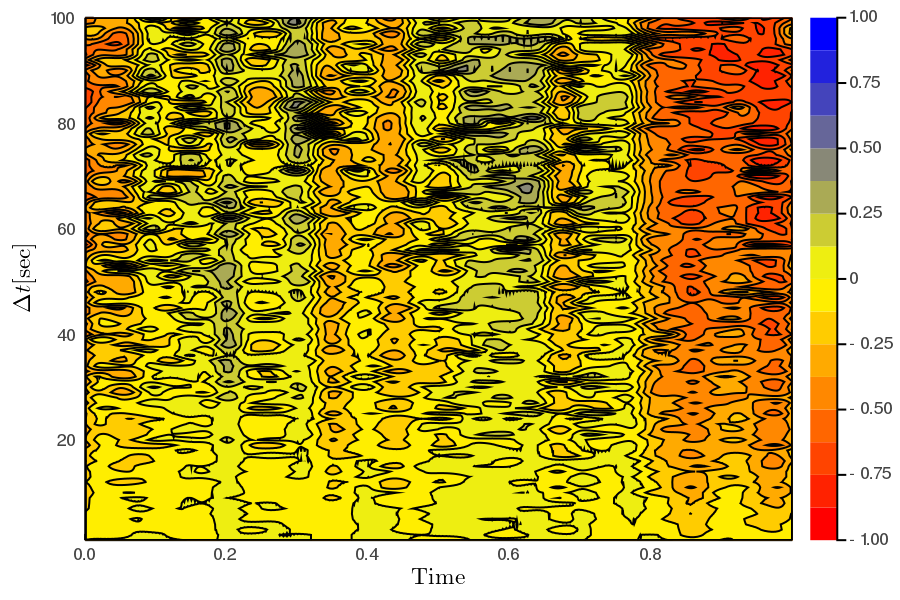}}
    \subfloat[CT, $M$=50]{\label{fig:CF_N:l}\includegraphics[width=0.33\textwidth]{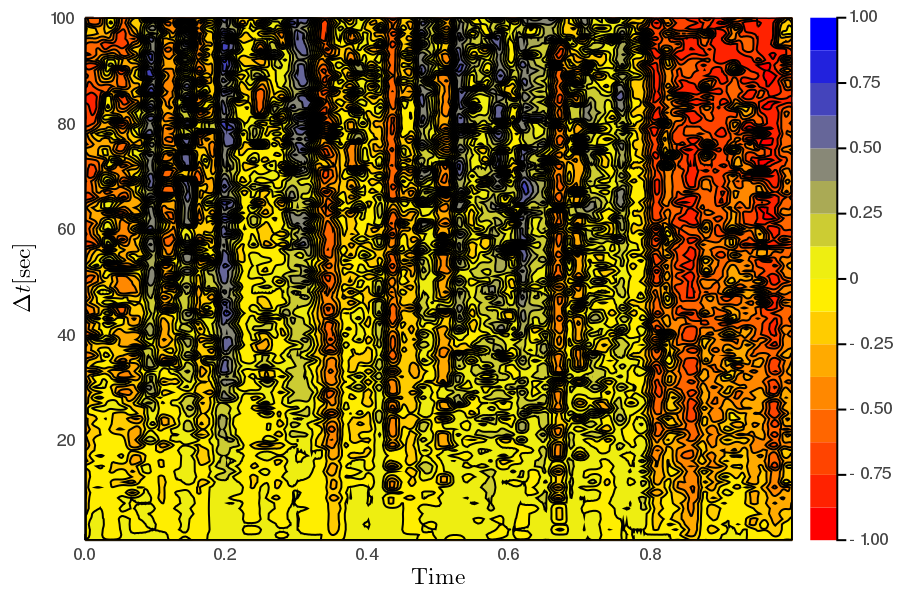}} 
    \caption{Here we investigate the impact of $\Delta t$ ranging from 1 to 100 on the instantaneous correlation of a Heston model with parameters in \Cref{tab:param2} under the presence of asynchrony. We simulate $n = 28,800$ synchronous grid points which is then sampled with an exponential inter-arrival with a mean of 30 seconds, resulting in $n_i \approx n/\lambda_i$. The three columns of the figures are $M=10,20$ and $50$ while the first two rows are the Malliavin-Mancino estimates visualised as surface and contour plots; likewise, the last two rows are that of the Cuchiero-Teichmann estimates. The time-scale $\Delta t$ is controlled using \cref{eq:comp:6} for the Malliavin-Mancino estimates while the time-scale for the Cuchiero-Teichmann estimates are controlled using the previous tick interpolation. We see the visualisation of the instantaneous Epps effect and notice that the Cuchiero-Teichmann estimates are not stable for different values of $\Delta t$ for fixed value of time $t$.}
\label{fig:CF_N}
\end{figure*}

\subsection{Impact of N}\label{subsec:cutfreqN}

The cutting frequency $N$ is the number of Fourier coefficients of the price process used to estimate the Fourier coefficients of the volatility. For the synchronous case, \cite{MRS2017} have argued that without the presence of asynchrony or microstructure noise the Nyquist frequency for $N$ is the best choice. Here we want to investigate the impact that the Epps effect arising from asynchrony has on the instantaneous correlation through the implied time-scale $\Delta t$ by choice of $N$, analogous to the case of integrated correlation in \cite{PCEPTG2020a,RENO2001}. In the case of the Cuchiero-Teichmann estimator, we will use the previous tick interpolation to synchronise the asynchronous samples to a particular time-scale. To this end, we simulate $n = 28,800$ grid points using a Heston model with parameters in \Cref{tab:param2}. The synchronous process is then each sampled with an exponential inter-arrival with a mean of 30 seconds giving $n_i \approx n / \lambda_i$.

\Cref{fig:CF_N} investigates $\Delta t$ ranging from 1 to 100 seconds for three cases of $M$. Concretely, the columns of the figures are $M = 10, 20$ and 50 respectively while the first two rows are the Malliavin-Mancino estimates and the last two rows are the Cuchiero-Teichmann estimates. There are three things to notice in these figures. First, we see that when $\Delta t$ is small the correlation remains around zero; while the correlations starts to emerge as $\Delta t$ increases for the various choices of $M$. This is a demonstration of the instantaneous Epps effect. Second, $M$ is severely affected by asynchrony. The instantaneous correlation presents the zigzagging behaviour for much smaller choices of $M$ relative to the synchronous case. For example, the choice of $M=100$ in \Cref{fig:CF_M:e} resulted in a relatively smooth plot; while for the asynchronous case $M=50$ in \Cref{fig:CF_N:c,fig:CF_N:i} presents more rapid fluctuations. This is especially true for the Cuchiero-Teichmann estimates which leads to the third point. Here the Cuchiero-Teichmann estimates do not only present rapid fluctuations for fixed $\Delta t$ and varying time $t$ (caused by larger $M$), it also presents rapid fluctuations for fixed $t$ and varying $\Delta t$. This means the Cuchiero-Teichmann estimates are highly unstable under the presence of asynchrony for different values of $\Delta t$. This can be seen by the horizontal black marks on the contour plots. These black marks are caused by sudden changes in values and the horizontal nature means that for a specific time $t$ in \cref{eq:instantMM:8}, the estimates can take on very different values for similar values of $\Delta t$. This is different to the black marks of the Malliavin-Mancino estimates in \Cref{fig:CF_N:f} where the marks run vertically down the contour plot. These changes in estimates are from large $M$ rather than the instability from different $\Delta t$ values. This means that for a fixed point in time $t$ of \cref{eq:instantMM:4}, various values of $\Delta t$ do not cause sudden changes in the estimates. This provides an interesting insight into the two methods dealing with asynchrony. Through bypassing the time-domain the Malliavin-Mancino estimator is able to produce stable estimates while the previous tick interpolation produces highly unstable estimates for various $\Delta t$. This is due to the fact that under asynchrony, the Malliavin-Mancino uses all the available observations which are fixed. In the case of using the previous tick interpolation, the synchronised sample path can change under various choices of $\Delta t$. These changes in the synchronised price paths are more apparent for larger $\Delta t$ which is where the instabilities are occurring in \Cref{fig:CF_N:j,fig:CF_N:k,fig:CF_N:l}. This is the source causing the instability. The spot estimates provide an interesting perspective on this because this is not easily seen in the integrated estimates as it gets hidden away in the averaging.

\subsection{Dealing with asynchrony}\label{subsec:cutfreq_OT}

\begin{figure*}[p]
    \centering
    \subfloat[MM, $1/\lambda = 10$]{\label{fig:OT_Asyn:a}\includegraphics[width=0.33\textwidth]{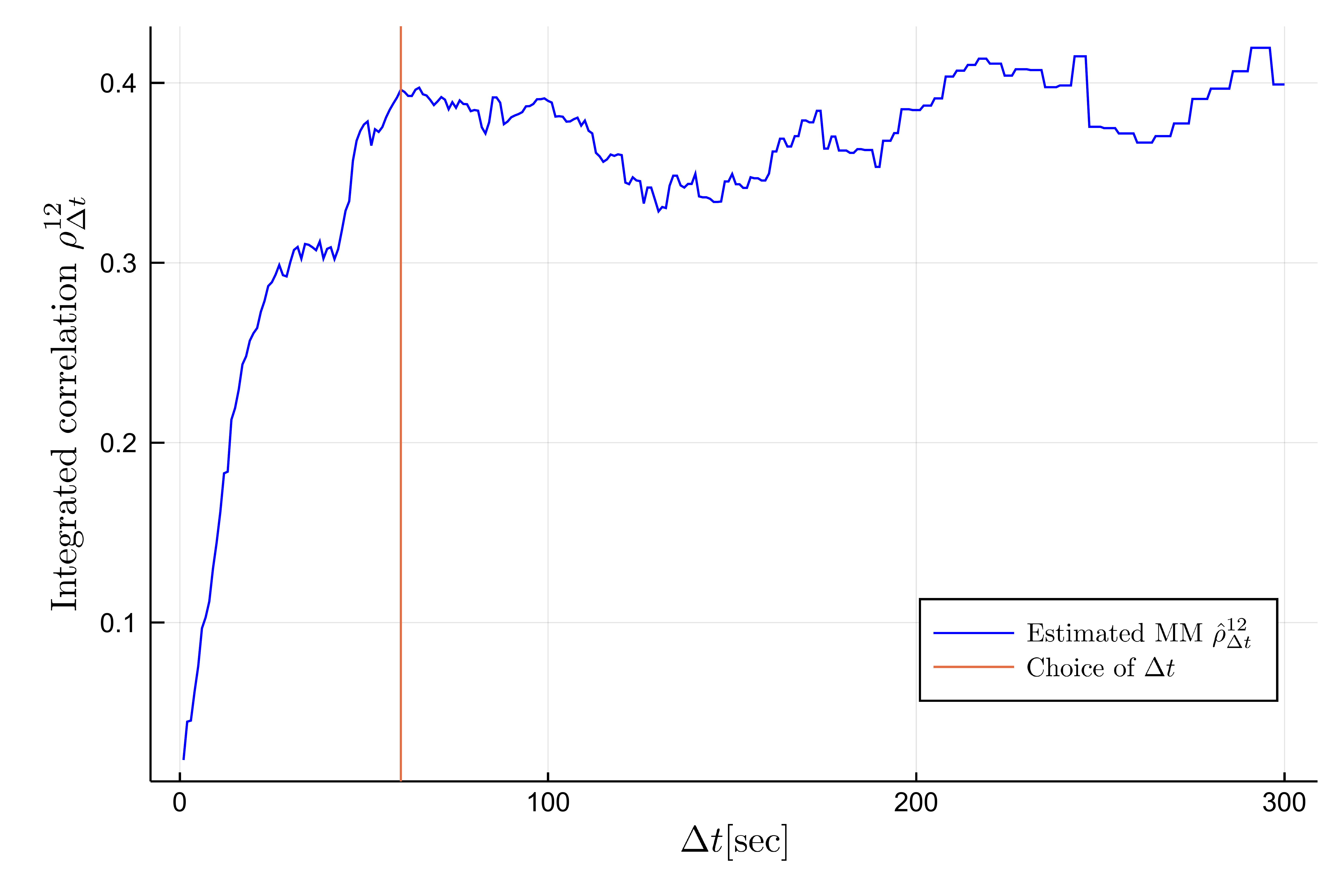}}  
    \subfloat[MM, $1/\lambda = 20$]{\label{fig:OT_Asyn:b}\includegraphics[width=0.33\textwidth]{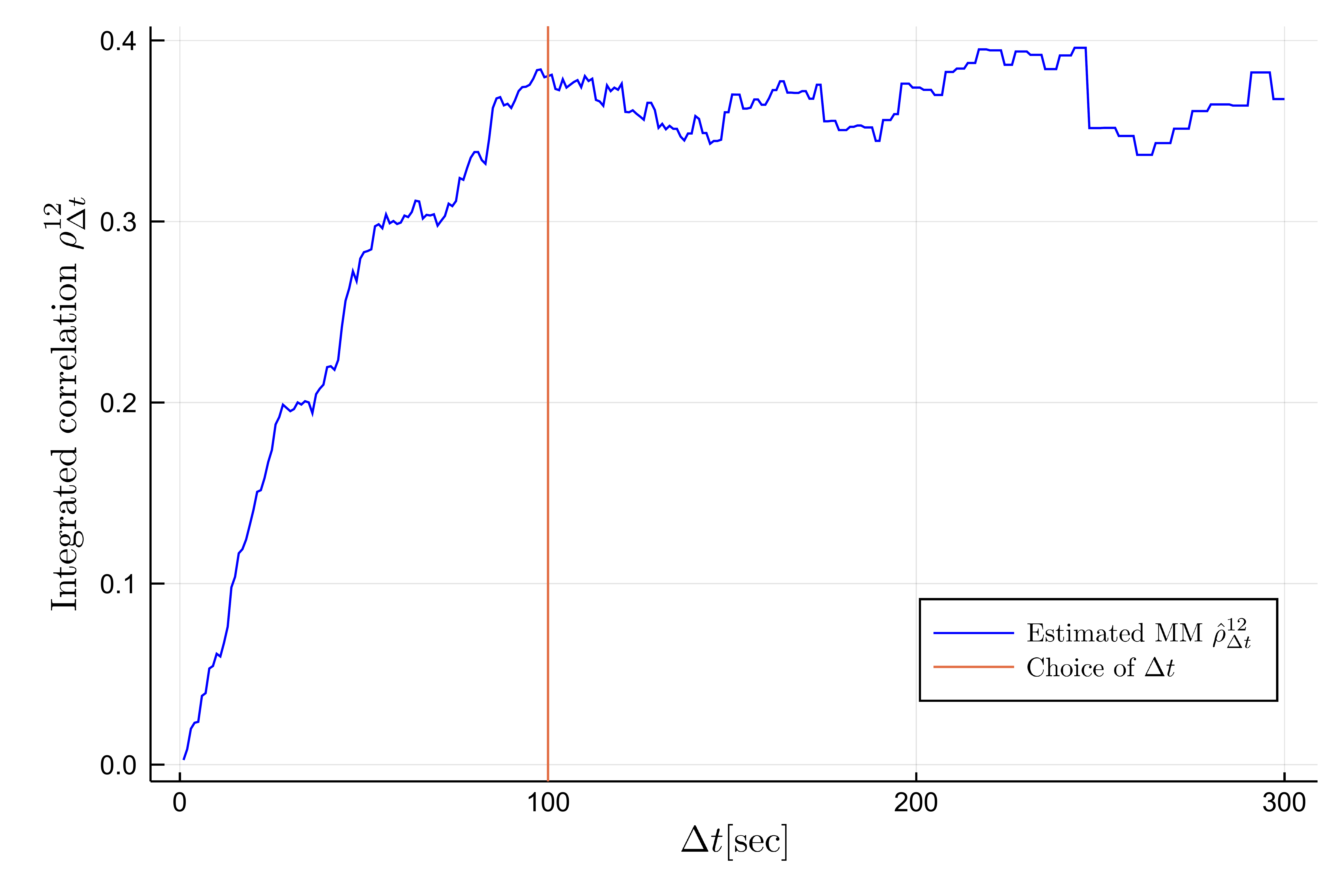}}
    \subfloat[MM, $1/\lambda = 50$]{\label{fig:OT_Asyn:c}\includegraphics[width=0.33\textwidth]{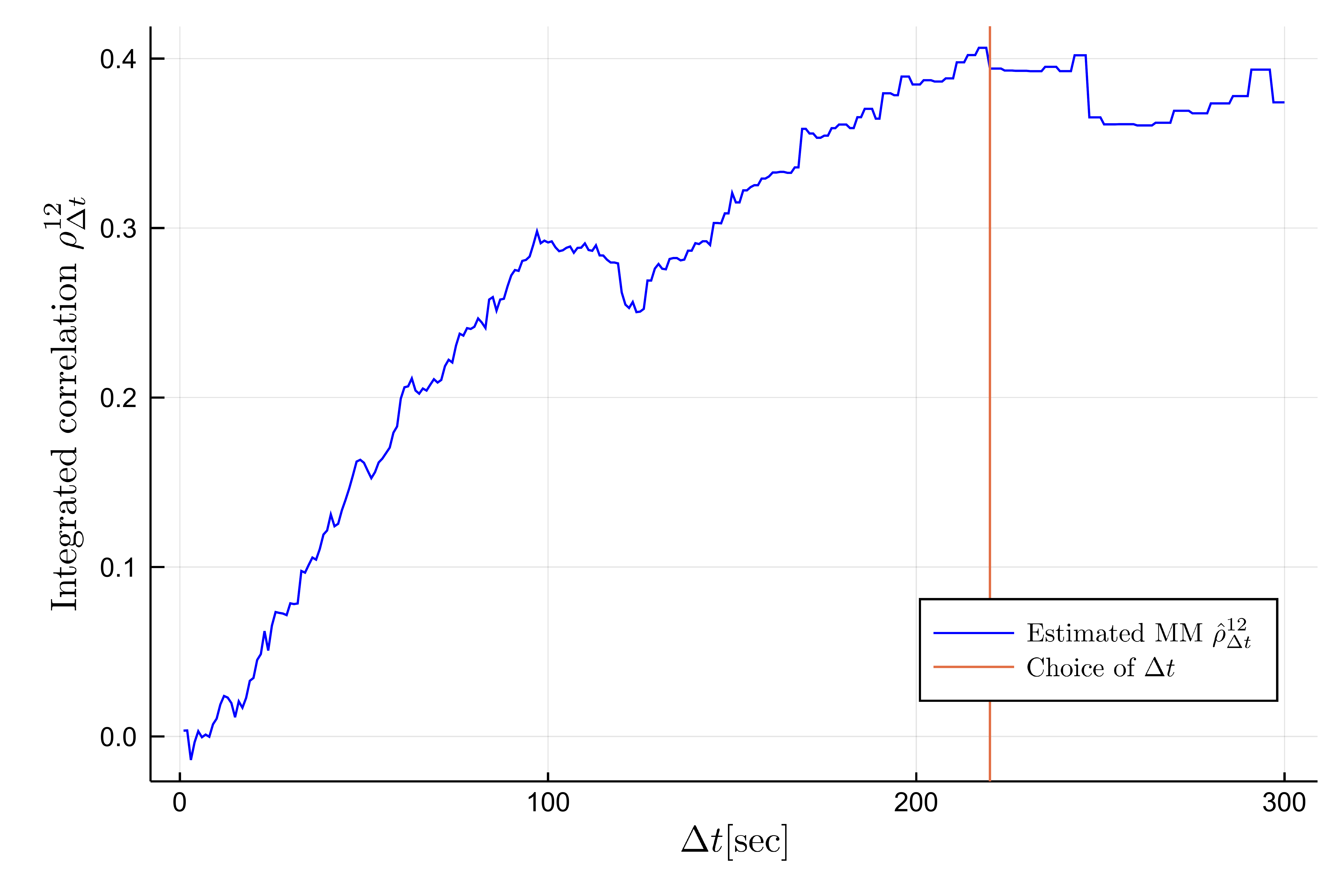}}   \\
    \subfloat[MM, $1/\lambda = 10$, $\Delta t = 60$]{\label{fig:OT_Asyn:d}\includegraphics[width=0.33\textwidth]{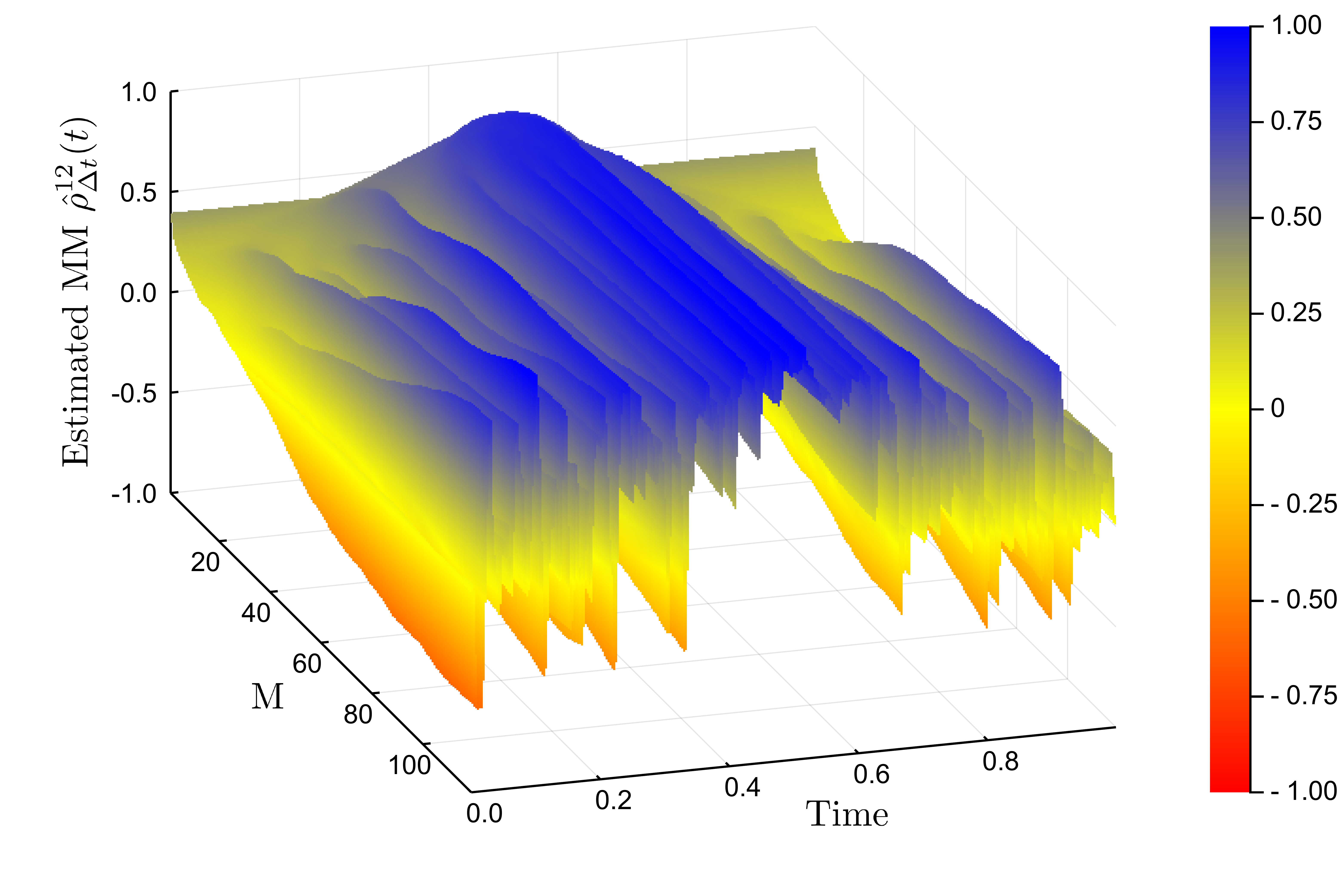}}  
    \subfloat[MM, , $1/\lambda = 20$, $\Delta t = 100$]{\label{fig:OT_Asyn:e}\includegraphics[width=0.33\textwidth]{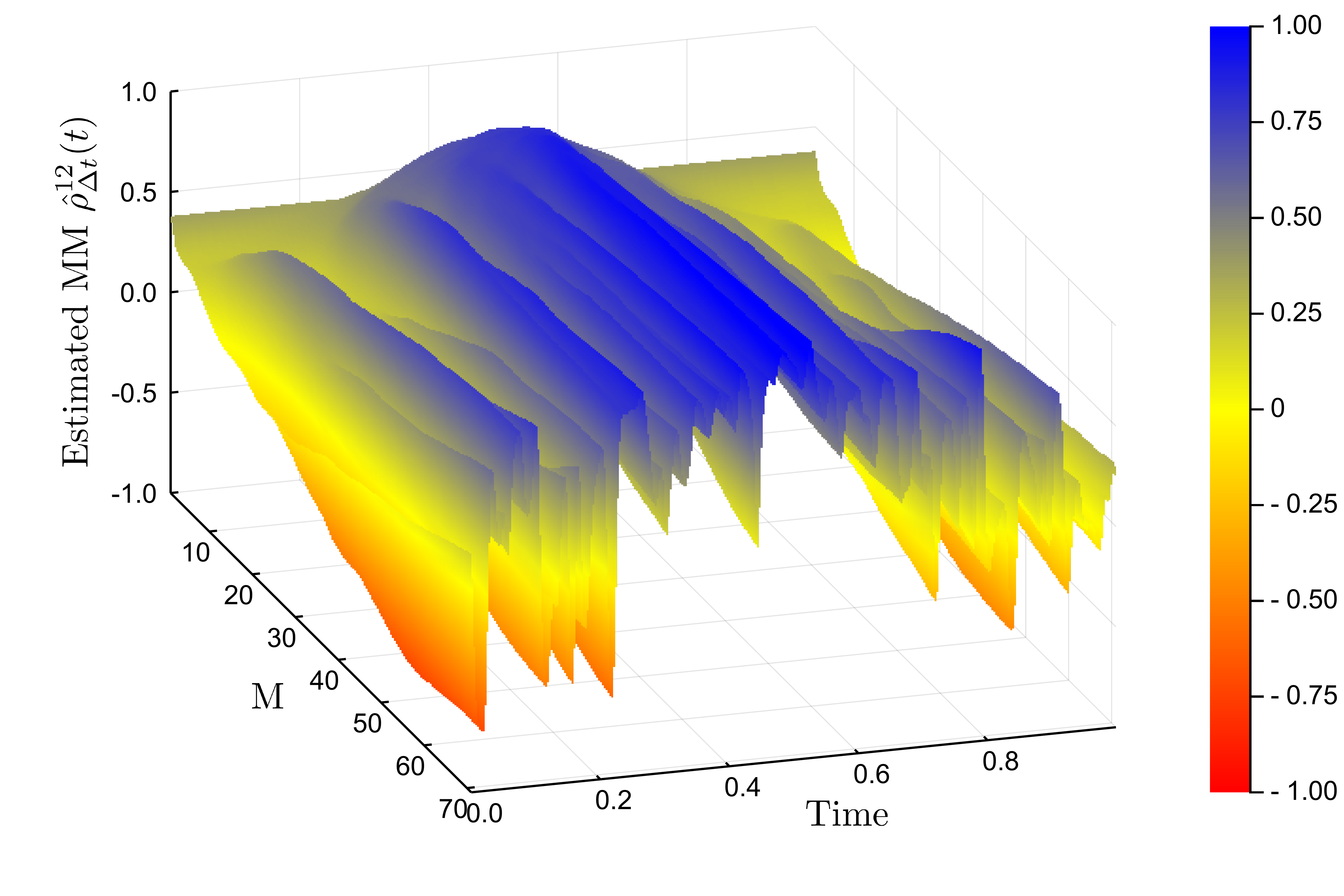}}
    \subfloat[MM, $1/\lambda = 50$, $\Delta t = 220$]{\label{fig:OT_Asyn:f}\includegraphics[width=0.33\textwidth]{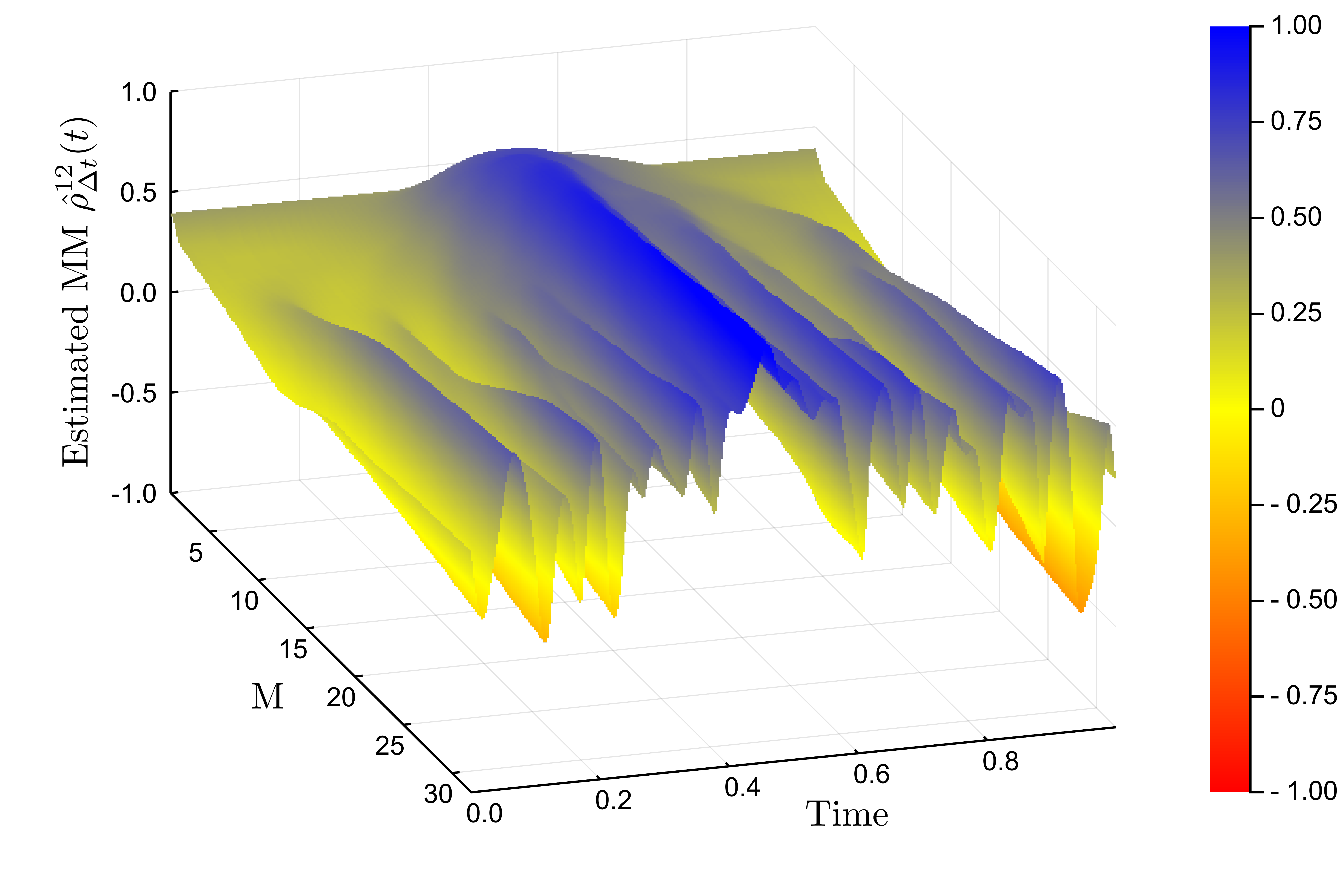}}   \\
    \subfloat[MM, $1/\lambda = 10$, $\Delta t = 60$]{\label{fig:OT_Asyn:g}\includegraphics[width=0.33\textwidth]{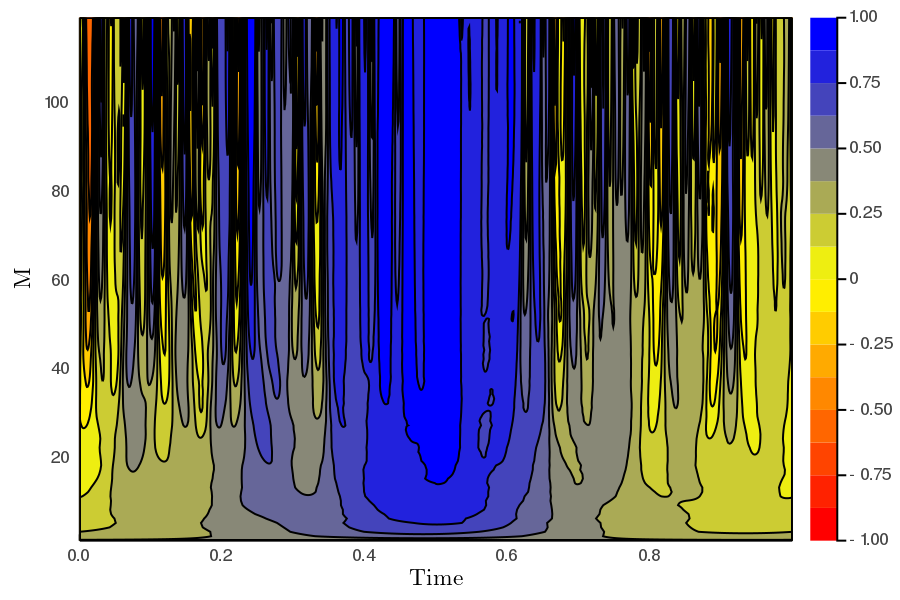}}  
    \subfloat[MM, , $1/\lambda = 20$, $\Delta t = 100$]{\label{fig:OT_Asyn:h}\includegraphics[width=0.33\textwidth]{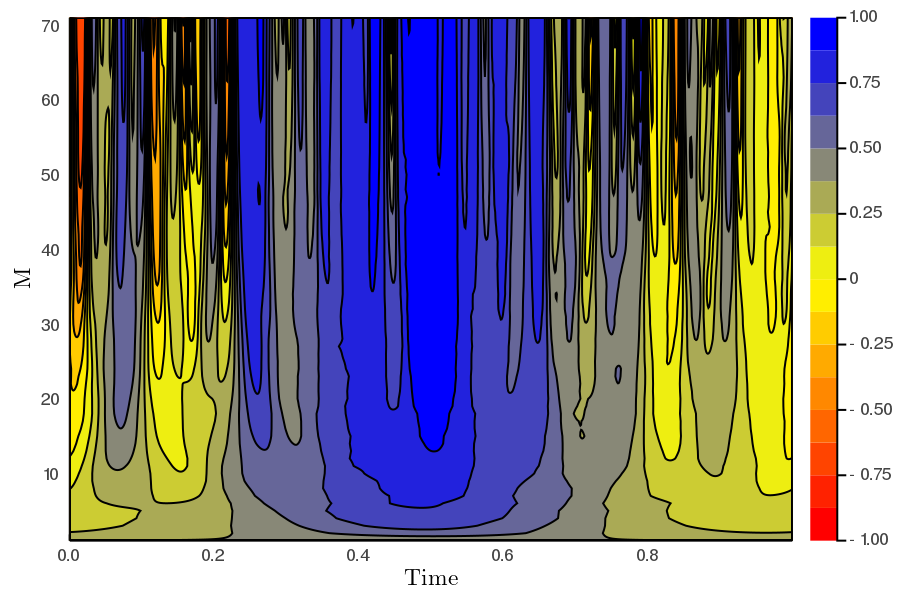}}
    \subfloat[MM, $1/\lambda = 50$, $\Delta t = 220$]{\label{fig:OT_Asyn:i}\includegraphics[width=0.33\textwidth]{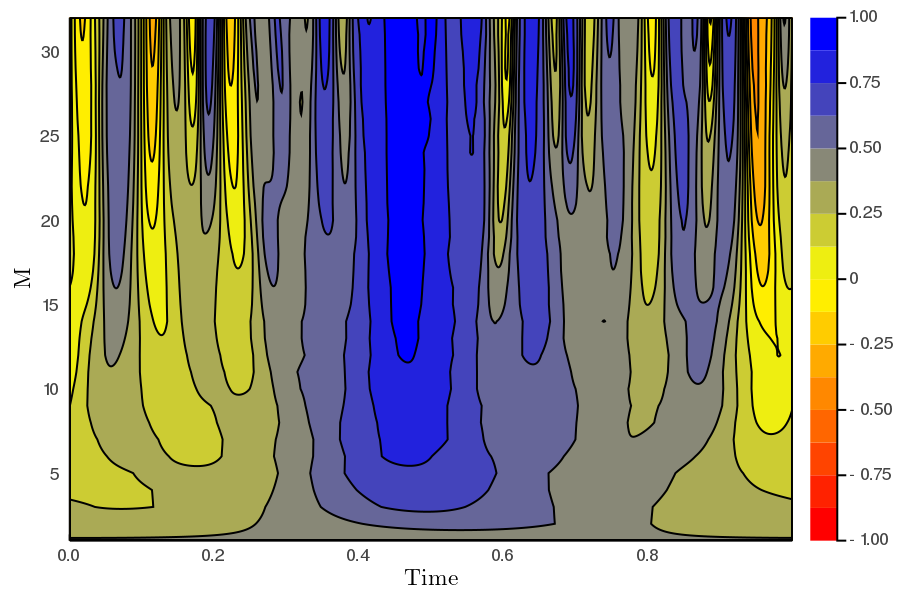}}   \\
    \subfloat[MM, $1/\lambda = 10$, $M$ = 15, $\Delta t = 60$]{\label{fig:OT_Asyn:j}\includegraphics[width=0.33\textwidth]{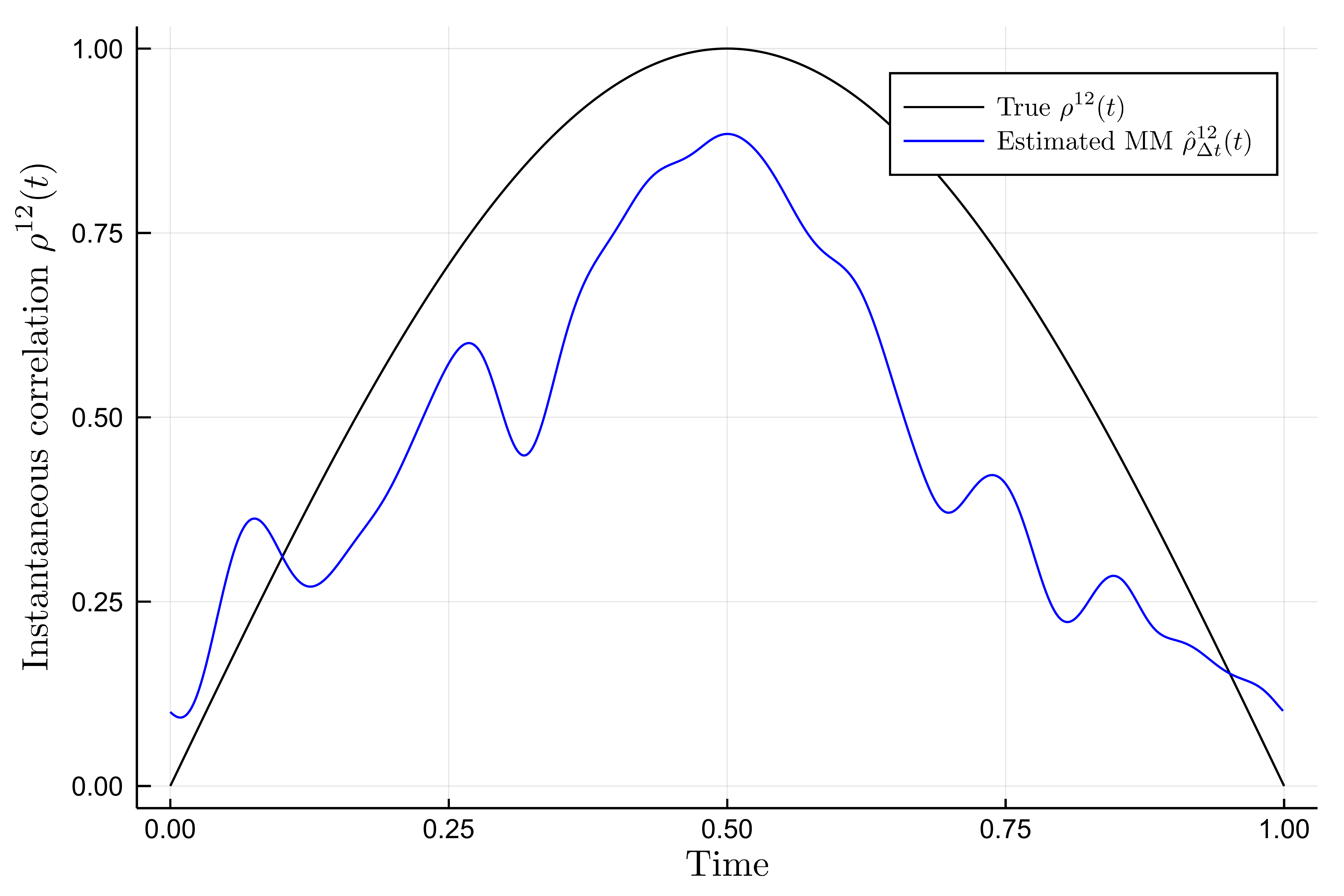}}  
    \subfloat[MM, $1/\lambda = 20$, $M$ = 11, $\Delta t = 100$]{\label{fig:OT_Asyn:k}\includegraphics[width=0.33\textwidth]{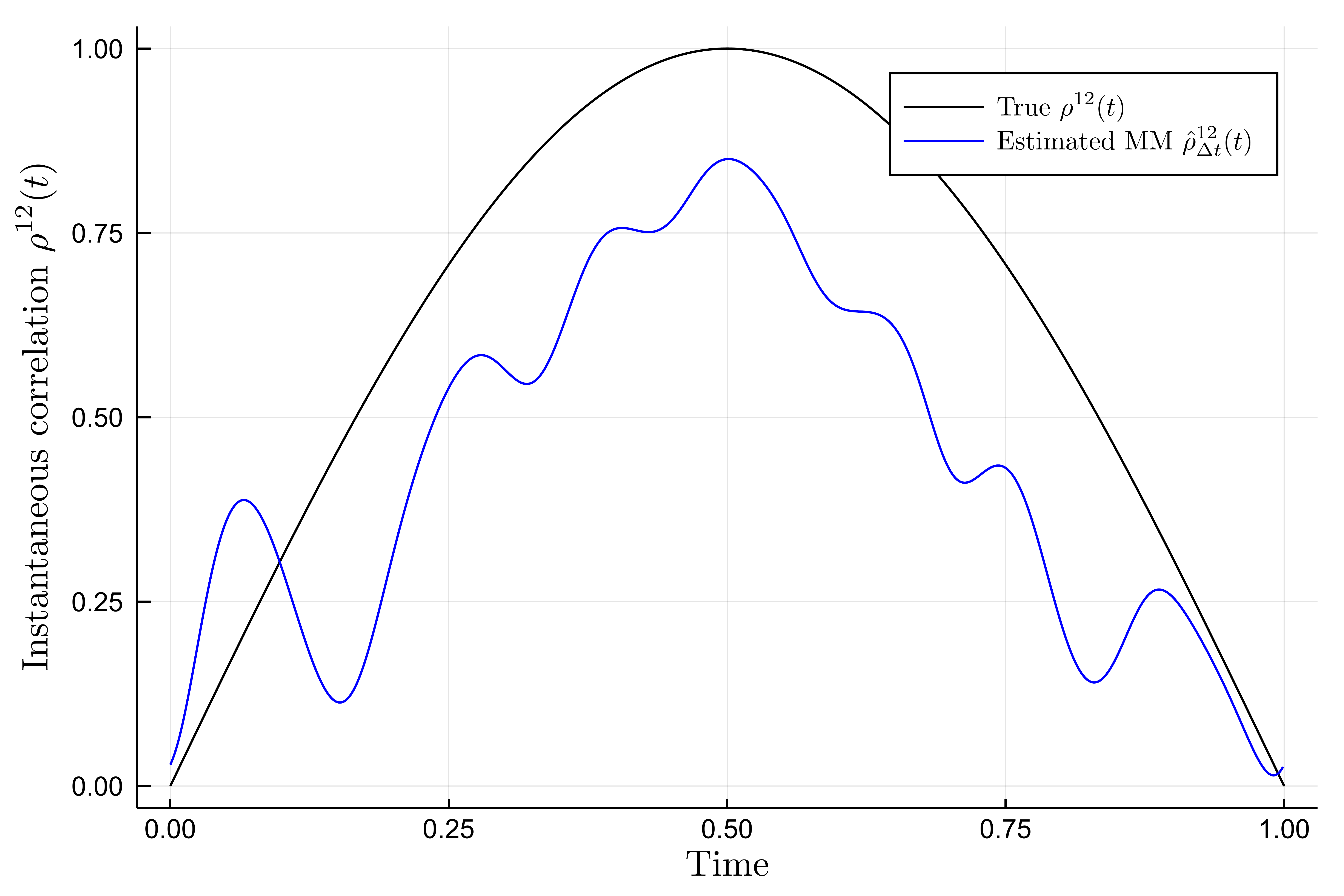}}
    \subfloat[MM, $1/\lambda = 50$, $M$ = 10, $\Delta t = 220$]{\label{fig:OT_Asyn:l}\includegraphics[width=0.33\textwidth]{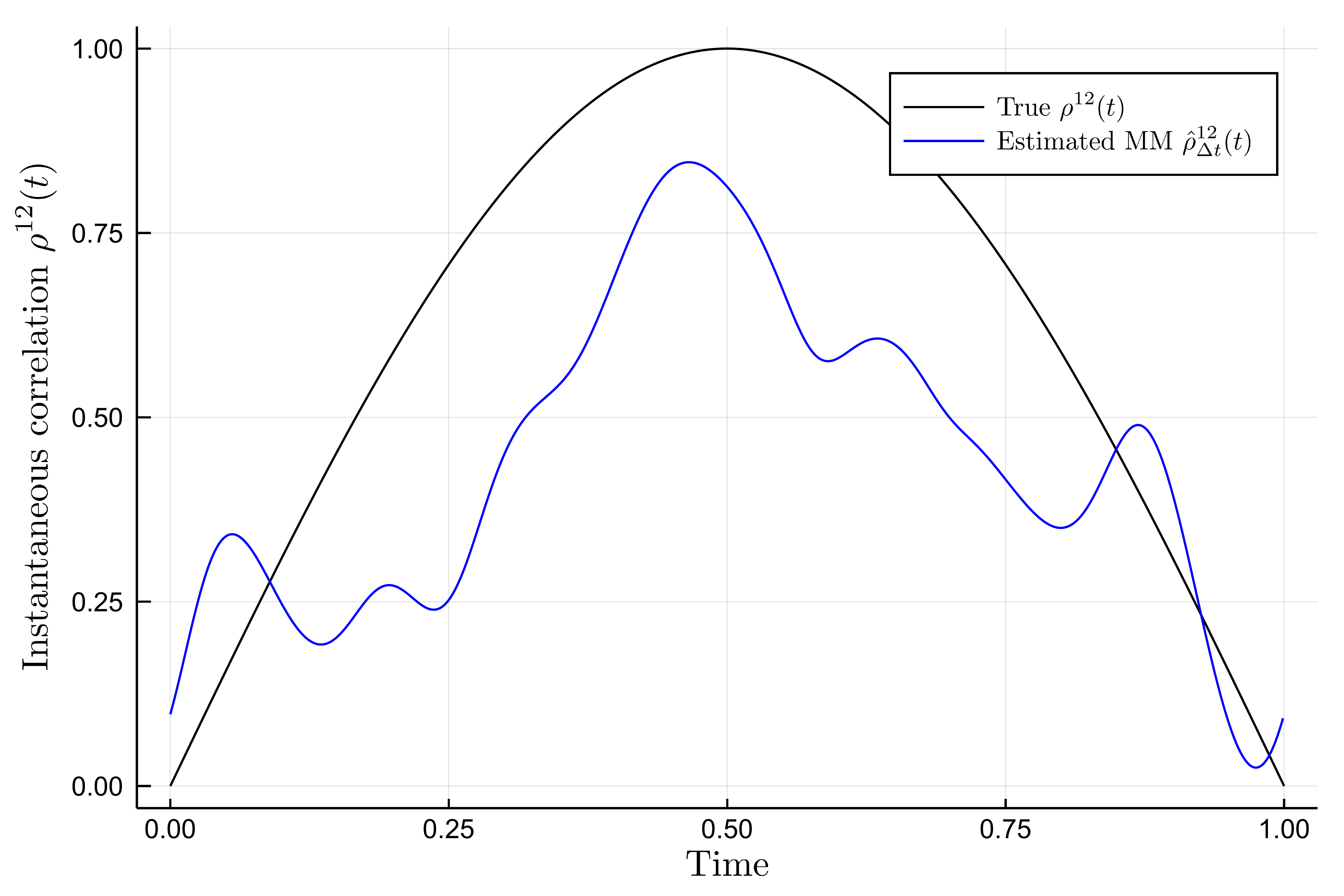}} 
    \caption{Here we demonstrate how to pick $N$ on a case-by-case basis. The simple diffusion model with deterministic correlation in \cref{eq:CF:1,eq:CF:2} is simulated for $n=28,800$ synchronous grid points. Three cases of asynchrony is then introduced by sampling each of the synchronous grid points with an exponential inter-arrival with mean 10, 20 and 50 for the three cases. These are columns one to three respectively. \Cref{fig:OT_Asyn:a,fig:OT_Asyn:b,fig:OT_Asyn:c} is the Malliavin-Mancino integrated correlation with the Dirichlet kernel plotted as a function of the time-scale $\Delta t$. We pick $N$ based on $\Delta t$ which ameliorates the Epps effect, resulting in $\Delta t = 60, 100$ and 220 for the three cases. \Cref{fig:OT_Asyn:d,fig:OT_Asyn:e,fig:OT_Asyn:f} and \Cref{fig:OT_Asyn:g,fig:OT_Asyn:h,fig:OT_Asyn:i} are surface and contour plots for the Malliavin-Mancino spot estimates with $M$ ranging from 1 to 119, 71 and 32 respectively. \Cref{fig:OT_Asyn:j,fig:OT_Asyn:k,fig:OT_Asyn:l} we compare the Malliavin-Mancino spot estimates (blue line with label ``Estimated MM'') with $M = 15, 11$ and 10 for the respective three cases against the true instantaneous correlation (black line with label ``True'') from \cref{eq:CF:2}. We see that in a simulation setting, the instantaneous correlations can be recovered under the presence of asynchrony.} 
\label{fig:OT_Asyn}
\end{figure*}

\cite{Chen2019} provided a break down for the impact of asynchrony. There is a trade-off between the rate of convergence and the bias caused by asynchrony which he calls the ``curse of asynchrony''. Moreover, he provides a sufficient (not necessary) condition to ameliorate the effect. Here we look at this trade-off through the lens of the Epps effect. This downward bias in the Epps effect is caused by the fact that the underlying co-variation is extracted when the asynchronous observations overlap \citep{MSG2011}. There are several methods to correct for this in the case of integrated covariances such as using the Hayashi-Yoshida estimator \citep{HY2005}, directly accounting for the non-overlapping effects at a particular time-scale \citep{PCEPTG2020b, MSG2011}, or simply investigating larger time-scales \citep{RENO2001}. In the case of the instantaneous estimates, it is not clear how the first two correction methods can be applied. Therefore, the most feasible approach is to follow \cite{RENO2001} and find a small $N$ which corrects for the Epps effect. With this in mind, the choice of $N$ should not be chosen as a one size fits all choice. This is because different types of asynchrony and different levels of inhomogeneity lead to different Epps curves.\footnote{Note that the Epps curves refer to the integrated correlation plotted as a function of the sampling interval $\Delta t$.} These different curves require different time-scales to remove the decay. This can be seen in \Cref{fig:OT_Asyn:a,fig:OT_Asyn:b,fig:OT_Asyn:c}. Therefore, the choice of $N$ should depend on the specific Epps curve from the specific case of asynchrony.

A property of the Epps curves is that after the time-scale $\Delta t$ has reached the saturation level, the correlations remain at that level for even larger $\Delta t$.\footnote{Note that for the Malliavin-Mancino estimates, larger $\Delta t$ can present a deviation from the saturation level because of small $N$ which increases the variability of the estimates. This can be seen in for example \cite{RENO2001} when the cutting frequency is too small.} This allows us to find an appropriate time-scale without needing knowledge of the ``true'' underlying correlation levels such as the case of picking $N$ to minimize the error with respect to the MSE. Concretely, we can pick $N$ based on the minimum $\Delta t$ which achieves the saturation level in the Epps curves. This allows the decay in correlation caused by the Epps effect to be removed while providing the largest $N$ possible to provide a good approximation of \cref{eq:instantMM:3} using the Bohr convolution product. It remains unclear how one should pick the appropriate $M$ after a suitable choice of $N$. However, as a consequence of the sampling theorem, $M$ can only be investigated for $M \leq N/2$ to avoid aliasing in the reconstruction of the spot estimates. The non-uniform fast Fourier implementation by \cite{PCEPTG2020a} provides a fast method to compute the integrated correlations for the Epps curves and to investigate the spot estimates for various choices of $M$ once $N$ has been chosen. This procedure is a purely pragmatic approach based on the understanding of the Epps effect to deal with the issue of asynchrony.

% These choices can be further adjusted to satisfy conditions in \cite{MRS2017} and \cite{Chen2019} for asymptotic properties if required.

To demonstrate this pragmatic approach, let us consider a simple bivariate diffusion defined as:
\begin{equation} \label{eq:CF:1}
    dX^i_t = \sigma_i dW_i(t), \quad i = 1, 2,
\end{equation}
where the Brownian motions $W_1$ and $W_2$ have a deterministic instantaneous correlation given by:
\begin{equation} \label{eq:CF:2}
    \rho^{12}(t) = \sin \left( t \pi T \right), 
\end{equation}
for $t \in [0, 1]$. The process is simulated for $n = 28,800$ grid points, $T=1$ trading day, and $(\sigma_1^2, \sigma_2^2) = (0.1, 0.2)$. Here we demonstrate the procedure using the Malliavin-Mancino estimates as they are more stable for various choices of $\Delta t$. \Cref{fig:OT_Syn} we establish the ground truth to see what choices of $M$ recover a good approximation of \cref{eq:CF:2}. \Cref{fig:OT_Syn:a,fig:OT_Syn:b} visualises the Malliavin-Mancino spot correlation estimates for various values of $M$ with the Nyquist choice for $N$ as the surface and contour plots respectively. \Cref{fig:OT_Syn:c} compares the Malliavin-Mancino (blue line) correlation spot estimates for $M=20$ and $N=$ Nyquist against the theoretical correlation (black line). We see that for the simple deterministic correlation, a small value of $M$ is sufficient in approximating the spot correlation. Moreover, additional harmonics $M$ presents fluctuations in the estimates through the adding redundant frequencies.

To demonstrate the need to pick $N$ based on the Epps curves, three cases of asynchronous sampling is used. The synchronous grid is sampled using an exponential inter-arrival with mean 10, 20 and 50 for the three cases. Each case yielding $n_i \approx n / \lambda$ for each asset. By plotting the integrated correlation as a function of the time-scale $\Delta t$ in \Cref{fig:OT_Asyn:a,fig:OT_Asyn:b,fig:OT_Asyn:c}, we see that different levels of asynchrony reach the saturation level at different time-scales. Therefore, the time-scale required to ameliorate the Epps effect should be checked on a case-by-case basis. Here we use the Malliavin-Mancino integrated volatility/co-volatility estimates using the Dirichlet kernel to estimate the integrated correlation. We use the Dirichlet kernel because it better recovers the theoretical Epps effect arising from asynchrony \citep{PCEPTG2020a}. The three time-scales (orange vertical line) which remove the Epps effect while preserving the largest $N$ are $\Delta t = 60, 100$ and 220 seconds. These choices result in $N = 239, 143$ and $64$ respectively. \Cref{fig:OT_Asyn:d,fig:OT_Asyn:e,fig:OT_Asyn:f} and \Cref{fig:OT_Asyn:g,fig:OT_Asyn:h,fig:OT_Asyn:i} are surface and contour plots for the Malliavin-Mancino spot estimates with $M$ ranging from 1 to 119, 71 and 32 respectively. We see that these choices of $N$ allow us to recover correlations that are no longer around zero. However, the rapid fluctuation from large $M$ occur for much smaller values of $M$ in the asynchronous case. The conundrum of picking an appropriate $M$ is further exacerbated under asynchronous observations. In the synchronous case, $M=100$ presented fluctuations but they were relatively small in size. Here in the asynchronous case, massive fluctuations start appearing for $M$ larger than 20 even after choosing an appropriate $N$. It remains unclear as to why this happens, but it is clear that under the presence of asynchrony $M$ needs to be smaller than the synchronous case. This can be problematic if one is trying to approximate more complex instantaneous dynamics such as the instantaneous correlation from the Heston model as small $M$ does not provide enough harmonics for an accurate approximation, but larger $M$ is also not an appropriate option. \Cref{fig:OT_Asyn:j,fig:OT_Asyn:k,fig:OT_Asyn:l} we compare the Malliavin-Mancino spot estimates (blue line) with $M = 15, 11$ and 10 for the respective three cases against the true instantaneous correlation (black line) from \cref{eq:CF:2}. These estimates are able recover the true correlations to a certain degree but are unsatisfactory compared to the synchronous case. This is only achievable under simulation conditions because we know the ground truth so that the choice of $M$ can be adjusted for a satisfactory recovery. How one should pick $M$ for the empirical data remains unclear, the only conditions we know is that $M$ must be small and must be less than $N/2$.

Apart from the issue of not having the ground truth instantaneous correlation for comparison to pick appropriate $M$ with empirical data, there is another issue with picking small $N$ to remove the Epps effect. By choosing large $\Delta t$ to remove the Epps effect hides the genuine effects resulting in a decay of correlations. This does not present an issue when the underlying correlation does not depend on $\Delta t$ such as diffusion models or when there is no lead-lag. However, when these genuine effects are present the better approach is to disentangle statistical effects causing the Epps effect from the genuine effects at various $\Delta t$. In the case of the integrated correlation, disentangling lead-lag from asynchrony has been done by \cite{MMZ2011}, and disentangling correlations that dependent on $\Delta t$ from asynchrony has been done by \cite{PCEPTG2020b}. As a preliminary investigation into the instantaneous Epps effect, here we will only consider the impact from changing $\Delta t$ rather than trying to disentangle the various effects.

% \newpage
\section{Empirical}\label{sec:empirical}

Here we aim to demonstrate the instantaneous Epps effect and ameliorate the effect with empirical Trade and Quote (TAQ) data. To this end, we obtain two banking equities from the Johannesburg Stock Exchange (JSE) for the week from 24/06/2019 to 28/06/2019. The two equities were extracted from Bloomberg Pro and processed to remove repeated time stamps by aggregating trades with the same time stamp using a volume weighted average. The equities considered are: Standard Bank Group Ltd (SBK) and FirstRand Limited (FSR). Each trading day is seven hours and 50 minutes (eight hour trading day and 10 minutes for the closing auction), thus $T = 1$ day equals $28,200$ seconds. Here we investigate each day separately because the trading days are not connected which can affect the spot estimates as they are built upon harmonics.

\begin{figure}[t!]
    \centering
    \includegraphics[width=0.5\textwidth]{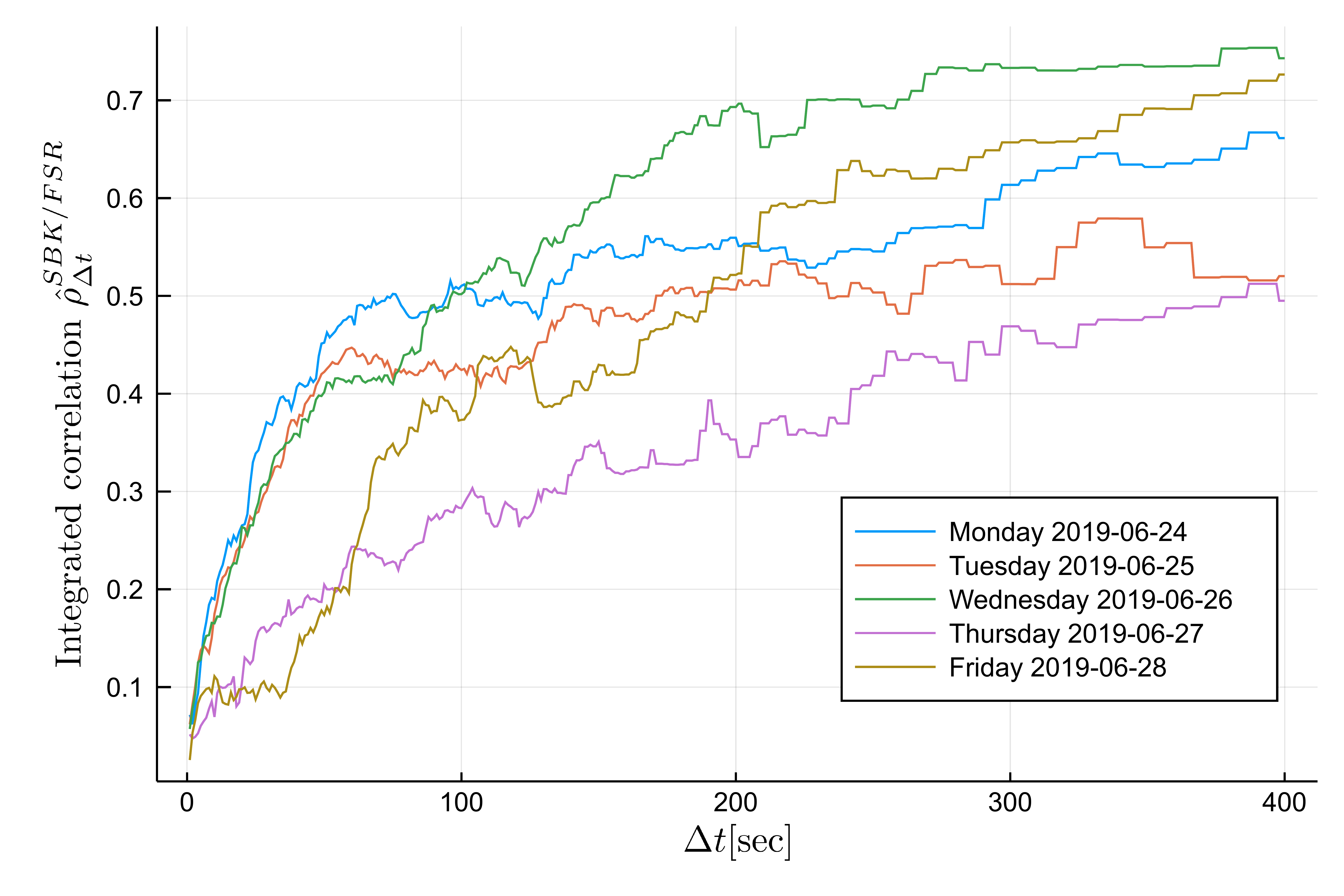}
    \caption{Here we plot the integrated correlation estimates for the equity pair SBK/FSR at various time-scales $\Delta t$. The estimates are obtained using the Malliavin-Mancino integrated volatility/co-volatility estimate with the Dirichlet kernel. The time-scale is adjusted using \cref{eq:comp:6} and ranges from 1 to 400 seconds. The Epps curves are plotted for each day of the week from 24/06/2019 to 28/06/2019. As per the legend, Monday the 24/06 is the blue line, Tuesday the 25/06 is the orange line, Wednesday the 26/06 is the green line line, Thursday the 27/06 is the purple line, and Friday the 28/06 is the dark-green line.}
    \label{fig:Emp_Int_EppsCurves}
\end{figure}

We begin by plotting the integrated correlation as a function of the sampling interval $\Delta t$ for each day using the Malliavin-Mancino integrated volatility/co-volatility with the Dirichlet kernel. The time-scale is adjusted using \cref{eq:comp:6} and ranges from 1 to 400 seconds. We see in \Cref{fig:Emp_Int_EppsCurves} that the integrated correlation between the days for the same equity pair can vary considerably, but they all exhibit the Epps effect. To avoid excessive figures, we will investigate the instantaneous correlation for Tuesday 25/06/2019 (orange line) and Wednesday 26/06/2019 (green line) as they both reach the saturation level around $\Delta t = 300$ seconds and they present different saturation levels.

\begin{figure*}[p]
    \centering
    \subfloat[MM, $M=10$, 25/06]{\label{fig:Emp_Inst_Epps:a}\includegraphics[width=0.245\textwidth]{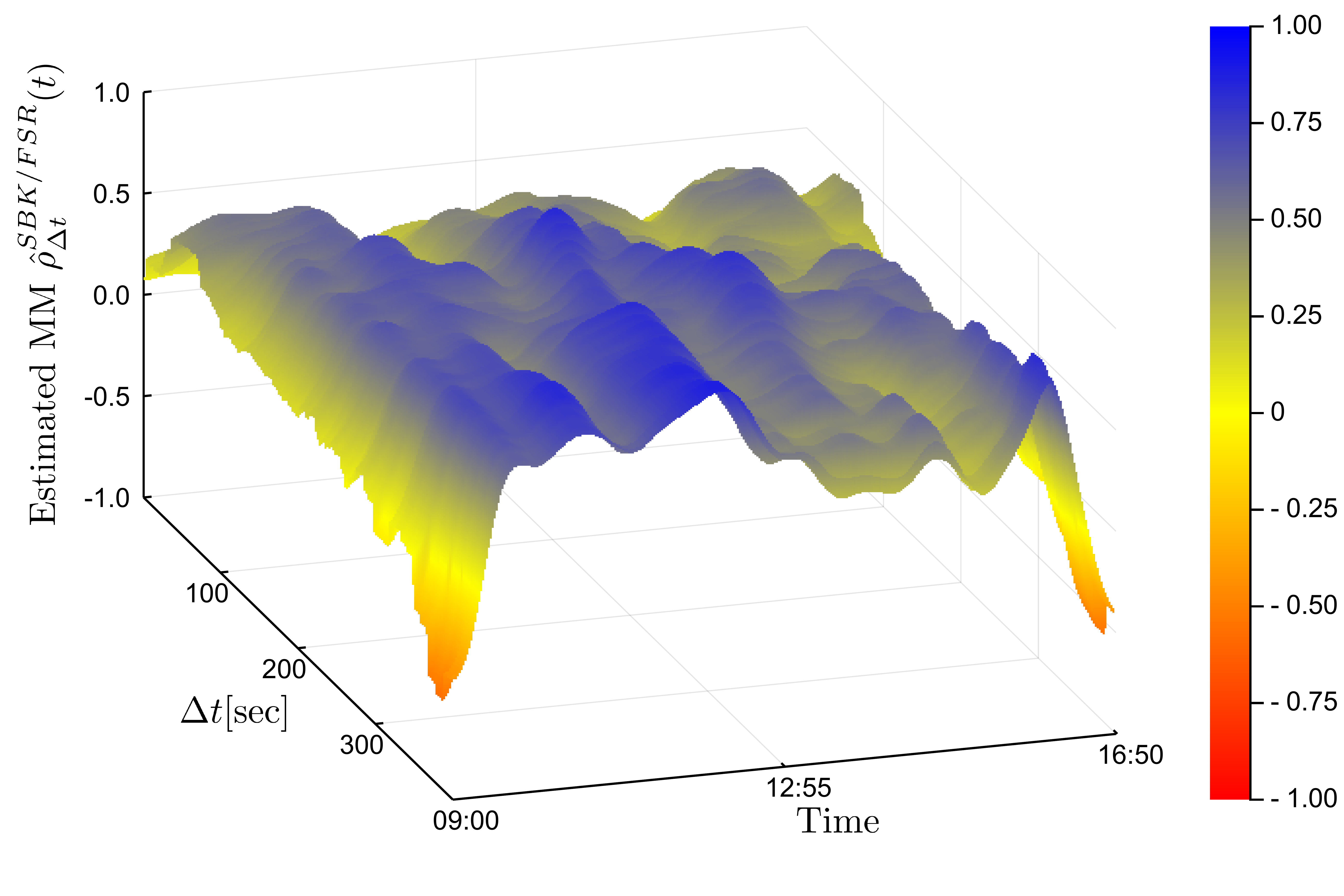}}  
    \subfloat[MM, $M=10$, 25/06]{\label{fig:Emp_Inst_Epps:b}\includegraphics[width=0.245\textwidth]{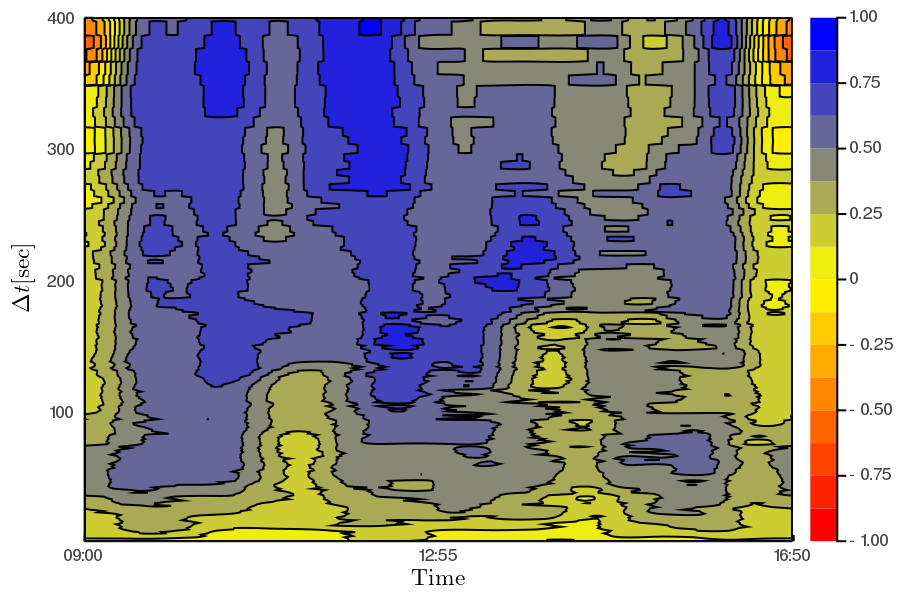}}
    \subfloat[MM, $M=10$, 26/06]{\label{fig:Emp_Inst_Epps:c}\includegraphics[width=0.245\textwidth]{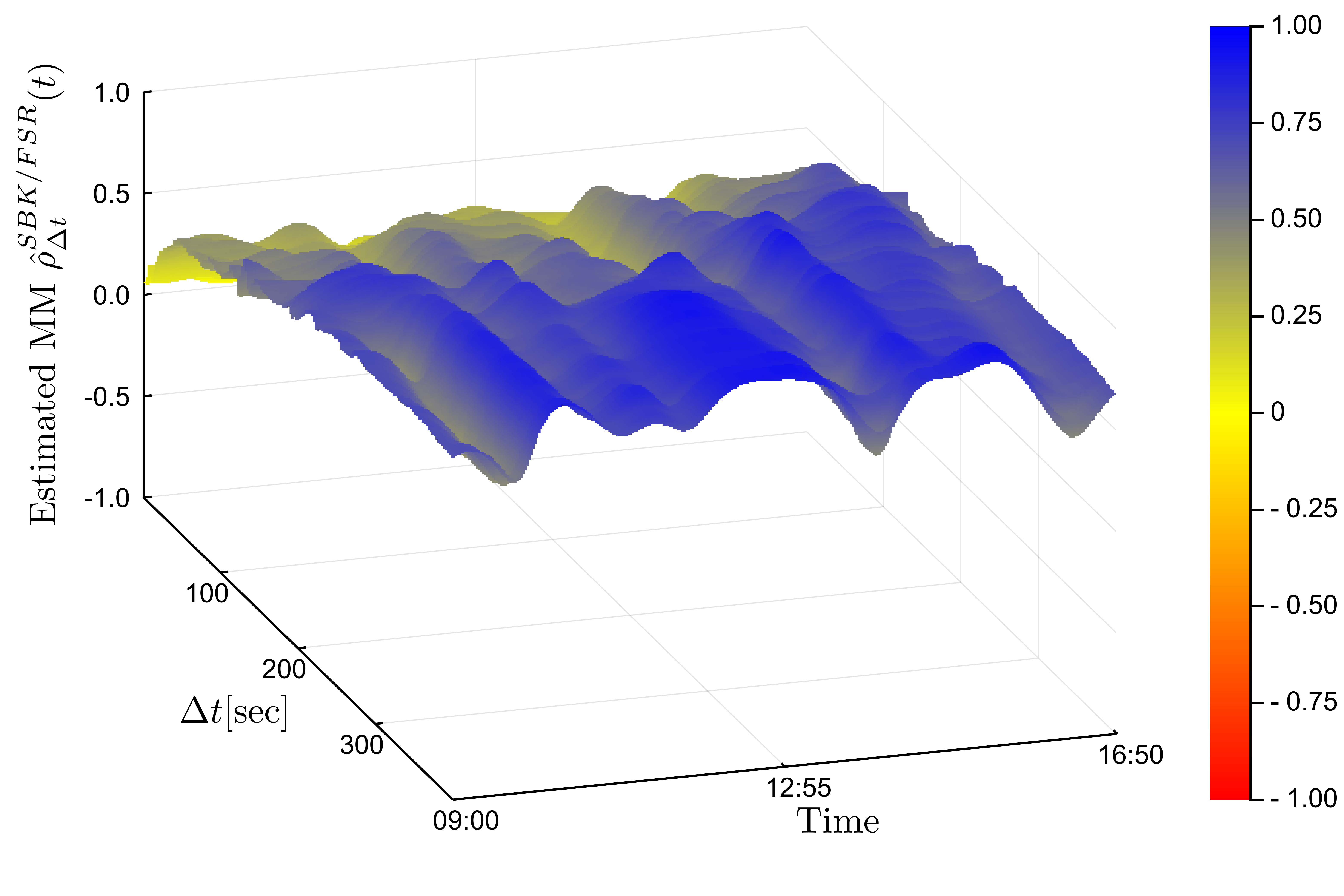}}  
    \subfloat[MM, $M=10$, 26/06]{\label{fig:Emp_Inst_Epps:d}\includegraphics[width=0.245\textwidth]{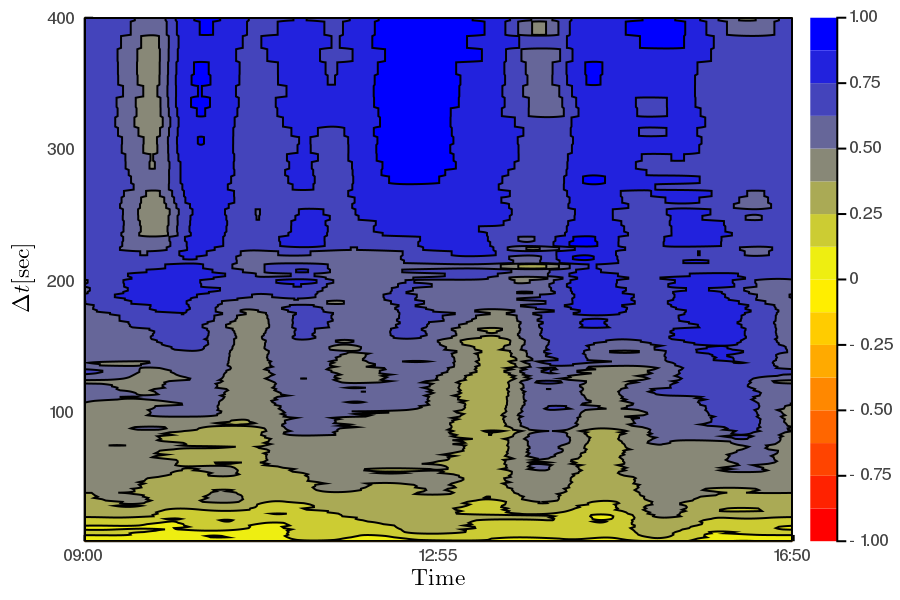}}    \\
    % \subfloat[MM, $M=20$, 25/06]{\label{fig:Emp_Inst_Epps:e}\includegraphics[width=0.245\textwidth]{Figures/Emp_MM_Inst_Surf_Epps_M20_Tues.pdf}}  
    % \subfloat[MM, $M=20$, 25/06]{\label{fig:Emp_Inst_Epps:f}\includegraphics[width=0.245\textwidth]{Figures/Emp_MM_Inst_HM_Epps_M20_Tues.png}}
    % \subfloat[MM, $M=20$, 26/06]{\label{fig:Emp_Inst_Epps:g}\includegraphics[width=0.245\textwidth]{Figures/Emp_MM_Inst_Surf_Epps_M20_W.pdf}}  
    % \subfloat[MM, $M=20$, 26/06]{\label{fig:Emp_Inst_Epps:h}\includegraphics[width=0.245\textwidth]{Figures/Emp_MM_Inst_HM_Epps_M20_W.png}}    \\
    \subfloat[CT, $M=10$, 25/06]{\label{fig:Emp_Inst_Epps:e}\includegraphics[width=0.245\textwidth]{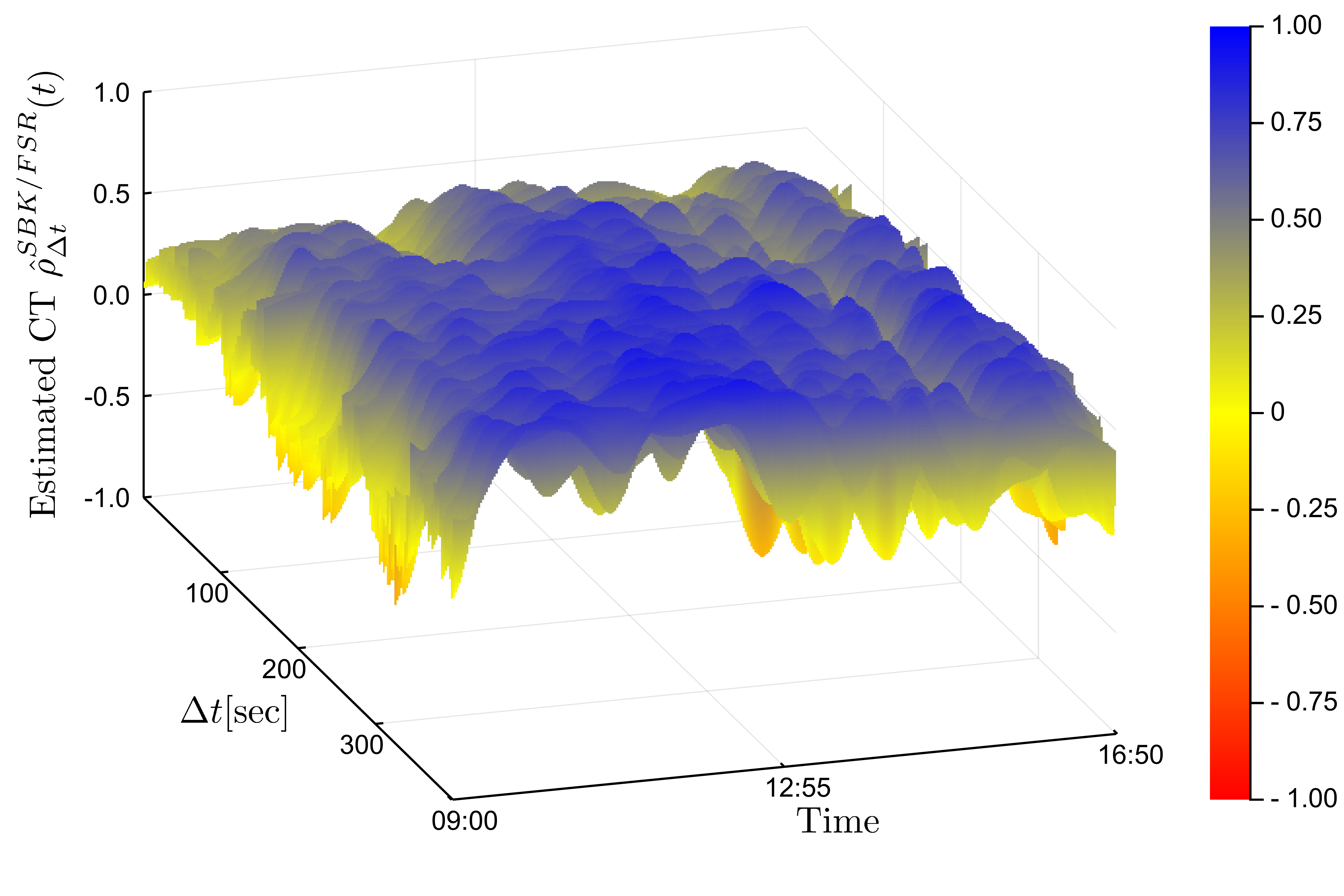}}  
    \subfloat[CT, $M=10$, 25/06]{\label{fig:Emp_Inst_Epps:f}\includegraphics[width=0.245\textwidth]{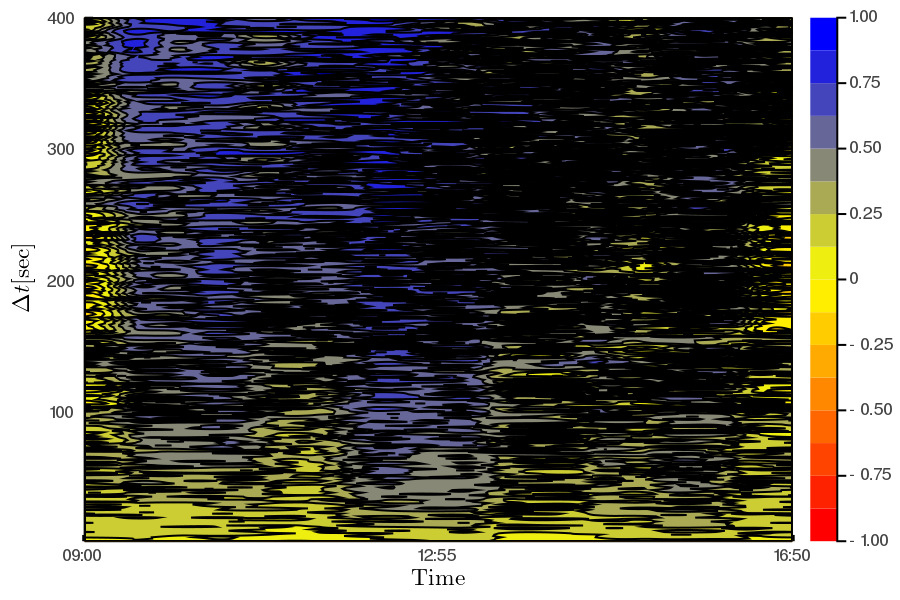}}
    \subfloat[CT, $M=10$, 26/06]{\label{fig:Emp_Inst_Epps:g}\includegraphics[width=0.245\textwidth]{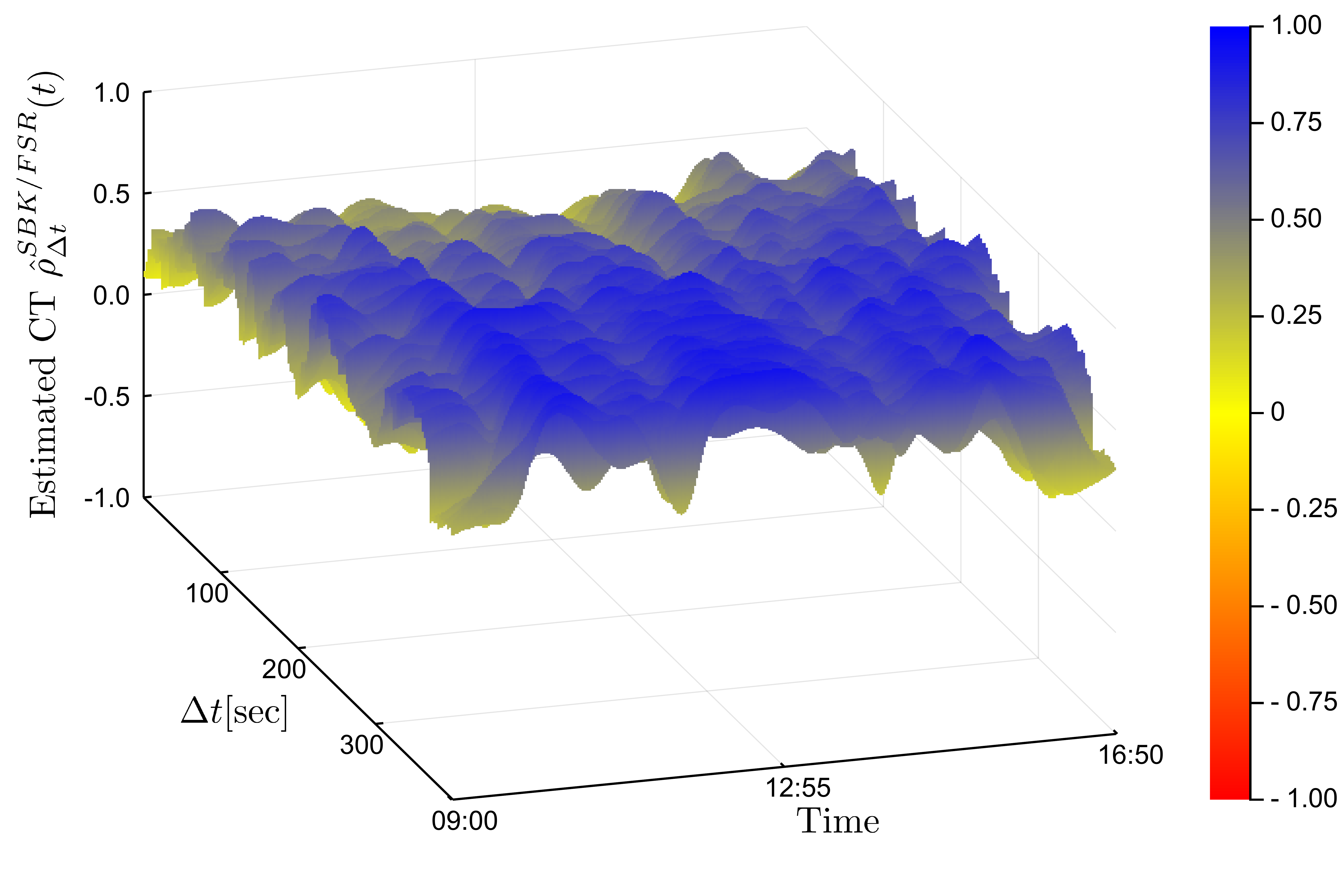}}  
    \subfloat[CT, $M=10$, 26/06]{\label{fig:Emp_Inst_Epps:h}\includegraphics[width=0.245\textwidth]{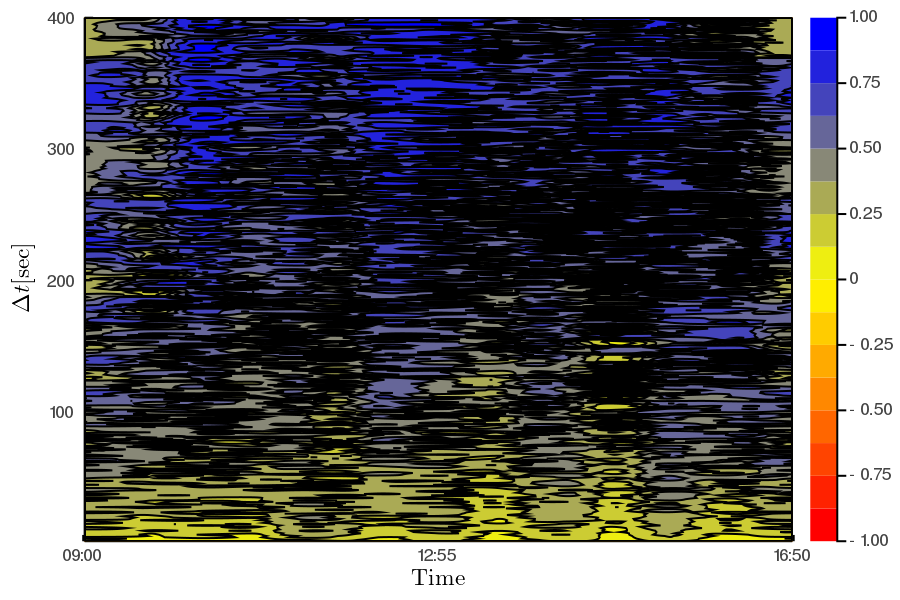}}    %\\
    % \subfloat[CT, $M=20$, 25/06]{\label{fig:Emp_Inst_Epps:m}\includegraphics[width=0.245\textwidth]{Figures/Emp_CT_Inst_Surf_Epps_M20_Tues.pdf}}  
    % \subfloat[CT, $M=20$, 25/06]{\label{fig:Emp_Inst_Epps:n}\includegraphics[width=0.245\textwidth]{Figures/Emp_CT_Inst_HM_Epps_M20_Tues.png}}
    % \subfloat[CT, $M=20$, 26/06]{\label{fig:Emp_Inst_Epps:o}\includegraphics[width=0.245\textwidth]{Figures/Emp_CT_Inst_Surf_Epps_M20_W.pdf}}  
    % \subfloat[CT, $M=20$, 26/06]{\label{fig:Emp_Inst_Epps:p}\includegraphics[width=0.245\textwidth]{Figures/Emp_CT_Inst_HM_Epps_M20_W.png}}
    \caption{Here we demonstrate the instantaneous Epps effect for empirical data. The first and second row plots the surface and contour plots of the Malliavin-Mancino and Cuchiero-Teichmann for fixed $M=10$ and varying $\Delta t$ respectively. The first and second columns are the surface and contour plot for Tuesday the 25/06/2019; similarly, the third and fourth columns are the surface and contour plot for Wednesday the 26/06/2019. The time-scale $\Delta t$ is controlled using \cref{eq:comp:6} for the Malliavin-Mancino estimator, while the Cuchiero-Teichmann estimator adjusts the time-scale using the previous tick interpolation. First, we see the Epps effect is present for the instantaneous correlations; second, the Cuchiero-Teichmann presents instabilities in the estimates for varying $\Delta t$.} 
\label{fig:Emp_Inst_Epps}
\end{figure*}

Next we demonstrate that the instantaneous Epps effect is present with empirical data. The first and second row in \Cref{fig:Emp_Inst_Epps} plots the surface and contour plots of the Malliavin-Mancino and Cuchiero-Teichmann for fixed $M=10$ and varying $\Delta t$ respectively. The first and second columns are the surface and contour plot for Tuesday the 25/06/2019 respectively; similarly, the third and fourth columns are the surface and contour plot for Wednesday the 26/06/2019 respectively. The time-scale $\Delta t$ is controlled using \cref{eq:comp:6} for the Malliavin-Mancino estimator, while the Cuchiero-Teichmann estimator adjusts the time-scale using the previous tick interpolation.\footnote{To start the previous tick interpolation for each day, we set $t=0$ once the equity pair has each made their first trade.} First, we see that the instantaneous Epps effect is present, as in the spot estimates are around zero for small $\Delta t$ and increase when $\Delta t$ increases. Second, as with the simulation experiments, the previous tick interpolation presents instabilities in the Cuchiero-Teichmann estimates for fixed time $t$ and varying $\Delta t$ seen as the horizontal black marks in the contour plots. Third, the intraday correlation structure can vary between the days for the same equity pair. In the Malliavin-Mancino estimates, we see that on Tuesday the 25/06/2019 there are large dips in correlations around the open and close of the trading day and also smaller dips in the afternoon but on Wednesday the 26/06/2019 the correlation remains stable around the same level throughout the whole day. The dips are aggregated and hidden into the integrated correlation in \Cref{fig:Emp_Int_EppsCurves} as a lower saturation level.

\begin{figure}[t!]
    \centering
    \includegraphics[width=0.5\textwidth]{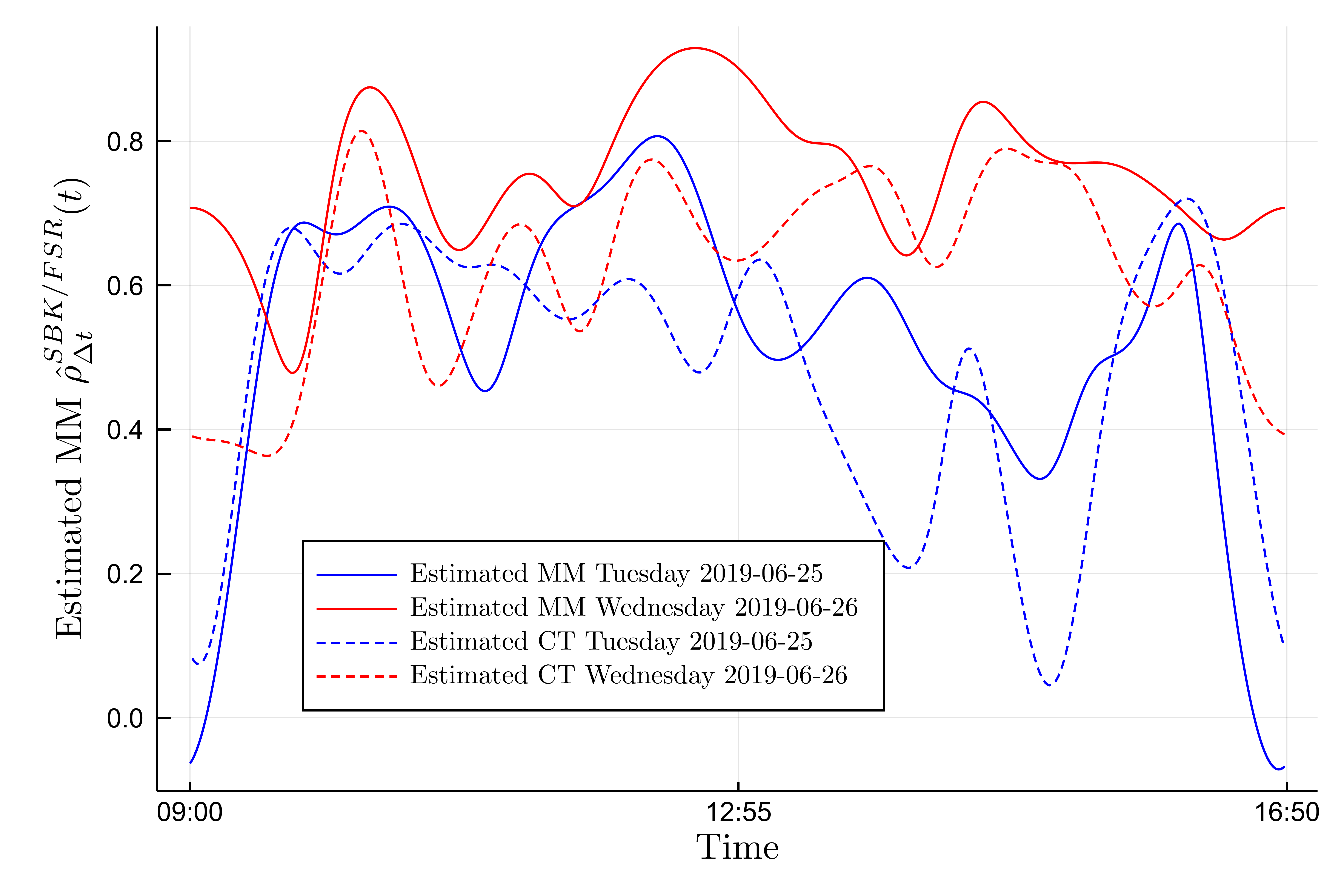}
    \caption{Here we compare the Malliavin-Mancino (solid lines) and Cuchiero-Teichmann (dashed lines) instantaneous estimates for $M=10$ and $\Delta t = 300$. Tuesday the 25/06/2019 are in blue while Wednesday the 26/06/2019 are in red. We see that the correlation structure can vary quite a bit between the days and that the two estimators recover similar intraday correlation dynamics.}
    \label{fig:Emp_Inst_Epps_Comp_dt300_M10}
\end{figure}

\begin{figure*}[p]
    \centering
    \subfloat[MM, $\Delta t = 300$, 25/06]{\label{fig:Emp_Inst_Epps_dt300:a}\includegraphics[width=0.245\textwidth]{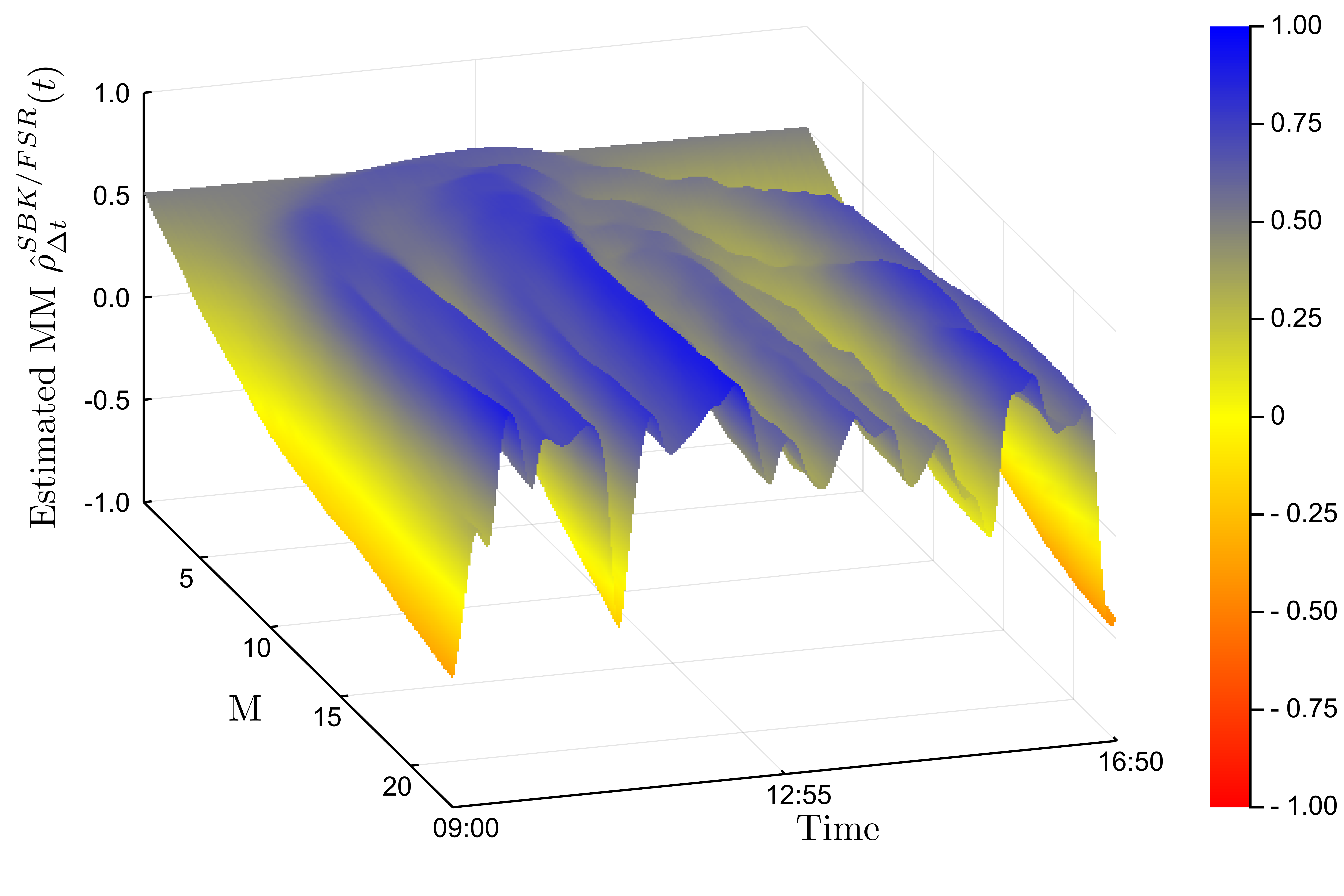}}  
    \subfloat[MM, $\Delta t = 300$, 25/06]{\label{fig:Emp_Inst_Epps_dt300:b}\includegraphics[width=0.245\textwidth]{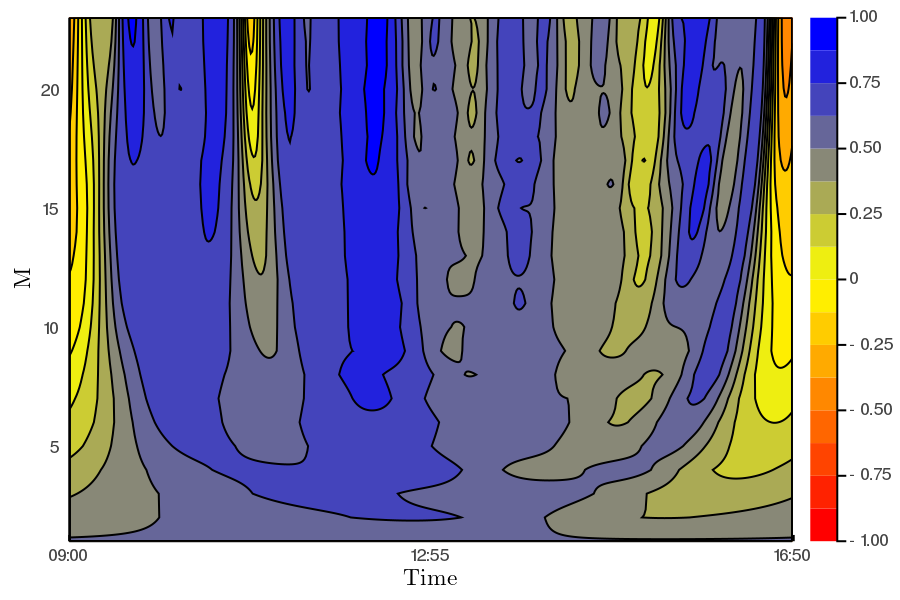}}
    \subfloat[MM, $\Delta t = 300$, 26/06]{\label{fig:Emp_Inst_Epps_dt300:c}\includegraphics[width=0.245\textwidth]{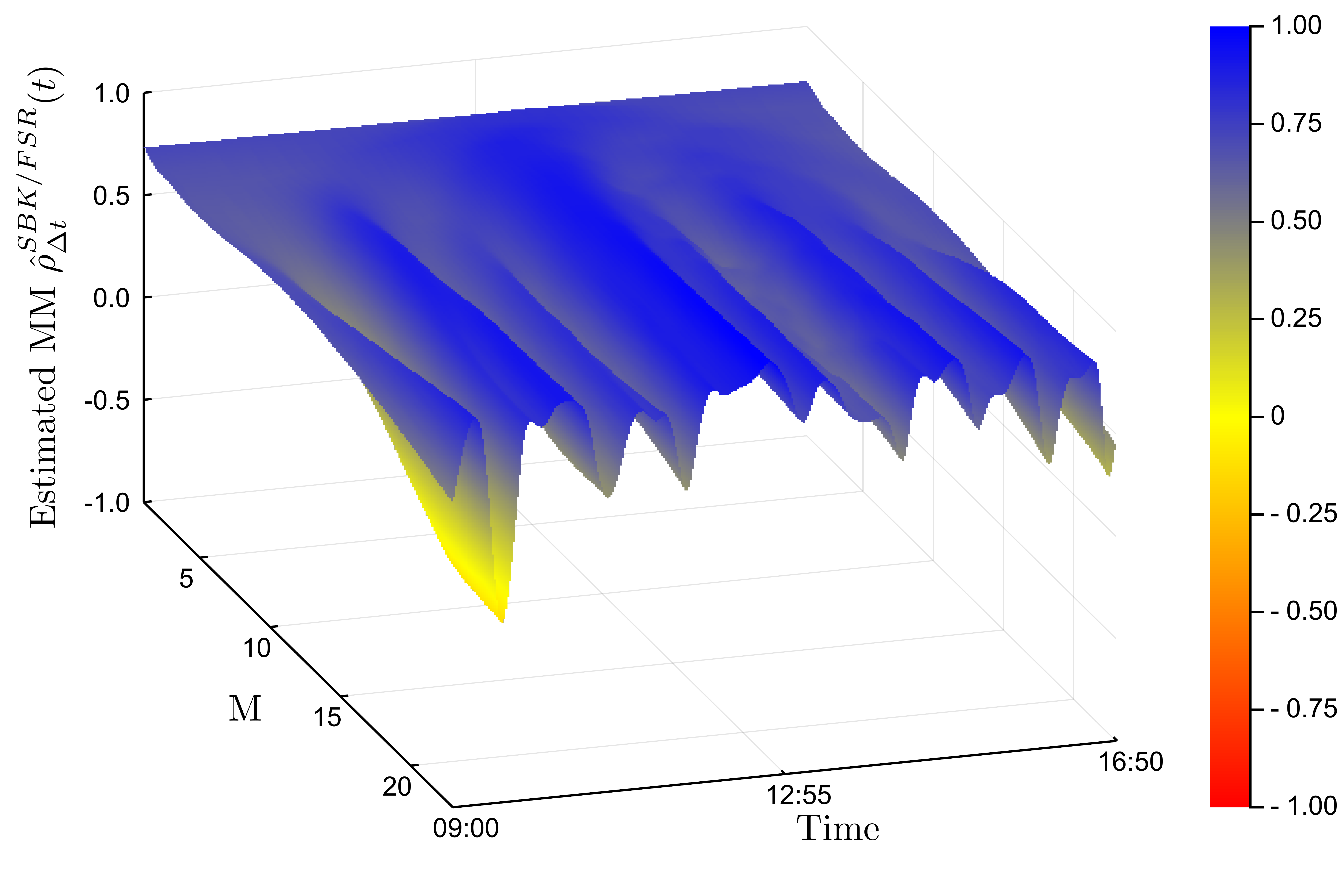}}  
    \subfloat[MM, $\Delta t = 300$, 26/06]{\label{fig:Emp_Inst_Epps_dt300:d}\includegraphics[width=0.245\textwidth]{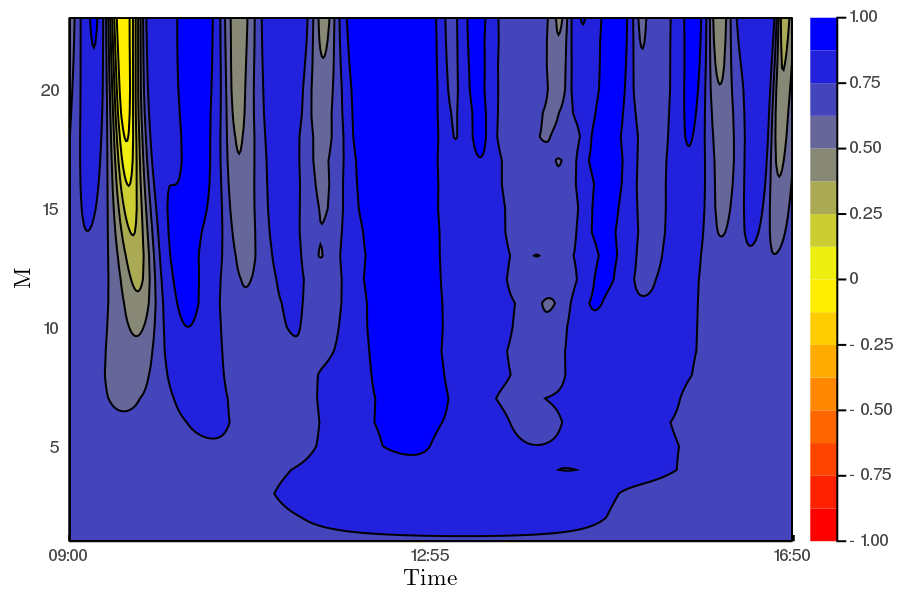}}    \\
    \subfloat[CT, $\Delta t = 300$, 25/06]{\label{fig:Emp_Inst_Epps_dt300:e}\includegraphics[width=0.245\textwidth]{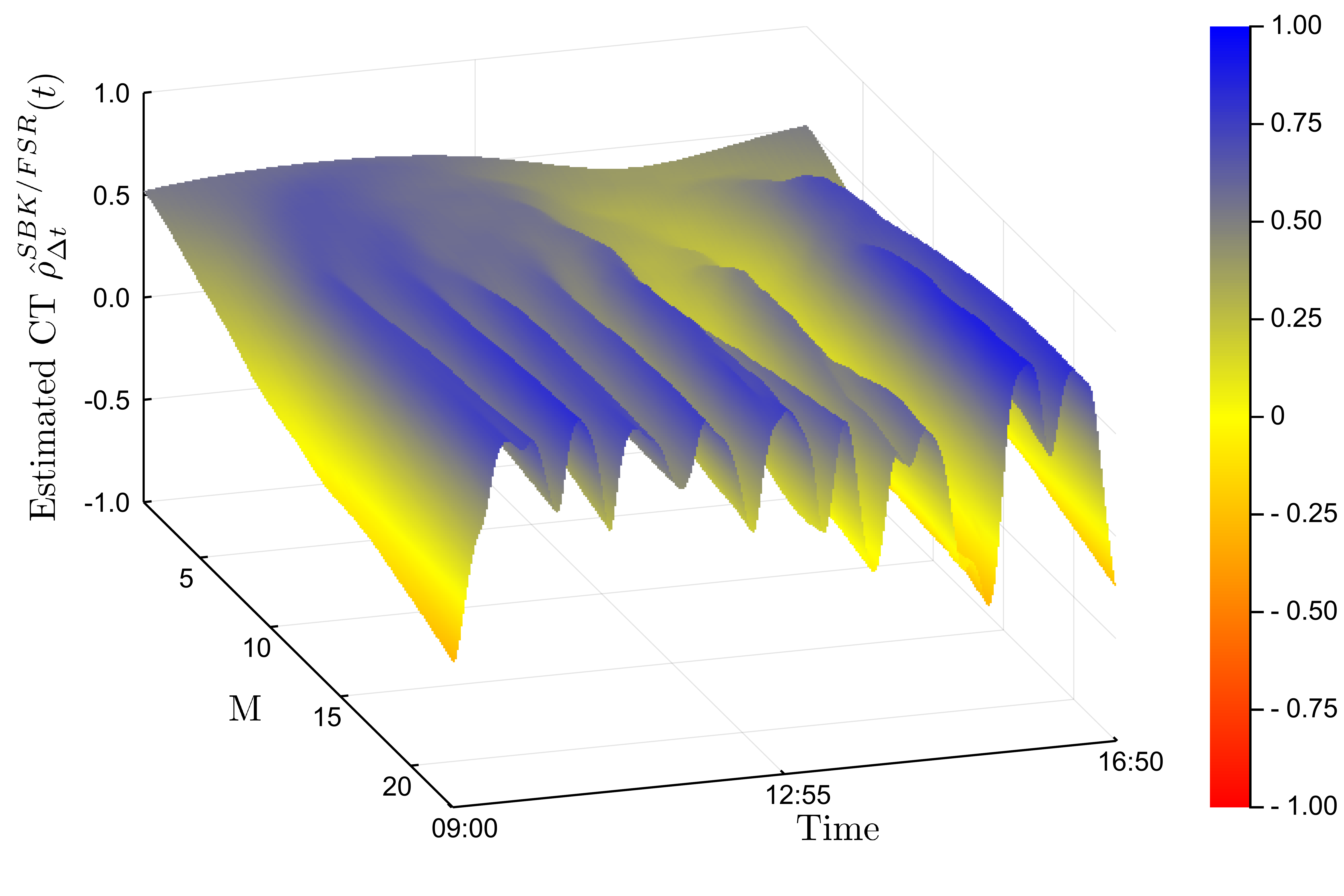}}  
    \subfloat[CT, $\Delta t = 300$, 25/06]{\label{fig:Emp_Inst_Epps_dt300:f}\includegraphics[width=0.245\textwidth]{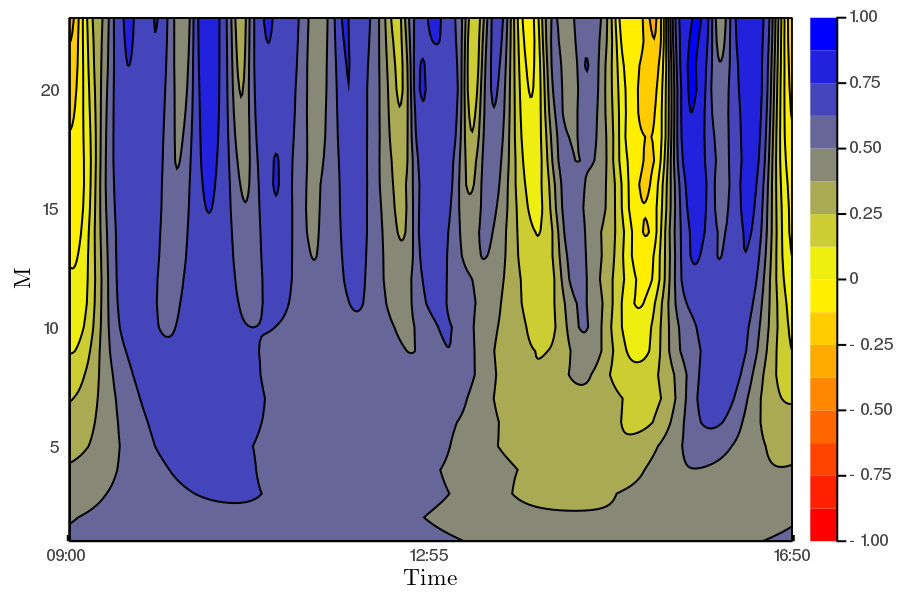}}
    \subfloat[CT, $\Delta t = 300$, 26/06]{\label{fig:Emp_Inst_Epps_dt300:g}\includegraphics[width=0.245\textwidth]{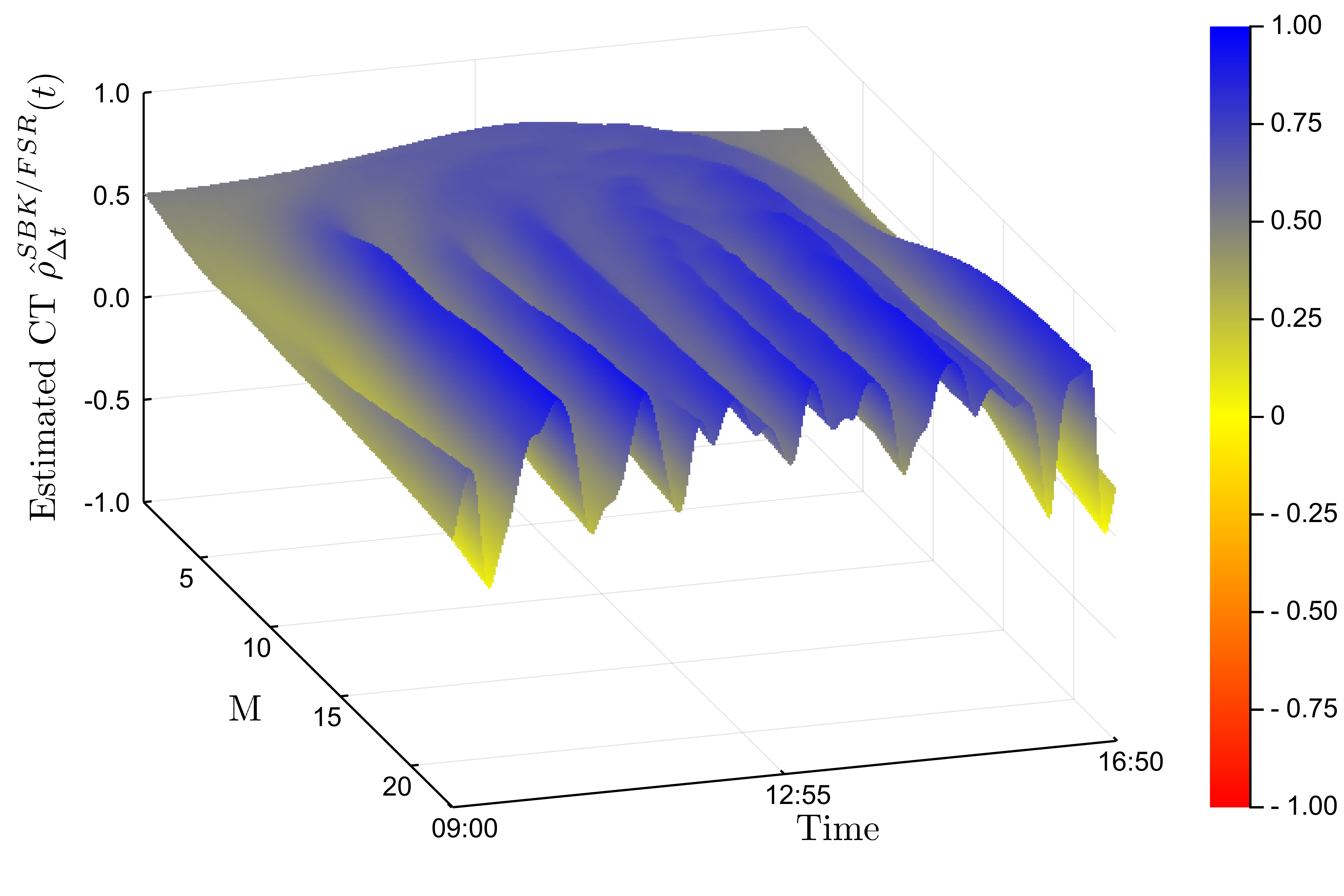}}  
    \subfloat[CT, $\Delta t = 300$, 26/06]{\label{fig:Emp_Inst_Epps_dt300:h}\includegraphics[width=0.245\textwidth]{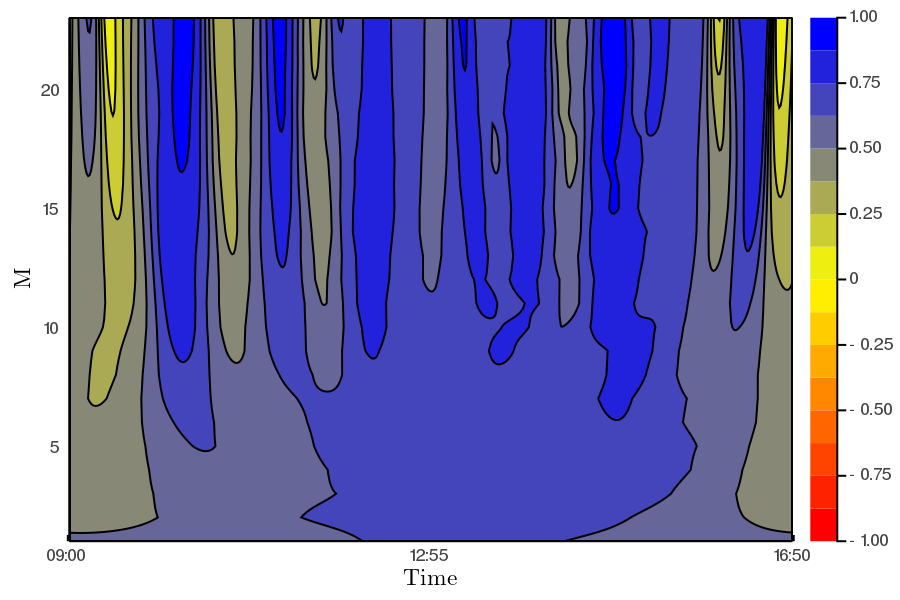}}
    \caption{Here we investigate various values of $M$ ranging from 1 to 23 after accounting for the Epps effect by picking $\Delta t = 300$. The first and second row plots the surface and contour plots of the Malliavin-Mancino and Cuchiero-Teichmann for fixed $\Delta t = 300$ and varying $M$ respectively. The first and second columns are the surface and contour plot for Tuesday the 25/06/2019; similarly, the third and fourth columns are the surface and contour plot for Wednesday the 26/06/2019. We see that larger $M$ allows us to achieve higher peaks and lower troughs but at the cost of additional fluctuation.} 
\label{fig:Emp_Inst_Epps_dt300}
\end{figure*}

The Epps curves in \Cref{fig:Emp_Int_EppsCurves} for Tuesday the 25/06/2019 and Wednesday the 26/06/2019 reach the saturation level around $\Delta t = 300$. This results in $N = 46$ and therefore the range of $M$ to investigate is from 1 to 23. The first and second row of \Cref{fig:Emp_Inst_Epps_dt300} plots the surface and contour plots of the Malliavin-Mancino and Cuchiero-Teichmann for fixed $\Delta t = 300$ and varying $M$ respectively. The first and second columns are the surface and contour plot for Tuesday the 25/06/2019 respectively; similarly, the third and fourth columns are the surface and contour plot for Wednesday the 26/06/2019 respectively. As mentioned before, since we have no knowledge of what the true correlation structure should resemble, it is unclear how to pick an appropriate $M$. What we do notice is that additional $M$ builds upon the harmonics which increases the fluctuation in the estimates but also allows the estimates to reach higher peaks and lower troughs.

Without a clear method to pick $M$, we simply compare the two estimators for $M=10$. \Cref{fig:Emp_Inst_Epps_Comp_dt300_M10} compares the Malliavin-Mancino (solid lines) and Cuchiero-Teichmann (dashed lines) estimates for $M=10$ and $\Delta t = 300$. Tuesday the 25/06/2019 are in blue while Wednesday the 26/06/2019 are in red. The two estimators obtain relatively similar estimates and both capture the same correlation dynamics but are not the same. First, the Cuchiero-Teichmann estimates are less stable for similar choices of $\Delta t$ meaning the dynamics obtained here can completely change for similar $\Delta t$. Second, without some resolution for the instantaneous Epps effect it remains unclear how one can examine the effect of jumps in the estimators under the presence of asynchrony.

At this stage, the Malliavin-Mancino estimator is the preferred estimator for ultra-high frequency finance. At this scale, the impact of asynchrony takes outweighs the impact of jumps. Therefore, the ability to bypass the time domain and avoiding the need to impute data so that the estimates are more stable makes the Malliavin-Mancino estimator more attractive.

\section{Conclusion}\label{sec:conclusion}

In this paper, we compared the Malliavin-Mancino and Cuchiero-Teichmann Fourier spot volatility estimators for various Stochastic models with and without jumps. In the synchronous case, both estimators recover the entire volatility matrix with high fidelity when there are no jumps. With jumps, both estimates are able to recover the co-volatility but only the Cuchiero-Teichmann faithfully recovers the volatility. In the asynchronous case and investigating small time-scales $\Delta t = 1$, only the Malliavin-Mancino estimator recovers the volatility and both estimators obtain estimates around zero for the co-volatility due to the Epps effect.

% We find that under synchronous observations with no jumps, both estimators recover the entire volatility matrix with high fidelity. Under the presence of jumps, both estimates are able to recover the co-volatility. For the volatility estimates, the Malliavin-Mancino estimates present spikes in volatility caused by jumps while the Cuchiero-Teichmann faithfully recovers the volatility. With the addition of asynchrony using a Poissonian sampling method and investigating small time-scales $\Delta t = 1$, only the Malliavin-Mancino estimator recovers the volatility while the Cuchiero-Teichmann under-estimates the volatility. For the co-volatility, both estimators obtain estimates around zero which is attributed to the Epps effect.

We investigated the impact of various cutting frequencies $M$ and $N$ in the estimators. We find that the reconstruction frequency $M$ plays a key role in the accuracy of the approximation for the spot estimates and that the choice of $M$ should depend on the composition of the true spot parameter of interest. The choice of $M$ needs to achieve a subtle balance between achieving a good approximation without adding too many redundant frequencies. This problem is further exacerbated under the presence of asynchrony, where the large fluctuations appears for smaller choices of $M$. There is currently no clear choice for $M$ under asynchrony, but the choice should be small and must satisfy $M \leq N/2$ to avoid aliasing. 

We demonstrated the instantaneous Epps effect arising from asynchrony under simulation. The time-scales are controlled through $N$ for the Malliavin-Mancino estimator while the previous tick interpolation is applied for the Cuchiero-Teichmann estimator. We find that the Malliavin-Mancino estimates produce stable estimates for different $\Delta t$ because it bypasses the need for interpolation in the time domain; while the Cuchiero-Teichmann estimates are unstable for varying $\Delta t$, because of their use of previous tick interpolation. 

We provide an {\it ad hoc} approach to deal with asynchrony by picking $N$ such that we have a time-scale that reaches the saturation level of the Epps curves. Although the approach is more versatile than the choices of $N$ presented in the literature, which are given for specific cases of asynchrony; it still remains naive (as an approach to correct for the Epps effect) because it can still conceal genuine causes of the Epps effect from statistical ones.

% Although the approach is not as arbitrary for the choices of $N$ presented in the literature, at least for specific cases of asynchrony;

Finally, we apply the estimators to Trade and Quote data from the Johannesburg Stock Exchange for two banking equities. We demonstrate the instantaneous Epps effect in the South African equity market and once again find that the Malliavin-Mancino estimator obtains stable estimates for various $\Delta t$ compared to the Cuchiero-Teichmann estimator. We find that the intraday correlation dynamics between days for the same equity pair can vary significantly. Moreover, interesting dynamics can occur within the day which gets concealed through averaging in the integrated estimates.

We argue the Malliavin-Mancino estimator is the preferred estimator for ultra-high frequency finance when the impact of asynchrony is one of the main concerns, even in the presence of jumps. Crucially, we think that the literature may benefit from an estimator which is robust to jumps, as with the Cuchiero-Teichmann estimator, but implemented in a manner also able to bypass the time domain like the Malliavin-Mancino estimator. This may produce stable estimates under asynchrony that are robust to the impact of jumps. 

Other avenues for future research involve obtaining a better understanding of the instantaneous Epps effect so that a correction for asynchrony may be applied at various time-scales, such as the correction by \cite{PCEPTG2020b} and \cite{MSG2011}, but for the case of the instantaneous correlation. Another potentially interesting area may be to apply the instantaneous correlation estimates into clustering algorithms such as the Agglomerative Super-Paramagnetic Clustering algorithm \citep{LYTG2019} to see if the variation in the intraday correlation dynamics result in different market states compared to the static viewpoint traditionally used.

\section*{Reproducing the Research}

The data used in the research can be found in \cite{PCEPTG2020DATAb}. The results can be reproduced by running the script files in the GitHub resource \citep{PCEPTG2020CODEc}. Additional GIFs for the surface and contour plots can also be found in the GitHub resource.

\section*{Acknowledgements}

The author would like to thank Tim Gebbie and Etienne Pienaar for various valuable conversations and critique. In particular I would like to thank Tim Gebbie for introducing me to this area of research and the various related research problems, and for many detailed technical conversations and help with preparing the paper. All remaining typographical and technical errors are mine.

\bibliographystyle{elsarticle-harv}

\bibliography{PCTG-FIE-2020.bib}

\end{document}